\documentclass[a4paper, 12pt, reqno]{article}
\usepackage{etoolbox} 
\usepackage{etex}
\usepackage{amssymb, amsmath, amsthm, amsfonts, amscd}
\usepackage[dvips]{graphicx}
\usepackage[natbibapa]{apacite}
\usepackage{natbib}
\usepackage{geometry}
\usepackage{pstricks}
\usepackage{pst-grad}
\usepackage{breqn}
\usepackage{egameps}
\usepackage{sgame}
\usepackage{setspace}
\usepackage{psfrag}
\usepackage{url}
\usepackage{ifthen}
\usepackage{verbatim}
\usepackage{eurosym}
\usepackage{fancybox}
\usepackage{enumerate}
\usepackage{wasysym}
\usepackage{datetime}
\usepackage{booktabs}
\usepackage{dsfont}
\usepackage{accents}
\usepackage{enumitem}
\usepackage{titlesec}
\usepackage{soul}
\usepackage{mathbbol}
\usepackage[driverfallback=dvips]{hyperref}
\usepackage{breakcites}
\usepackage{bm}
\usepackage{multirow, makecell, caption}
\usepackage{tikz}

\newcommand{\FO}{\Exp[\Theta_{(1:2)}]}
\newcommand{\SO}{\Exp[\Theta_{(2:2)}]}

\usepackage{pifont}
\newcommand{\cmark}{\ding{51}}%
\newcommand{\xmark}{\ding{55}}%

\usetikzlibrary{decorations.pathreplacing}
\newif\ifnodefs
\nodefstrue %comment out to use private definitions later

\newboolean{inarticle}
\setboolean{inarticle}{true}

\newsavebox{\fsavebox}
\newsavebox{\gsavebox}
\theoremstyle{plain}
\date{}

\renewcommand{\dateseparator}{-}
\newcommand{\todayiso}{\twodigit\day \dateseparator \shortmonthname[\month]\dateseparator \the\year }

\xdefinecolor{DarkBlue}{rgb}{0,0.08,0.45}
\xdefinecolor{lavender}{rgb}{0.8,0.6,1}
\xdefinecolor{olive}{cmyk}{0.64,0,0.95,0.4}
\xdefinecolor{majorelleblue}{rgb}{0.38, 0.31, 0.86}
\xdefinecolor{munsell}{rgb}{0.0, 0.5, 0.69}

\newtheorem{t_1}{Theorem}
\newtheorem{t_2}{Theorem}

\newtheorem{t_4}{Theorem}
\newtheorem{t_5}{Theorem}

\newtheorem{t_7}{Theorem}

\newtheorem{remark}[t_2]{Remark}

\newtheorem{definition}[t_4]{Definition}
\newtheorem{lemma}[t_5]{Lemma}
\newtheorem{proposition}[t_1]{Proposition}
\newtheorem*{proposition*}{Proposition}
\newtheorem*{lemma*}{Lemma}
\newtheorem{corollary}[t_7]{Corollary}
\newtheorem*{corollary*}{Corollary}

\definecolor{tomgreen}{rgb}{.0,.5,.0}
\newcommand{\fric}[2]{\,\!\ensuremath{^{#1}\!/_{#2}}}

\newcommand{\frad}[2]{\displaystyle\frac{#1}{#2}}

\newcommand{\dint}{\displaystyle\int}

\cornersize{.1}

\newcounter{ex_1}

\definecolor{LightRed}{rgb}{1.0,0.5,0.5}

\newcommand{\eps}{\varepsilon}

\DeclareMathOperator{\Exp}{\mathds{E}}%
\DeclareMathOperator*{\argmax}{arg\,max}

\makeatletter
\newlength{\myFootnoteWidth}
\newlength{\myFootnoteLabel}
\renewcommand{\@makefntext}[1]{%
  \setlength{\myFootnoteWidth}{\columnwidth}%
  \addtolength{\myFootnoteWidth}{-\myFootnoteLabel}%
  \noindent\makebox[\myFootnoteLabel][r]{\@makefnmark\ }%
  \parbox[t]{\myFootnoteWidth}{#1}%
} \makeatother

\newcommand{\captionfonts}{\small}

\makeatletter  % Allow the use of @ in command names
\long\def\@makecaption#1#2{%
  \vskip\abovecaptionskip
  \sbox\@tempboxa{{\captionfonts #1: #2}}%
  \ifdim \wd\@tempboxa >\hsize
    {\captionfonts #1: #2\par}
  \else
    \hbox to\hsize{\hfil\box\@tempboxa\hfil}%
  \fi
  \vskip\belowcaptionskip}
\makeatother   % Cancel the effect of \makeatletter

\renewcommand{\emph}[1]{{\it{#1}}}

\makeatletter
\patchcmd{\NAT@test}{\else \NAT@nm}{\else \NAT@nmfmt{\NAT@nm}}{}{}

\DeclareRobustCommand\citepos
  {\begingroup
   \let\NAT@nmfmt\NAT@posfmt% ...except with a different name format
   \NAT@swafalse\let\NAT@ctype\z@\NAT@partrue
   \@ifstar{\NAT@fulltrue\NAT@citetp}{\NAT@fullfalse\NAT@citetp}}

\let\NAT@orig@nmfmt\NAT@nmfmt
\def\NAT@posfmt#1{\NAT@orig@nmfmt{#1's}}

\makeatother

\newcommand{\gcol}[1]{{\color{gray}{#1}}}

\begin{document}\onehalfspacing
%%%----------------------------------------------------------------
\title{{\huge Contesting fake news \thanks{Thanks for helpful comments to Pierre Fleckinger, J{\"{o}}rg Franke, Qiang Fu, Alex Gershkov, Rudi Kerschbamer, Bettina Klose, Dan Kovenock, Jianpei Li, Ella Segev, Alberto Vesperoni, Jasmin Wachter, Huseyin Yildirim, and conference/workshop/seminar participants in Austria, Chile, China, France, Germany, Israel, Portugal, Sweden, and the United Kingdom.  $^{\S }$University of Vienna, Department of Business Decisions and Analytics, 1090 Vienna, Austria, \texttt{daniel.rehsmann@univie.ac.at}. $^{\dag}$Universit\'{e} de Grenoble, GAEL, 38040 Grenoble Cedex 9, France, \texttt{beatrice.roussillon@univ-grenoble-alpes.fr}. $^{\natural }$University of Klagenfurt, Department of Economics, 9020 Klagenfurt, Austria, \texttt{paul.schweinzer@aau.at}. (\todayiso).}}}
\author{Daniel Rehsmann$^\S$ \and B{\'{e}}atrice Roussillon$^\dag$ \and Paul Schweinzer$^\natural$}
\maketitle\thispagestyle{empty}
%%%----------------------------------------------------------------
\begin{abstract}\singlespacing
    \noindent We model competition on a credence goods market governed by an imperfect label, signaling high quality, as a rank-order tournament between firms.  In this market interaction, asymmetric firms  jointly and competitively control the aggregate precision of a label ranking the competitors' qualities by releasing individual information.  While the labels and the aggregated information they are based on can be seen as a public good guiding the consumers' purchasing decisions, individual firms have incentives to strategically amplify or counteract the competitors' information emission, thereby manipulating the aggregate precision of product labeling, i.e., the underlying ranking's discriminatory power.  Elements of the introduced theory are applicable to several (credence-good) industries that employ labels or rankings, including academic departments, ``green'' certification, movies, and investment opportunities.\\
    
    \noindent JEL: \textit{C7, D7, H4, M3.}\\
    
    \noindent Keywords: \textit{Labeling, Credence goods, Contests, Product differentiation.}
\end{abstract}
%%%---------------------------------------------------------------------------------------

\section{Introduction}\onehalfspacing
Contests and tournaments are typically modeled such that a given ``black box'' technology ranks some dimension of the contestants' competitive efforts, expenditures, or qualities.  The exact properties of this ranking technology, such as the precision with which the underlying strategic variables are translated into ranks, are usually kept exogenous.  This paper formalizes a novel type of interaction that allows contestants to strategically and jointly control the precision of the employed ranking technology through costly ``marketing'' outlays.  Intuitively, creating an imprecise ranking is viewed as less costly than the assembly of the perfect ranking embodied in, for instance, auction models, which can translate even slight differences in monetary bids into significant changes in assignment probabilities.

To put this idea to work, we consider a market of vertically differentiated goods or services with a credence or experience aspect.\footnote{``Credence qualities are those which, although worthwhile, \emph{cannot be evaluated in normal use}.  Instead, the assessment of their value requires additional costly information.  In contrast to \emph{experience goods}, this utility is difficult to gauge even after consumption.  An example would be the claimed advantages of the removal of an appendix, which will be correct or not according to whether the organ is diseased.'' \citep{Darby_Karni73}.  In \citepos{Dulleck_Kerschbamer_Sutter09} taxonomy of credence goods, this paper models the case of \emph{label credence goods} in which ``\emph{consumers know what they want or need, but observe neither what they get nor the utility derived from what they get.''}}  In this market, consumers cannot discern different qualities of some credence good without further information.  A simple, ordinal ranking (i.e., labeling) of these qualities or goods would, however, be helpful to inform the public's consumption decisions.

In our model, we assume that the precision of such a ranking, i.e., the probability with which the best product is ranked first, is controlled by the \emph{aggregated information} released by the competing firms.
Such a ranking benefits the firms and is assumed to spontaneously arise as firms release potentially conflicting, individual information.  Formally, this ranking translates market-provided information into probabilities of which product quality is ranked first, relative to the other products.  Hence, in the modeled strategic interaction, a firm's decision is how much costly information to release to positively or negatively manipulate the aggregate precision of the product ranking, i.e., the industry's labeling ``standard.''

In terms of novel economic results, we show \emph{first} that although the firms' individual information emission depends on the (consumer-unknown) product qualities, the total sum of market-provided information is constant, irrespective of the firms' product qualities, if information emission cost is quadratic.  Hence, in this case, consumers cannot learn about firms' qualities from the observed aggregated information.  We restrict attention to this quadratic cost of information provision for most of the paper.
\emph{Second}, for any interaction on the defined labeled credence goods market, a threshold measure of consumer heterogeneity exists below which both firms supply ranking-precision enhancing information.  Beyond this threshold, the lower-quality firm turns to obfuscation and supplies what we interpret as ``fake news.''
\emph{Third}, we introduce and discriminate between two effects of increased labeling precision: The first \emph{differentiation effect} allows firms to vertically differentiate and escape Bertrand competition.  The second \emph{categorization effect} makes mislabeling less likely to benefit (hurt) the higher (lower) quality firm.  The interplay between these two effects determines the firms' interaction on label credence goods markets.
\emph{Fourth}, we provide a set of welfare results showing under which circumstances the introduction of credence good labeling benefits (which) market participants.  Finally, we characterize a simple regulatory pricing scheme that ensures that adopting labels unambiguously improves the welfare of all market participants.

The rest of this introduction provides a motivating example in which aggregated market information emission affects a product-ranking standard.  Consider a ranking of academic departments to guide, for instance, ex-ante uninformed students on the relative merits of potential future almae matres.\footnote{The United Kingdom's \href{https://www.ref.ac.uk}{Research Excellence Framework} ranks individual departments based on criteria established by consultation of Universities UK and the University \& College Union (among others).  Hence, the information provided by universities is aggregated into a common ranking standard.}  Two departments are asked on the basis of which factors they wish to be ranked.  Suppose one department is highly successful in research. Consequently, it may insist that the number of top-five publications enters the ranking prominently. In contrast, another department that excels in student satisfaction will want this information included in a nationwide ranking of departments.  If the ranking is supposed to capture research quality, the number of high-quality publications is more informative than student satisfaction.  Hence, a ranking based exclusively on the number of top publications is ``more precise'' in picking out a high research-quality department.  Conversely, if educational experiences are to be evaluated, the satisfaction data may be more helpful, and including the number of top-five publications makes such a ranking ``less precise.''

We view the contribution of this paper as twofold.  On the theoretical side, we introduce a novel family of contests that endogenize the degree of discrimination on which the underlying relative ranking is built.  This makes applications to (partial) credence or experience goods markets possible in which consumer-impenetrable product descriptions are translated into a simple ordinal ranking or label.  While the credence-goods literature, in general, is well-developed, very few papers study this ubiquitous case of labeled credence goods.  In our stylized application, we integrate both market sides into a simple standard equilibrium model of vertical product differentiation in which consumer demand reacts endogenously to the firms' aggregated choice of information release.  This allows us to analyze the competitive effects of the variation of ranking precision in label credence markets.

\subsection*{Related literature}

The idea that, in many circumstances, rewards based on a mere ranking of competitors can achieve socially beneficial outcomes is due to \citet{Lazear_Rosen81} and the contest literature they initiated.  For detailed and recent surveys of this literature, see, for instance, \citet{Corchon_Serena18} and \cite{Fu_Wu19}.  To our knowledge, the present study is the first contribution that allows contestants to endogenously control the precision of the underlying ranking technology in a strategic fashion.

However, a small set of papers endogenize some aspects of a ranking's precision into a contest.
\citet{Michaels88} is, as far as we know, the first to allow a ``monopoly politician'' the ability to set the discriminatory power of the Tullock contest success function to extract rents from symmetric constituents optimally.
Into a similar setup, \citet{Dasgupta_Nti98} add a designer's intrinsic valuation of the prize in addition to the valuation of the competitors.
\citet{Wang10} allows for two asymmetric contestants, deriving the designer's optimal choice of accuracy, depending on the asymmetric contestants' ability spread.
\citet{Yildirim15} models accuracy as the elasticity of contestants' efforts and derives comparative statics related to the heterogeneous players' payoffs.
\citet{Ewerhart17b} derives further revenue rankings for asymmetric Tullock contests, depending on an employed decisiveness (or discriminatory power) parameter.
\citet{Bruckner_Sahm23} explore the optimal accuracy choice problem in multistage political competitions, finding that ``a decisive primary might actually decrease the chances of winning the general election.''
\citet{Deng_Gafni_Lavi_Lin_Ling23} allow for the use of contest precision as a competition instrument between contest organizers who compete to attract contestants among their potentially heterogeneous contests.
%While all of the above relate precision to the discriminatory power of the Tullock success function, \citet{Polishchuk_Tonis13} allow for a larger class of rankings to show that the Tullock contest success function are revenue-optimal provided that contestants possess independent private values.
None of these papers allows for the endogenous control of ranking precision as a competitive dimension between asymmetric players.  Allowing for a significantly richer class of ranking technologies than explored previously, we embed this strategic competition into a vertically differentiated, label credence goods market.

The industrial organization literature on (competitive) labeling is well-developed and has been surveyed by \citet{BergesSennouEA_04} and \citet{Sheldon17}.\footnote{More distantly, our paper also relates to the literature on information disclosure and unraveling \citep{Milgrom08}, as well as to the field of standard setting \citep{Yang23}.  Some questions asked in the literature on the control of information flows overlap with ours, albeit in the differential setting of political campaigning \citep{Gul_Pesendorfer12}.}  The papers closest to our idea of labeling are \citet{Roe_Sheldon07}, \citet{Bonroy_Constantatos15}, \citet{Baski_Bose17}, and \citet{Scott_Sesmero22}.\footnote{\citet{Lehmann-Grube97} is, in some sense, diametrically opposite to our paper in investigating pure quality competition in a vertically differentiated market.  The recent \citet{Bizzotto_Harstad23} discusses how labels and certification thresholds influence competition among firms that are able to adjust their capacities to achieve high quality before some certification threshold is set.}  They use a market-share approach which can be interpreted as a fixed-precision contest.  The link to our contribution is that the degree of label-induced vertical product differentiation depends on the information disclosed by the label.
\citet{Roe_Sheldon07} analyze how the practical implementation of a label (e.g., mandatory/voluntary, continuous/discrete, using a private certifier or public agency) can affect the size and distribution of surplus created in a vertically differentiated credence market.\footnote{Recent interest in credence goods has been spurred by applications to competition policy, health care, and the regulation of legal counseling \citep{Chen_Li_Zhang22}.  Comprehensive surveys include \citet{Dulleck_Kerschbamer06}, recently updated by \citet{Balafoutas_Kerschbamer20}.  The latter refer to our model environment of goods or services with ``unobservable attributes that remain undetected even after consumption'' as \emph{label credence goods}.  We are unaware of previous applications of contest-driven consumer demand to the analysis of competitive (label) credence or experience goods markets.} In their models, governments can influence quality disclosure and the information a label communicates.  Hence, a firm's strategic choice is to decide whether or not to hire a private certifier on top of some existing governmental labeling. This differs from (and complements) the approach of the present paper, which varies the level of information a label transfers to the consumers.
\citet{Bonroy_Constantatos15} survey a variety of industrial organization models investigating how different label implementations affect welfare.  They discuss questions of labeling policy with implications on firms' lobbying activities and incentives to develop labels of particular forms and stringency.
\citet{Baski_Bose17} discuss how consumers' quality overestimation affects firms' incentives to improve product quality in a vertically differentiated credence goods market governed by a perfect label.
\citet{Scott_Sesmero22} study the efficiency and distributional effects of consumers' misperception of product quality---contrasting with our imperfect rankings---in a vertically differentiated food market, both theoretically and empirically. They show that information-based policies aimed at curbing quality misperception (e.g., stricter labeling policies, nudging, changes in the labeling format) may have deleterious effects on efficiency and, perhaps most importantly, hurt the consumers they strive to protect.
The competitive ranking precision aspects introduced in the present paper are not included in any of these contributions.  They are, as far as we know, a novel approach to the labeling problem.

We define the formal structure of the market interaction in the following Section~\ref{model_sec} and characterize the firms' equilibrium behavior in Section~\ref{analysis_sec}, which also contains all main results including elements of a welfare analysis.  We illustrate our results through several examples in Appendix~\ref{example_sec}, equilibrium existence is established in Appendix \ref{existence_app}, and all proofs can be found in Appendix \ref{appendix}.

\section{Model of a labeled credence market}\label{model_sec}

\subsection{Supply side}

There are two risk-neutral firms, $\mathcal{N}=\{1,2\}$, each of which produces a good of quality $\theta_i$, $i \in \mathcal{N}$. These qualities are assumed to be independently distributed according to $\theta_i \sim F_{[0,\bar{\theta}]}$, $\bar{\theta} \in \mathds{R}_{++}$, with continuous and strictly positive density $f(\theta_i)$.  We write $\bm{\theta}=(\theta_1,\theta_2)$ for the quality vector and, without loss of generality, re-index firms such that $\theta_1\geq \theta_2$.  We assume that realizations of qualities are commonly known among the two firms but only the distribution of qualities $F$ is known to the consumers.
Denoting the mean and variance of the quality distribution $F$ by $(m,\sigma^2)$, we assume that the following dispersion property holds:\footnote{The standard restriction to the class of increasing failure rate distributions fails to provide sufficient conditions for our purposes of bounding ratios of expected order statistics. \eqref{ass1} is satisfied for a broad class of distributions, as e.g., for the uniform distribution (as limiting case), for the standard triangular distribution,  for standard values of the truncated normal distribution, %with $m=\bar{\theta}/2$ and $\sigma>\bar{\theta}/5$,
the log-normal distribution with appropriately shifted mean-variance pair, and the beta distribution with parameters $\alpha,\beta \in [1,2)$. %$1\leq\alpha<2$, $1\leq\beta<2$.
It is violated by U-shaped distributions as, e.g., the beta distribution with $\alpha,\beta <1$.} %$\alpha<1$, $\beta<1$.}
\begin{dmath}[number={A1}]\label{ass1}
    \frac{m+\sigma/\sqrt{3}}{m-\sigma/\sqrt{3}}\leq 2. %\tag{A1}
\end{dmath}
For simplicity, production and distribution of the goods are assumed to be costless.  Once the good is produced, there is nothing a firm can do to alter its quality.
While qualities are fixed, firms can release (dis-)information $\rho_i \in \mathds{R}$ allowing the two products to be ranked by means of a label.  The absolute sum $r=|\rho_1+\rho_2|$ of this emitted information is observed by the consumers and determines the aggregate precision used to rank the products.\footnote{\citet{Dulleck_Kerschbamer06,Chen_Li_Zhang22} discuss the incentives of strategic experts in credence markets.  Since our focus is on the firms' information emission, we view the ranking as emerging spontaneously through the firms' activities.  \citet{White18} provides an overview of the literature on rating agencies, including several examples in which rankings arise from the (strategic) operations of market intermediaries.}$^,$\footnote{Allowing consumers to observe the sign of the aggregate information introduces additional antisymmetric equilibria on the negative information orthant (in which rational consumers invert the observed rankings) but otherwise adds little insight to the analysis.}
This assumption effectively creates a (labeled) credence goods market and reflects the idea that, although consumers may be unable to directly observe or credibly interpret the individual product information provided by firms, $\rho_i$, they may have access to aggregations, news reports, or summarily market statistics which accurately represent the overall state of the industry concerned.\footnote{An example of a real-world aggregation mechanism fitting the strategic information transmission aspect of our model is the London Inter-Bank Offered Rate (Libor).  The British Banking Association's short-term interest rate Libor was compiled daily as a trimmed average from banks' individual reports on their costs to borrow funds from each other.  In use since 1969, its importance diminished sharply after the 2012 Libor Scandal which exposed systematic fraud and collusion resulting in strategic rate manipulation \citep{Hou_Skeie13}.  Prior to that, the press habitually referred to the Libor as ``the most important number in the world.'' (\href{https://www.reuters.com/markets/rates-bonds/curb-use-dollar-libor-alternatives-fed-rate-says-watchdog-2023-07-03/}{Reuters, 3-July-2023})}

The aggregate observed market information, $r$, affects the extent of product differentiation in the consumer market and, thus, the firms' expected rank-dependent revenues represented by the winner's prize, $P^{1}(r)$, and loser's prize, $P^{2}(r)$, in the contest for being ranked first.
More precisely, we assume that firm $i \in \mathcal{N}$ maximizes
\begin{dmath}\label{gen-max3_equ}
    \max_{\rho_i}\ u_i(\bm{\theta},r) = q_i(\bm{\theta},r) P^1(r) + (1-q_i(\bm{\theta},r)) P^2(r) - c(|\rho_i|)
\end{dmath}
in which $q_i(\bm{\theta},r = |\rho_1+\rho_2|)$ is player $i$'s probability of being ranked first. The firms' information dissemination cost, $c(|\rho_i|)$, is assumed to be symmetric, strictly increasing, and strictly convex with $c(0)=0$.  

For $i\in \mathcal{N}$ and $j=3-i$, we make the following assumptions on the imperfect ranking of the firms' qualities $q_i(\bm{\theta},r)$ which we collectively refer to as $\mathcal{Q}$:\footnote{\label{fn:fn-q-def}An example of such a function $q(\bm{\theta},r)$ is the Logistic form:
\begin{dmath*}
    q(\bm{\theta}, r) = \frac{1}{1 + \exp^{-\kappa(\theta_1, \theta_2; r)}}
\end{dmath*}
in which $\kappa(\theta_1, \theta_2; r)$ can implement various well-known contest success functions.  For instance, ratio-based rankings with $\kappa_1 = r(\log(\theta_1) - \log(\theta_2))$ \citep{Tullock80}, difference-based rankings with $\kappa_2 = 2 \operatorname{arctanh}(2(\theta_1-\theta_2) r)$ for $0\leq \fric{1}{2}+r(\theta_1-\theta_2)\leq 1$ \citep{Che00}, a piecewise-constant version of the all-pay auction with $\kappa_3 = \operatorname{sgn}(\theta_1-\theta_2)\log(r+1)$, and a noise-based ranking with $\kappa_4 = \log (2/(1 + \operatorname{erf}((-r(\theta_1-\theta_2))/(\sqrt{2}))) - 1)$ \citep{Lazear_Rosen81}.  Concrete numerical examples of these are explored in Appendix \ref{example_sec}.
    }
    \begin{enumerate}[label=(Q\arabic*), align=left, leftmargin=4em]
        \item the ranking of firms' qualities $\bm{\theta}$ is observable and verifiable with $q_i(\bm{\theta},r)+q_j(\bm{\theta},r)=1$;
              
        \item for $r>0$, $q_i(\bm{\theta},r)$ is strictly increasing in $\theta_i$;
              
        \item $q_i(\bm{\theta},r)$ is continuous, strictly increasing in $r=|\rho_i+\rho_j|$ if $\theta_i>\theta_j$, and $q_i(\bm{\theta},0)=\fric{1}{2}$;
              
        \item $q_i(\bm{\theta},r)=\fric{1}{2}$ for $\theta_i=\theta_j$; hence, $\partial q_i(\theta_1=\theta_2,r)/\partial r = 0$;
              
        \item $q_i(\bm{\theta},r)$ is sufficiently continuously differentiable in $\theta_i$ and has at most one inflection point at fixed $\theta_j>0$, with $\partial^2 q_i/\partial \theta_i^2 \geq 0$ for $\theta_i \leq \theta_j$ and $\partial^2 q_i/\partial \theta_i^2 \leq 0$ for $\theta_i \geq \theta_j$;\footnote{We adopt the convention that curvature changes at an inflection point.  Therefore, the piecewise linear functions we define in appendix-subsection \ref{Stepwise-constant_example_sec} have no inflection point in the ranges of interest.}
              
        \item $q_i(\bm{\theta},r)$ is sufficiently continuously differentiable in $r$ and satisfies $\partial^2 q_i/\partial r^2 \leq 0$ for $\theta_i \geq \theta_j$.
    \end{enumerate}
    
    Appendix \ref{example_sec} explores several examples of ranking technologies satisfying these properties.  Some assumptions are relaxed to accommodate our examples of a difference-based ranking of subsection \ref{linear_example_sec} as well as the piecewise constant form of subsection \ref{Stepwise-constant_example_sec}.  
    Moreover, section \ref{others_example_sec} provides examples for the wider applicability of the idea of precision contests in a simpler setting, beyond the endogenised market structure defined in the following subsections.

    \subsection{Demand side}
    
    A unit mass of consumers, each with demand for a single good, is represented through a distribution of valuations $\mu \sim G_{[0,s]}$, $s\in \mathds{R}_{++}$, with continuous and strictly positive density $g(\mu)>0$.\footnote{In general, there are technical problems associated with the use of a continuum of independent random variables.  These play no role in our analysis and could be resolved along the lines discussed by \citet{Lang19}.}  Throughout the analysis of the labeling application, we restrict attention to the case in which $G_{[0,s]}$ follows a uniform distribution.  We interpret the upper bound, $s$, as a measure of the consumers' preference  heterogeneity.
    The utility of a type-$\mu$ consumer is assumed to be quasi-linear with
    \begin{dmath}\label{consumer-utility_equ}
        v(\mu,\theta) \hiderel{=} \mu \tilde{\theta} - \tilde{p} %\hiderel{\geq} 0
    \end{dmath}
    in which the price $\tilde{p}$ is paid for a product of (expected) quality $\tilde{\theta}$.  Outside options are zero.
    
    Apart from these individual preferences, the main element informing consumer demand is the commonly known outcome of a public ranking of the qualities $\theta_1$ and $\theta_2$, arising spontaneously following the firms' observed release of information $r=|\rho_1+\rho_2|$.\footnote{In this paper, consumers know that the assigned ranks or labels are correct only with some probability $q(\bm{\theta},r)$.  This contrasts with \citet{Scott_Sesmero22} who allow for consumers to \emph{misperceive} labels, i.e., confuse the first with the second label, resulting in suboptimal purchasing decisions.} In the absence of a ranking for the underlying credence good (or the case of observed $r=0$), consumers cannot distinguish between products.  Products of identical expected qualities are in equilibrium then sold under Bertrand competition at the same price with the firms sharing expected profits equally.
    
    Consumers do not know the realization of product qualities but form expectations of these, based on the commonly known distribution $F$.  They observe the absolute market information $r=|\rho_1+\rho_2|$ which they know to correspond to the ranking precision, i.e., the probability with which the first-ranked (or labeled) good actually has the higher quality.  Consumers cannot observe the individually emitted components $\rho_i$.%\footnote{Such a incredible or impenetrable nature of individual information releases can be observed, e.g., in the aggregation mechanism underlying the Libor interest rate. Reuters Group PLC which is in charge of aggregating the banks individual interest rate reports, verifies the credibility of reports only in an aggregate way, i.e., by expelling outliers before calculating the average. The aggregation mechanism therefore crucially depends on ``banks to tell the truth about their borrowing rates'', since `` actual rates at which banks borrow from each other are known only to the lenders and borrowers, and possibly to their brokers.''(\href{https://www.wsj.com/articles/SB120831164167818299}{The Wall Street Journal, 16-April-2008}).}

    \subsection{Labeling contest}
    
    In our labeling contest, a firm's prize for coming first (second) is the expected revenue resulting from consumer demand captured by the first-labeled (second-labeled) product, given the observed ranking governed by $\left(q_1(\bm{\theta},r), 1-q_1(\bm{\theta},r)=q_2(\bm{\theta},r)\right)$.  To determine this demand, we use a standard vertical product differentiation model in which the expected quality is signaled through the product rank.\footnote{The classic vertical-differentiation reference is \citet{Gabszewicz_Shaked_Sutton_Thisse81}, succinctly summarized by \citet{Tirole88}.}  Given their mutually known qualities, first, each firm determines (unannounced) optimal pricing strategies $p^1(r), p^2(r)$, with superscripts corresponding to possible ranks.\footnote{Since consumers' valuations only depend on the product ranking---everything else being uninformative---each firm faces the same optimization problem for choosing rank-dependent pricing strategies and thus selects the same symmetric equilibrium vector of conditional product prices.  Hence, we drop firm subscripts on prices and revenues.  Optimal pricing strategies only depend on consumer expectations and are independent of individual qualities.}  This defines the labeling contest, with prizes $P^1(r)$ and $P^2(r)$, in which the firms choose individual information release, $\rho_i$.
    
    It is useful to define the consumers' expectation of the first- and second-ranked product qualities, given an observed ranking of precision $r$. This precision-dependent expectation is denoted by:
    \begin{equation}\label{cons-exp2_equ_dr}\parbox{.9\linewidth}{%
            \setlength{\belowdisplayskip}{0pt} \setlength{\belowdisplayshortskip}{0pt}
            \setlength{\abovedisplayskip}{0pt} \setlength{\abovedisplayshortskip}{0pt}
            \begin{dgroup*}\begin{dmath*}
                    \Lambda^1(r) = \Exp\left[ q_1 (\bm{\theta},r) \theta_1 + (1-q_1 (\bm{\theta},r)) \theta_2 \hiderel{\mid} r \right],
                \end{dmath*}
                \begin{dmath*}
                    \Lambda^2(r) = \Exp\left[ q_1 (\bm{\theta},r) \theta_2 + (1-q_1 (\bm{\theta},r)) \theta_1 \hiderel{\mid} r \right].
                \end{dmath*}\end{dgroup*}}
    \end{equation}
    These expectations are based on consumers' beliefs about qualities, conditional on the observed aggregate market information $r$.\footnote{In equilibrium, however, aggregate information $r^\ast$ turns out to be constant with respect to qualities by virtue of Proposition \ref {proposition_1} and an additional assumption on costs \eqref{ass:q-costs}.  Therefore, conumers cannot update their beliefs, and we simplify notation by using unconditional beliefs if there is no danger of confusion.}  Using the \emph{unconditional, prior} joint order probability density, $f_{(1,2:2)}(\bm{\theta})$ of randomly drawn quality $\theta_1$ exceeding quality $\theta_2$, these can be written as:
    \begin{equation}\label{cons-exp1_equ_dr}\parbox{.9\linewidth}{%
            \setlength{\belowdisplayskip}{0pt} \setlength{\belowdisplayshortskip}{0pt}
            \setlength{\abovedisplayskip}{0pt} \setlength{\abovedisplayshortskip}{0pt}
            \begin{dgroup*}\begin{dmath*}
                    \Lambda^1(r) = \int_0^{\bar{\theta}} \int_0^{\tilde{\theta}_1 }
                    \left( q_1(\tilde{\bm{\theta}},r) \tilde{\theta}_1 + (1-q_1(\tilde{\bm{\theta}},r)) \tilde{\theta}_2 \right)
                    f_{(1,2:2)}(\tilde{\bm{\theta}}) d\tilde{\theta}_2  d\tilde{\theta}_1,
                \end{dmath*}
                \begin{dmath*}
                    \Lambda^2(r) = \int_0^{\bar{\theta}} \int_0^{\tilde{\theta}_1}
                    \left( q_2(\tilde{\bm{\theta}},r) \tilde{\theta}_1 + (1-q_2(\tilde{\bm{\theta}},r)) \tilde{\theta}_2 \right)
                    f_{(1,2:2)}(\tilde{\bm{\theta}}) d\tilde{\theta}_2  d\tilde{\theta}_1.
                \end{dmath*}\end{dgroup*}}
    \end{equation}
    In our independent qualities setting, the relevant joint order densities are:
    \begin{dmath}\label{eq-jointorder_highlow-simplified}
        f_{(1,2:2)}(\bm{\theta}) = 2 f(\theta_1) f(\theta_2),\ \theta_1 \hiderel{\geq} \theta_2 \hiderel{\in} [0,\bar{\theta}].
    \end{dmath}
    Given consumers' prior expectations of the two firms' products being ranked first
    \begin{dmath}\label{equ:priors}
        q_1(\bm{\theta},r),\ 1-q_1(\bm{\theta},r)
    \end{dmath}
    and symmetric rank- and precision-dependent pricing strategies $p^1(r),p^2(r)$, a marginal consumer of valuation $\mu$ is, for a particular observation of aggregate information, $r$, indifferent between buying the first- and second-ranked products if
    \begin{dmath}
        \mu \Lambda^1(r) - p^1 = \mu \Lambda^2(r) - p^2,
    \end{dmath}
    resulting in the vector of cutoffs:
    \begin{dmath}\label{sep-types_equ2}
        \hat{\mu} \hiderel{=} \biggl(\hat{\mu}_1^0 \hiderel{=} s,\
        \hat{\mu}_{2}^1 \hiderel{=} \frac{p^1 - p^2}{\Lambda^1(r) - \Lambda^2(r)},\
        \hat{\mu}_{3}^2 \hiderel{=} \frad{p^2}{\Lambda^2(r)}\biggr).
    \end{dmath}
    Because $\hat{\mu}_{3}^2 \geq 0$, the market is not generally fully served. Given these cutoffs, the firms maximize their rank-dependent, \emph{identity-independent} revenues, by choosing optimal pricing strategies %$\hat{p}^1(r)$ and $\hat{p}^2(r)$
    \begin{equation}\label{eqm-prices-def}
        \begin{array}{c}
            p^1(r)^\ast = \displaystyle \argmax_{p^1(r)} \dint^{\hat{\mu}_1^0}_{\hat{\mu}_2^1(p^1(r), p^2(r)^\ast)} g(\mu) d\mu, \\[9pt]
            p^2(r)^\ast = \displaystyle \argmax_{p^2(r)} \dint_{\hat{\mu}_3^2(p^2(r))}^{\hat{\mu}_2^1(p^1(r)^\ast,p^2(r))} g(\mu) d\mu
        \end{array}
    \end{equation}
    giving contest endogenous contest prizes as
    \begin{equation}\label{eqm-profits-def}
        \begin{array}{c}
            %				P^1(r) = p^1(r)^\ast G\left(\hat{\mu}_1^0\right)-p^1(r)^\ast G\left(\hat{\mu}_{2}^1(p^1(r)^\ast, p^2(r)^\ast)\right)
            P^1(r) = p^1(r)^\ast \dint^{\hat{\mu}_1^0}_{\hat{\mu}_2^1(p^1(r)^\ast, p^2(r)^\ast)} g(\mu) d\mu \hiderel{=}
            p^1(r)^\ast\left(G(\hat{\mu}_1^0)-G(\hat{\mu}_2^1(p^1(r)^\ast, p^2(r)^\ast))\right), \\[9pt]
            %				P^2(r) = p^2(r)^\ast G\left(\hat{\mu}_2^1(p^1(r)^\ast, p^2(r)^\ast)\right)-p^2(r)^\ast G\left(\hat{\mu}_{3}^2(p^2(r)^\ast)\right).
            P^2(r) = p^2(r)^\ast \dint_{\hat{\mu}_3^2(p^2(r)^\ast)}^{\hat{\mu}_2^1(p^1(r)^\ast,p^2(r)^\ast)} g(\mu) d\mu \hiderel{=}
            p^2(r)^\ast\left(G(\hat{\mu}_2^1(p^1(r)^\ast, p^2(r)^\ast))-G(\hat{\mu}_3^2(p^2(r)^\ast))\right).
        \end{array}
    \end{equation}
    Notice that these problems are symmetric and firms choose the same rank-dependent pricing strategies in equilibrium.  In the interaction defined in subsection~\ref{time_sec}, firms announce rank-dependent prices ${p^1}$ and ${p^2}$ after consumers observe realized precision $r$ and the product ranks.
    
    A result due to \citet{Gabszewicz_Shaked_Sutton_Thisse81} establishes that such an equilibrium pricing vector exists and has the required properties.
    
    \begin{proposition}\label{comp_prop}
        For any market information $r>0$ and any distribution of consumer tastes $G$ with strictly positive and weakly concave density $g$, there exists an equilibrium vector of pricing strategies $p^1(r)^\ast > p^2(r)^\ast > 0$, provided that the failure rate
        \begin{dmath}\label{ex-cond_equ}
            \frac{g(\mu)}{1-G(\mu)}\text{ is strictly increasing.}
        \end{dmath}
    \end{proposition}
    As a consequence, $P^1(r)$ and $P^2(r)$ are well-defined in \eqref{eqm-profits-def} and can be interpreted as endogenized prizes in a labeling contest in which firms manipulate the precision of the labels which partially inform consumers on the offered qualities.

    \subsection{Timing and information}\label{time_sec}
    
    Both prizes $P^k(r)$ and optimal pricing strategies $p^k(r)^\ast$, $k=1,2$, are functions of the available information, i.e., the ranking precision $r=|\rho_1+\rho_2|$.  We are looking for asymmetric, pure-strategy Bayesian Perfect Nash equilibria in which each firm $i=1,2$ chooses pairs of emitted information and rank-$k$-dependent prices $\left( \rho_i, p^k=p^k(r)^\ast \right)$.  Hence, the interaction timing is:\vspace{6mm}
    
    \begin{center}
        \begin{tikzpicture}[%
                every node/.style={
                        font=\scriptsize,
                        align=center,
                        % Better alignment, see https://tex.stackexchange.com/questions/315075
                        text height=1ex,
                        text depth=.25ex,
                    },
            ]
            % draw horizontal line
            \draw[->] (0,0) -- (13.5,0);
            
            % add vertical ticks
            \foreach \x in {0,2.5,5.5,9,12.5}{
                    \draw (\x cm,3pt) -- (\x cm,0pt);
                }
            
            % place axis labels below
            \node[anchor=north] at (0,0) {$t=0$};
            \node[anchor=north] at (2.5,0) {$t=1$};
            \node[anchor=north] at (5.5,0) {$t=2$};
            \node[anchor=north] at (9,0) {$t=3$};
            \node[anchor=north] at (12.5,0) {$t=4$};
            \node at (14,0) {time};
            
            % add action labels above
            \node[anchor=south,above=1pt] at (0,0) {qualities $\theta_i$\\[-4pt] are drawn};
            \node[anchor=south,above=1pt] at (2.5,0) {firms emit\\[-4pt] information $\rho_i(\boldsymbol{\theta})$};
            \node[anchor=south,above=1pt] at (5.5,0) {ranking $q_i(\boldsymbol{\theta},r)$\\[-3pt] realizes};
            \node[anchor=south,above=1pt] at (9,0) {firms announce rank-\\[-4pt] dependent prices $p^k$};
            \node[anchor=south,above=1pt] at (12.5,0) {consumption \&\\[-4pt] profits realize};
            
        \end{tikzpicture}
    \end{center}
    
    Consumers cannot observe (or credibly interpret) the firms' individually released information, $\rho_i$, but they can observe the absolute amount of aggregated information $r=|\rho_1+\rho_2|$.  Consumers realize that this aggregate information determines the precision with which the ranking of product labels is correct, i.e., corresponds to the true order of qualities.
    Firms, by contrast, are assumed to fully understand the technology the industry is based on and therefore know each others' product qualities $\theta_i$.
    After consumers observe the ranking's aggregate precision and it's realization, firms announce rank-$k$-dependent prices $p^k$.

    \subsection{Off-equilibrium path beliefs}\label{beliefs_sec}
    
    In principle, consumers may observe aggregate information $r$ which is incompatible with expected equilibrium play, i.e., $r \neq r^\ast$.  Similarly, firms' pricing announcements my deviate from equilibrium prices \eqref{eqm-prices-def}, i.e., $\bm{p}\neq \bm{p}^\ast$.
    For these cases, we define consumers' conditional beliefs\footnote{We are grateful to an anonymous referee for pointing out this off-equlibrium belief structure which generalizes over our previous formulation.  The defined beliefs are discussed in relation to our existence arguments in Remark \ref{rem:off-eqm-beliefs}.}
    \begin{dmath}\label{equ:beliefs-gen}
        \tilde{q}(\bm{\theta},r,\bm{p}).
    \end{dmath}
    On the equilibrium path, both prices and beliefs are independent of qualities $\bm{\theta}$ and these beliefs reduce to the priors $q(\bm{\theta},r^\ast)$ from \eqref{equ:priors} as shown in Proposition \ref {proposition_1}, given our assumption on costs \eqref{ass:q-costs}.  Hence, we simplify notation and use the priors \eqref{equ:priors} throughout.
    
    We use the same belief structure also off the equilibrium path, i.e., consumers ignore attempted price and precision manipulations because they know that, in equilibrium, these cannot transfer information about qualities $\bm{\theta}$.  Therefore, we assume that consumers resort to their priors $q(\bm{\theta},r^\ast)$, \emph{whatever} the observed aggregate precision $r$ and announced prices $\bm{p}$.

    \section{Results}\label{analysis_sec}
    
    \subsection{Preliminaries}
    
    We begin the analysis by collecting some simple results which will be used and referred to repeatedly in the subsequent reasoning.
    We denote expected orders by:
    \begin{dmath}\label{def-exp-orders}
        \FO \hiderel{=} \dint_0^{\bar{\theta}} \dint_0^{\tilde{\theta}_1 }
        \tilde{\theta}_1 f_{(1,2:2)}(\tilde{\bm{\theta}})
        d\tilde{\theta}_2  d\tilde{\theta}_1,\
        \SO \hiderel{=} \dint_0^{\bar{\theta}} \dint_0^{\tilde{\theta}_1 }
        \tilde{\theta}_2 f_{(1,2:2)}(\tilde{\bm{\theta}})
        d\tilde{\theta}_2  d\tilde{\theta}_1
    \end{dmath}
    and their sum of by:
    \begin{dmath}\label{def-sum-exp-orders}
        \hat{\theta} = \Exp[\Theta_{(1:2)}]+\Exp[\Theta_{(2:2)}]\hiderel{=}2m
    \end{dmath}
    in which $m$ denotes the mean of the quality distribution $F$.\footnote{Note that $\hat{\theta} = \bar{\theta}$ for all symmetric distributions.}
    
    \begin{lemma}\label{lemma_consumerexp}
        For any distribution of product qualities $F$ with associated positive density $f$, the following ``bookkeeping'' results hold for all $r>0$:
        \begin{dmath}\label{lemma-1-1}
            \Lambda^1(r)+\Lambda^2(r) \hiderel{=} \Exp[\Theta_{(1:2)}+\Theta_{(2:2)}] \hiderel{=} \hat{\theta},
        \end{dmath}
        
        \begin{dmath}\label{lemma-1-2}
            \FO \hiderel{\geq} \Lambda^1(r) \hiderel{>} \hat{\theta}/2 \hiderel{>} \Lambda^2(r) \hiderel{\geq} \SO \hiderel{>} 0,
        \end{dmath}
        
        \begin{dmath}\label{lemma-1-3}
            \lim_{r\to\infty}\Lambda^1(r)\hiderel{=}\FO,\
            \lim_{r\to\infty}\Lambda^2(r)\hiderel{=}\SO,
        \end{dmath}
        
        \begin{dmath}\label{lemma-1-4}
            (\Lambda^1)^{(n)}(r) =\displaystyle\int_0^{\bar{\theta}} \displaystyle\int_0^{\tilde{\theta}_1 }
            (\tilde{\theta}_1-\tilde{\theta}_2)\frad{\partial^n q_1(\tilde{\bm{\theta}},r)}{\partial r^n}
            f_{(1,2:2)}(\tilde{\bm{\theta}}) d\tilde{\theta}_2  d\tilde{\theta_1}\ \forall n \hiderel{\in} \mathds{N}^+.
        \end{dmath}
        Moreover, in case of $r=0$, we have
        \begin{dmath}\label{lemma-1-5}
            \Lambda^1(0)=\Lambda^2(0)\hiderel{=}\hat{\theta}/2.
        \end{dmath}
    \end{lemma}
    
    The following inequality links properties of the underlying quality distribution with (spacings of) expected values of statistical orders and is repeatedly used in our analysis.
    
    \begin{lemma}\label{lemma_spacings}
        For any continuous quality distribution $F$ with mean $m$ and variance $\sigma^2$ bounded on a positive interval, the ratio of expected order statistics satisfies:
        \begin{dmath}\label{lemma_spacings_equ1}
            \frac{\FO}{\SO}\hiderel{\leq}\frac{m+\sigma/\sqrt{3}}{m-\sigma/\sqrt{3}},
        \end{dmath}
        in which $\sigma$ denotes the standard deviation.
    \end{lemma}
    This result motivates assumption \ref{ass1}.  The bound \eqref{lemma_spacings_equ1} is tight for the case of the standard uniform distribution.

    \subsection{The labeling application}\label{labeling_subsec}
    
    To allow for the explicit determination of market prices, cutoffs, and prizes, we restrict attention to uniformly distributed consumer preferences, i.e., $G(\mu)=\mu/s$, for the rest of the paper.
    For these preferences, the maximization problems \eqref{eqm-prices-def} simplify to the optimal, rank-dependent pricing strategies:
    \begin{dmath}\label{ex-prices_equ}
        p^1(r)^\ast \hiderel{=} 2s\dfrac{\Lambda^1(r) \left(\Lambda^1(r)-\Lambda^2(r)\right)}{4\Lambda^1(r)-\Lambda^2(r)},\
        p^2(r)^\ast \hiderel{=} s\dfrac{\Lambda^2(r)\left(\Lambda^1(r)-\Lambda^2(r)\right)}{4 \Lambda^1(r)-\Lambda^2(r)}
    \end{dmath}
    resulting, from \eqref{sep-types_equ2}, in the equilibrium cutoffs:
    \begin{dmath}\label{uniform_cutoffs_equ}
        \hat{\mu}_1^0 \hiderel{=} s,\quad
        \hat{\mu}_2^1 \hiderel{=} s\frac{2\Lambda^1(r)-\Lambda^2(r)}{4\Lambda^1(r)-\Lambda^2(r)},\quad
        \hat{\mu}_3^2 \hiderel{=} s\frac{\Lambda^1(r)-\Lambda^2(r)}{4\Lambda^1(r)-\Lambda^2(r)}
    \end{dmath}
    giving, in turn, the rank-dependent contest prizes as functions of the available information as:
    \begin{equation}\begin{array}{rcl}\label{ex-prizes1_equ}
            P^1(r) & = & 4s\dfrac{\Lambda^1(r)^2\left(\Lambda^1(r)-\Lambda^2(r)\right)}{\left(\Lambda^2(r)-4\Lambda^1(r)\right)^2},           \\
            P^2(r) & = & s\dfrac{\Lambda^1(r)\Lambda^2(r)\left(\Lambda^1(r)-\Lambda^2(r)\right)}{\left(\Lambda^2(r)-4\Lambda^1(r)\right)^2}.
        \end{array}\end{equation}
    Since $\Lambda^1(r)+\Lambda^2(r)=\hat{\theta}$ from \eqref{def-sum-exp-orders} and Lemma~\ref{lemma_consumerexp}, \eqref{lemma-1-1}, this results in contest prizes capturing the labeled market segments:
    \begin{dmath}\label{ex-prizes0_equ}
        P^1(r)\hiderel{=}4s\frac{\Lambda^1(r)^2(2\Lambda^1(r)-\hat{\theta})}{(\hat{\theta}-5\Lambda^1(r))^2}, \quad P^2(r)\hiderel{=}s\frac{(\hat{\theta}-\Lambda^1(r))\Lambda^1(r)(2\Lambda^1(r)-\hat{\theta})}{(\hat{\theta}-5\Lambda^1(r))^2}.
    \end{dmath}
    
\begin{figure}[!ht]
        \begin{center}\footnotesize
            \psfrag{a}{\raisebox{1pt}{$\mu$}}
            \psfrag{b}{$\SO=\bar{\theta}/3$}
            \psfrag{c}{$\FO=2\bar{\theta}/3$}
            \psfrag{d}{$\hat{\mu}_2^1$}
            \psfrag{e}{$P^1$}
            \psfrag{f}{$\theta$}
            \psfrag{g}{$\hat{\mu}_3^2$}
            \psfrag{h}{$P^2$}
            \psfrag{i}{$\hat{\mu}_1^0$}
            \psfrag{j}{$\bar{\theta}$}
            \psfrag{k}{0}
            \scalebox{1}{\includegraphics{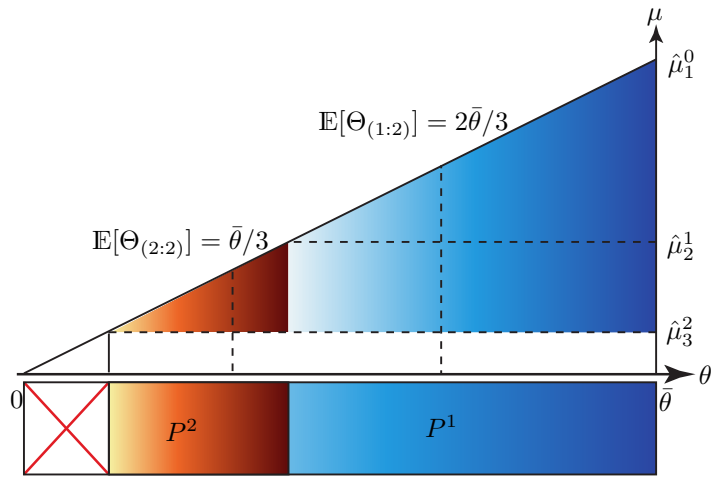}}
            \caption{The labeled consumer market for uniform product qualities.  The red-crossed consumer segment remains unserved.}\label{fig3}
        \end{center}
    \end{figure}

    The endogenous emergence of contest prizes resulting from consumer demand is illustrated in Figure~\ref{fig3}.  Lemma \ref{lemma_prizescon} establishes some properties of the labeled market-segment prizes \eqref{ex-prizes0_equ}.
    
    \begin{lemma}\label{lemma_prizescon}
        For any distribution $F$ of product qualities and uniformly distributed consumer tastes $\mu$, the following properties hold for $r>0$:
        \begin{dmath}\label{lemma-3-1}
            P^1(r) > P^2(r) \hiderel{>} 0,
        \end{dmath}
        
        \begin{dmath}\label{lemma-3-2}
            {P^1}'(r)>0, {P^1}''(r) \hiderel{<} 0 \text{ as well as } {P^1}'(r) \hiderel{>} {P^2}'(r),
        \end{dmath}
        
        \begin{dmath}\label{lemma-3-3}
            {P^2}'(r)>0, {P^2}''(r)\hiderel{<}0, \text{ as well as } {P^1}''(r) \hiderel{<} {P^2}''(r)
        \end{dmath}
        in which only \eqref{lemma-3-3} requires $F$ to satisfy \eqref{ass1}. Moreover, in case of $r=0$, $P^1(0)=P^2(0)=0$.
    \end{lemma}
    
    \begin{remark}[Price competition]
        Notice that, from \eqref{lemma-3-1}, the prizes $P^i(r)$ accruing to competing firms are zero for the (unlabeled) no-information case of $r=0$.  The reason is that Bertrand competition for identically perceived products (albeit of positive expected qualities) drives prices down to marginal cost which is zero in our environment.
    \end{remark}
    
    Notice that, for $r>0$, neither player leaves the market since expected payoffs \eqref{gen-max3_equ} are strictly positive in equilibrium.
    This follows from \eqref{ex-prizes0_equ}, continuity, and the fact that emission of no information is costless.
    
    We are now ready for the equilibrium characterization.
    \begin{proposition}\label{proposition_1}
        Consider player $i \in \mathcal{N}$ with objective \eqref{gen-max3_equ}.  A necessary condition for equilibrium is
        \begin{dmath}\label{lemma_1_equ}
            {P^1}'(\rho_1+\rho_2)+{P^2}'(\rho_1+\rho_2) = c'(|\rho_1|) + c'(|\rho_2|).
        \end{dmath}
    \end{proposition}
    
    \begin{corollary}\label{cor:quadratic}
        Since \eqref{lemma_1_equ} is independent of qualities $\bm{\theta}$, for quadratic costs
        \begin{dmath}\label{ass:q-costs}
            c(\rho) = (\delta/2) \rho^2,\ \delta\hiderel{>}0,
        \end{dmath}
        consumers observe the same aggregate information $r^\ast = \delta(\rho_1+\rho_2)$ in any equilibrium.
    \end{corollary}
    
    In order to prevent consumers from learning and updating their beliefs beliefs about qualities \eqref{cons-exp1_equ_dr} on the basis of the observed $r$, we assume for the rest of the paper that the firms' information emission costs are quadratic \eqref{ass:q-costs}.  As a result, consumers cannot learn about product qualities from the observed aggregated information.  Shutting down this indirect channel allows us to concentrate the subsequent analysis on learning through the realized ranking.
    
    In \eqref{lemma_1_equ}, aggregate information only depends on the ranking technology, consumers' expectations, and the sum of firms' marginal provision costs.  The left-hand side consumer perceptions and, therefore, their willingness to pay, only depend on the known ranking technology and the order statistics of the distribution of qualities, not on the realized draws (that consumers do not know).  For quadratic costs, the right-hand side sums directly to $\delta r$.  This simplifies the analysis considerably and would not generally be the case if, for instance, the firms' information costs would depend on the product qualities or consumers could directly observe or interpret (some statistic of) individually emitted $\rho_i$.  The same invariance does not hold for changes in preference heterogeneity, $s$, since prizes \eqref{ex-prizes0_equ}---and therefore the left-hand side of \eqref{lemma_1_equ}---respond directly to the induced changes in willingness to pay.
    
    The fact that, under quadratic costs, the level of information in the market stays constant, irrespective of the product quality spread, is in line with numerous studies demonstrating how difficult it is to reduce informational asymmetries in credence markets and is therefore interesting in its own right \citep{Dulleck_Kerschbamer06,Balafoutas_Kerschbamer20}.  Evidence shows that companies may want to limit the level of information they provide, and consumers may be willfully ignorant.  Our environment underscores difficulties in information provision in such markets, and the firms' interest to strategically manipulate information disclosure.
    
    \begin{remark}[Positive precision]
        The aggregate precision of the market-provided product ranking is positive in any equilibrium with
        \begin{dmath}\label{remark_3_equ}
            \rho_1 \geq |\rho_2|.
        \end{dmath}
        To see this, consider firm 1's first-order condition
        \begin{dmath}\label{gen-foc01}
            q_1(\bm{\theta},r) {P^1}'(r) + q_2(\bm{\theta},r){P^2}'(r) + \frac{\partial q_1(\bm{\theta},r)}{\partial \rho_1} \left(P^1(r)-P^2(r)\right) - c'(|\rho_1|)=0.
        \end{dmath}
        For $r>0$, we know that $q_1(\bm{\theta},r)>q_2(\bm{\theta},r)$ as well as $P^1(r)>P^2(r)>0$ and ${P^1}'(r)>{P^2}'(r)>0$ from Lemmata \ref{lemma_consumerexp} \& \ref{lemma_prizescon}.  Therefore, it must be the case that $c'(|\rho_1|) > c'(|\rho_2|)$.  Moreover, since prizes are positive and $q_1(\bm{\theta},r)>q_2(\bm{\theta},r)$, it must be the case that equilibrium $\rho_1 > 0$.
        Hence, in equilibrium, ``markets do not lie'' and competition results in the provision of helpful aggregate information to consumers.
    \end{remark}
    
    \begin{remark}[Total competition costs]\label{remark_7}
        The closer types are aligned in competitive situations, the higher are generally total expected contestants' efforts.  This is not the case in our setup because aggregate information is constant in the quality spread although the individual information emissions become more and more asymmetric as types diverge.\footnote{We are grateful to Florian Morath for relaying this observation.}  Since individual information cost is convex while aggregate information is constant, in the present setup, the firms' total competition-incurred information costs are \emph{minimized} with equal qualities.  Intuitively, since only the ranking precision and not the underlying qualities can be manipulated by the firms, no change in ranking precision will make a difference to the equal expected chances of being ranked first.
    \end{remark}
    
    Propositions \ref{existence_p1} and \ref{existence_p2} (in the appendix) establish sufficient conditions for the existence of equilibrium information release policies in the labeling application.  As the involved arguments are more technical in nature than the remainder of this exposition we transfer these deliberations---together with the discussion of appropriate belief structures---into Appendix \ref{existence_app}.
    
    We now establish an interesting economic property of the defined credence-market interaction.  The high-quality firm 1 always emits precision-enhancing information while, for any fixed quality spread $\theta_1>\theta_2$, there is a certain {degree of heterogeneity, $s$, at which the low-quality firm 2 switches to obfuscating the ranking (i.e., releasing precision-diminishing information), by releasing ``fake news.''
            
            \begin{proposition}\label{critical_prp}
                For sufficiently high preference heterogeneity $\tilde{s}$, there exists a set of parameters such that $\rho_1^\ast(\tilde{s})>0$ and $\rho_2^\ast(\tilde{s})=0$.
            \end{proposition}
            Intuitively, the proof utilizes the fact that benefits are linear in $s$ to establish the existence of a parameter pair $(\bm{\theta},s)$ at which firm 2's marginal benefits are zero, resulting in $\rho_2^\ast=0$.  We then apply Cauchy's mean value theorem to firm 2's benefit function to show that there exists a pair of product qualities $\bm{\theta}$, at which a maximum of the function $y(\bm{\theta}) = \left(1-q(\bm{\theta},\rho_1^\ast)\right){P^1}(\rho_1^\ast)/s$ in $\rho_1^\ast$ exceeds the limit of $P^2(r)$ at infinity.  The result then follows from the observation that this maximum of $y(\bm{\theta})$ increases with diminishing type spread $\bm{\theta}$, and corresponds to the limit of $P^1(r)/2$ at infinity, whenever $\theta_{2}\to \theta_{1}$.
            
            Our above results establish that similar-quality firms always jointly provide useful (i.e., precision-enhancing) information to the market.  Proposition \ref{critical_prp} shows that there exists a heterogeneity threshold $\tilde{s}$, such that for $s>\tilde{s}$, the lower-quality firm switches to strategic obfuscation (since firm 2's benefits are shown to be quasi-concave in Proposition \ref{existence_p2}). A sufficient degree of consumer heterogeneity in this case amplifies the spread in profits, $P^i$, such that the \emph{categorization effect} governed by the labeling-competition $q$ outstrips the beneficial \emph{differentiation effect} of information $r$ on profits.\footnote{In principle, a similar result could also be shown for fixed consumer heterogeneity, $s$, and increasing quality spreads but, in that case, the potentially restrictive upper bound on available qualities $\bar{\theta}$ implies that $\theta_1$ cannot generally be raised to the required extent.}
            At $\rho_2^\ast=0$, by continuity, the categorization effect outweighs the differentiation effect, thereby increasing the competitive pressure on firm 2.  Thus, even higher preference heterogeneity (or quality spread) will result in obfuscation.
            
            Proposition \ref{critical_prp} is what partly motivates our interest in ``fake news.''  Firms cooperate to improve consumer information for relatively similar quality types on ``homogeneous'' markets while they choose to go at loggerheads for larger type differences on more heterogeneous markets.  As a simple corollary it can be shown that aggregate market information, $r$, is improving with the heterogeneity of the consumer market.
            
            The above Propositions \ref{proposition_1} \& \ref{critical_prp} cannot generally accommodate ranking technologies $q_i(\bm{\theta},r)$ with non-differentiabilities in precision or quality.  Nevertheless, we show in our examples in Appendix \ref{example_sec} that obfuscation-equilibria may still exist in those cases (see \ref{linear_example_sec} and \ref{Stepwise-constant_example_sec}).
            
            \begin{remark}[Maximum product differentiation]
                The maximum differentiation principle---i.e., concavity of $P^1(r)$ and $P^2(r)$ from \eqref{lemma-3-2} and \eqref{lemma-3-3}---is a well-known result implying that firms differentiate qualities in order to avoid Bertrand price competition \citep{Economides86}.  Something similar is happening in the present model: for modest quality dispersion or low heterogeneity, both firms reinforce the market segmentation by improving information, exploiting the above-introduced differentiation effect.\footnote{Remark \ref{remark_7} suggests that the case of equal qualities exhibits the unencumbered differentiation effect since firms have the same chance of being labeled first, thus eliminating the categorization effect.}  But since the quality dispersion also enters the probability with which the assigned labels are (perceived as) correct, i.e., the above-defined categorization effect, there is a point at which the lower-quality firm starts to strategically obfuscate and to partially offset the better-quality firm's information release.
            \end{remark}
            
            \begin{remark}[Asymmetric costs]
                Proposition \ref{critical_prp} (and later Proposition \ref{existence_p2}) admit insights into the case of heterogeneous information dissemination costs.  Consider asymmetric cost functions of the form $c(|\rho_1|)$ and $\gamma c(|\rho_2|)$, $\gamma \in [0,1]$, for firms 1 and 2, respectively.  For a suitably chosen quality pair $\bm{\theta}$, the boundary case of $\gamma=0$ results in positive equilibrium information $r^\ast=\argmax_r u_2(\bm{\theta},r)$ because firm 2's benefits are single-peaked whenever \eqref{eq:IHR} is satisfied.  Reducing firm 1's costs to zero, by contrast, increases equilibrium information release without bound since firm 1's benefits are strictly increasing.
            \end{remark}

            \subsection{Welfare properties}
            
            We start with the observation that, in any equilibrium, the precision of the market-provided product ranking is higher than what a cartel consisting of the two firms provides.
            \begin{proposition}\label{no-cartel_prp}
                In any equilibrium of the labeling interaction, the precision of the market-provided product ranking \eqref{lemma_1_equ} is higher than what a cartel (or multi-product monopoly) would choose.
            \end{proposition}
            
            Intuitively, a single decision maker (or two cartelized firms) will optimally choose symmetric information emission costs and therefore $\rho_1=\rho_2$.  With strictly convex costs this leads to less precision than what the competitive duopoly provides.
            Consumer welfare for the labeled market results from \eqref{consumer-utility_equ} and \eqref{cons-exp1_equ_dr} as
            \begin{equation}\label{def:consumer-welfare-segments}\begin{array}{rcl}
                    W_H(r)= \dint_{\hat{\mu}_2^1}^{\hat{\mu}_1^0} \left(\mu \Lambda^1(r) - {p^1} \right) dG(\mu), \\
                    W_L(r)= \dint_{\hat{\mu}_3^2}^{\hat{\mu}_2^1} \left(\mu \Lambda^2(r) - {p^2} \right) dG(\mu)
                \end{array}\end{equation}
            summing to served consumer welfare
            \begin{dmath}\label{def:consumer-welfare}
                W_C(r)= W_H(r)+W_L(r)
            \end{dmath}
            with the measure of unserved consumer dead-weight loss given by
            \begin{dmath}
                W_0(r)= \int_{0}^{\hat{\mu}_3^2} \mu  dG(\mu).
            \end{dmath}
            We show that consumer welfare \eqref{def:consumer-welfare} is strictly decreasing in $r$, due to (i) firms recovering their costs of information by charging increased equilibrium prices $(p^i)^\ast$ and (ii) the served market segments monotonically decreasing with increasing information $r$.
            \begin{proposition}\label{consumer-welfare_prp}
                In the defined labeling interaction in which firms charge competitive duopoly prices \eqref{ex-prices_equ}, consumer welfare \eqref{def:consumer-welfare} is strictly decreasing in $r$.
            \end{proposition}
            
            Together with firm utilities defined in \eqref{gen-max3_equ}, we define total welfare as
            \begin{dmath}\label{def:total-welfare}
                W(r)= u_1(r)+u_2(r)+ W_C(r).
            \end{dmath}
            We compare this ranked welfare measure to the unlabeled market, characterized by Bertrand-competition and zero prices, resulting in
            \begin{dmath}\label{def:unlabeled-welfare}
                W_U=\int_{0}^{s} \left( \Exp[\Theta] \mu - 0\right) dG(\mu), \text{ with } \Exp[\Theta] \hiderel{=} \int_{0}^{\bar{\theta}} \theta dF(\theta).
            \end{dmath}
            Notice that dead-weight loss in the competitive, unranked case is zero; all consumers are served.  Moreover, in terms of welfare, the labeled market approaches the unlabeled case as information vanishes
            \begin{dmath}
                \lim_{r \to 0}\ W(r) = W_U.
            \end{dmath}
            Our next result shows that adding information to an unlabeled market increases total welfare, with the gain in producer surplus exceeding the loss in consumer welfare.
            
            \begin{proposition}\label{total-welfare_prp}
                Total welfare is strictly increasing in information release, $\rho_i$, at the unlabeled point $\rho_1=\rho_2=0=r$.
            \end{proposition}
            
            Our minimal welfare analysis shows that, if possible, profit opportunities induce both firms to escape Bertrand competition through the introduction of a label.  Therefore, it is in the firms' interests to implement our assumption of a ``spontaneously'' arising ranking.
            Moreover, since total welfare is increasing through the introduction of the label, (unserved) consumers could in principle be compensated while still allowing for vertical differentiation.
            
            \begin{remark}[Social planner benchmark]
                Consider a benevolent social planner who knows product qualities and is able to perfectly label product qualities with infinite precision.  Consumers therefore know that the two labeled products are of expected qualities \eqref{def-exp-orders}, with no danger of mislabeling.  Firms have no means of manipulating the infinitely precise ranking and compete costlessly on a vertically differentiated market with exogenous quality through the optimal choice of prices as in \citet{Gabszewicz_Shaked_Sutton_Thisse81}.  The resulting welfare is, for infinitely precise ranking in \eqref{def:total-welfare},
                \begin{dmath}
                    u_1 + u_2 + W_C =
                    4s\frac{(\Lambda^1)^2(2\Lambda^1-\hat{\theta})}{(\hat{\theta}-5\Lambda^1)^2} + s\frac{(\hat{\theta}-\Lambda^1)\Lambda^1(2\Lambda^1-\hat{\theta})}{(\hat{\theta}-5\Lambda^1)^2} + s\frac{(\Lambda^1)^2(5\hat{\theta}-\Lambda^1)}{2(\hat{\theta}-5\Lambda^1)^2}
                \end{dmath}
                simplifying to
                \begin{dmath}
                    W^\ast \hiderel{=} s\frac{\FO\left(11\FO^2+3\hat{\theta}\FO-2\hat{\theta}^2\right)}{2(\hat{\theta}-5\FO)^2}
                \end{dmath}
                which unambiguously improves over market welfare \eqref{def:total-welfare} for any finite (equilibrium) precision $r>0$.
            \end{remark}}
    
    Consider now the diametrically opposite case in which a regulator has the power to set prices but is otherwise restricted to using the same market-provided information-based product ranking that consumers have access to.  In this setting the following pricing scheme $\hat{\bm{p}}=(\hat{p}_1,\hat{p}_2)$ is improving consumer welfare over the unlabeled case.
    
    \begin{proposition}\label{regulatory-welfare_prp}
        Regulating the lower-ranked firm to marginal cost, $\hat{p}_2=0$, together with
        \begin{dmath}\label{regulatory-p1_equ}
            0 \hiderel{<} \hat{p}_1 \hiderel{<} \bar{p}_1^\ast \hiderel{<} p^1(r^\ast)^\ast
        \end{dmath}
        in which
        \begin{dmath}\label{regulatory-welfare_equ}
            \bar{p}_1^\ast \hiderel{=} \bar{p}_1(r^\ast) \hiderel{=} s\left(\Lambda^1(r)-\Lambda^2(r)-\sqrt{\left(\Exp[\Theta] -\Lambda^2(r)\right)\left(\Lambda^1(r)-\Lambda^2(r)\right)}\right)
        \end{dmath}
        improves consumer welfare over the unlabeled case while ensuring firm participation.
    \end{proposition}
    
    Intuitively, regulating the lower-ranked product price, $\hat{p}_2$, to equal marginal cost eliminates the dead-weight loss resulting from competitive, strictly positive prices \eqref{ex-prices_equ}.  The positive price on the remaining first-ranked product segment, $\hat{p}_1$, ensures the supply of non-zero aggregate information, $r$ (which the regulator knows in equilibrium from \eqref{lemma_1_equ}, irrespective of $\bm{\theta}$).  Since there is a positive probability of mislabeling, the lower-ranked firm also expects positive profits.
    
    Since total firm profits are zero in the unlabeled case and increasing in information, $r$, the result is unambiguously welfare-improving.\footnote{As the second-ranked firm is regulated to marginal cost while the first-ranked firm enjoys the fully differentiated prize mass, such a policy would create long-run incentives to improve quality.}

    \section{Concluding remarks}
    
    We study a novel class of integrated market interactions in which firms compete on the informational content of a ranking that labels otherwise indistinguishable products.  We establish a set of economically meaningful properties and show that equilibrium existence is unproblematic for a wide class of ubiquitous contest success functions.
    Besides defining a new interaction based on the precision of observed information and characterizing the ensuing equilibria, our principal results include: 1) The modeled label credence markets generally provide helpful overall information to the consumers.  2) Given quadratic costs, the amount of market-provided aggregate information is constant, irrespective of the quality spread between firms. 3) We introduce and discuss the competing economic effects of increased labeling precision in the form \emph{differentiation} and \emph{categorization} effects.  4) There are widely applicable conditions under which the lower-quality firm turns to strategic obfuscation (i.e., the release of ``fake news'') if the quality spread between products becomes too large or the heterogeneity among consumers too high.  5) The maximum differentiation principle is weakened if the underlying information can be used by consumers to differentiate between competing products.  6) Conditions on the circumstances under which credence good labeling enhances welfare over an unlabeled market for all market participants.  7) Virtually all forms of contest success functions typically employed in the literature are subsumed by our analysis.
    
    Other interesting applications match aspects of our model's structure besides the academic department-ranking example discussed in the Introduction.  One such example is the current debate around the \href{https://docs.score-environnemental.com/v/en/}{\emph{Eco-score}} food label, implemented by the French authorities with a consortium of socio-economic actors.  The Eco-score labels products A (preferred, green) to E (to be avoided) based on an aggregation of several environmental criteria (e.g., water or energy consumption, level of biodiversity conservation, pesticides, etc.).  The number of criteria used to calculate the Eco-score is shown on the label, but consumers cannot observe each criterion's evaluation process or individual results.  Consumers only observe the number of criteria underlying the ranking, i.e., the precision of the ranking and the realization of the ranking, i.e., the Eco-score the product receives.  The Eco-score introduction has seen considerable debate between various market players.  In particular, food companies have been reported to lobby for the inclusion of criteria that ensure intensive farming is given a good Eco-score \citep{StilettoEA23}.
    %https://help.yuka.io/l/en/article/4mt8nhb5fz-what-are-the-main-scientific-sources-for-nutri-score-evaluation
    %https://www.foodnavigator.com/Article/2023/05/11/nutri-score-fake-news-scientists-counter-frequent-misunderstandings-about-fop-label
    %https://www.ncbi.nlm.nih.gov/pmc/articles/PMC10421117/
    %https://reporterre.net/La-bataille-des-lobbies-pour-torpiller-le-score-environnemental-des-aliments
    
    A second example is the recent designation of nuclear energy and natural gas as sustainably ``green'' by the European Commission.  Costly political and industry lobbying activities based on which energy-form categorization  to assign ``green labels'' are well-documented, and several European Union member states sued the Commission over the implemented rules.\footnote{The European Commission (EC) explicitly states the anticipated consumer reaction as one of the central aims of the taxonomy, which it sees as a ``list of environmentally sustainable economic activities'' whose overall purpose is to provide a ``science-based classification system that allows financial and non-financial companies to share a common definition of sustainability when determining their investment choices'' (\href{https://ec.europa.eu/newsroom/fisma/items/738336/en}{EC press release 7-Mar-2022}).  The rejected ``amber category'' would have, in addition to the implemented green and red labels, resulted in a more precise ranking (\href{https://ec.europa.eu/commission/presscorner/detail/en/IP_22_2}{EC press release 1-Jan-2022}, \href{https://ec.europa.eu/commission/presscorner/detail/en/QANDA_22_712}{EC press release 2-Feb-2022}).}
    In these examples, consumer demand reacts endogenously to the changed informativeness or precision of the realized labeling standard---providing the original rationale for the firms' activities.
    
    Third, the organization of sports competitions yields examples of situations where ranking precision is a strategic choice. \citet{Wang10} reports on the motivation behind the change of the points scoring system by the International Table Tennis Federation from 21 to 11 as the ``domination of China meant that there was little incentive for the other teams. Reducing the accuracy level increases the chance that a team other than China will win, thus inducing more effort from the other teams. This increased competition could, in turn, result in greater effort from the Chinese team.''  \citet{Deng_Gafni_Lavi_Lin_Ling23} argue that similar reasoning has led to changing the best-of-three finals structure in the Israeli basketball league to a single game: ``From 1970 to 2006, the Maccabi Tel Aviv team lost only one championship, while after the change, it lost six.''  Likewise, \citet{Yildirim15} reports the slow adoption of obviously accuracy-improving video replay technology to support refereeing decisions in European soccer competitions as born by non-aligned competitor interests.
    
    More distant applications of ideas relating to our environment may include political campaigning and party fundraising, as well as pharmaceutical drug testing, marketing, computing platform security, and advertising.  In all these examples, the possibility for differentiation through some form of labeling seems crucial for a firm's ex-ante quality investment incentives, which otherwise may not be valued by the market.
    
    \appendix
    %\section*{Appendix}
    
    \section{Appendix: Examples of labeling precision contests\protect}\label{example_sec}
    
    We now illustrate the general, analytic results of the main paper by means of several, mostly numerical examples for the transformation of firms' emitted information into (credence good) market ranking precision, i.e., the mapping from quality difference to label probability assignments.  Our examples show that essentially \emph{all} standard contest success functions (and some more we introduce) lend themselves naturally for the precision interaction this paper introduces.  The Tullock-ratio, difference, piecewise-constant, and noise-based forms we use below to rank qualities $\bm{\theta} = (\theta_1,\theta_2)$ are illustrating the class $\mathcal{Q}$ on which we base the general analysis of the previous sections.  The common example properties are uniformly drawn qualities on $[0,\bar{\theta}>0]$, uniformly distributed consumer tastes on $[0,s>0]$, and quadratic costs of information emission $c(\rho) = \delta/2 \rho^2$ with $\delta=2$.
    The following Table \ref{exAss_tab} summarizes the relationship of our examples to the set of assumptions $\mathcal{Q}$:
    \begin{table}[h!]\centering \begin{tabular}{l|cccc}
            \multirow{2.5}{*}{} & \multirowcell{2.5}{Ratio-based                                           \\ \ref{tullock_example_sec}}
                                & \multirowcell{2.5}{Difference-based                                      \\ \ref{linear_example_sec}}
                                & \multirowcell{2.5}{Piecewise-constant                                    \\ \ref{Stepwise-constant_example_sec}}
                                & \multirowcell{2.5}{Noise-based                                           \\ \ref{lnr_example_sec}} \\ \\ \hline
            (Q1)                & \cmark                                & \cmark         & \cmark & \cmark \\
            (Q2)                & \cmark                                & locally \cmark & \xmark & \cmark \\
            (Q3)                & \cmark                                & locally \cmark & \cmark & \cmark \\
            (Q4)                & \cmark                                & locally \cmark & \cmark & \cmark \\
            (Q5)                & \cmark                                & locally \cmark & \xmark & \cmark \\
            (Q6)                & \cmark                                & locally \cmark & \cmark & \cmark \\ \hline
        \end{tabular}
        \caption{Matching assumptions to example properties.}\label{exAss_tab}
    \end{table}
    
    Despite the fact that some examples of ranking functions violate several of the assumed properties $\mathcal{Q}$, all examples verify the full set of properties derived in Propositions \ref{comp_prop}--\ref{critical_prp}, as well as existence, Propositions \ref{existence_p1} and \ref{existence_p2}.

    \subsection{Ratio-based ranking}\label{tullock_example_sec}
    
\begin{figure}[htb!]\large
        \begin{center}
            \begin{psfrags}
                \psfrag{a}{$q(\bm{\theta},r)$}
                \psfrag{b}{$\theta_{1}/\theta_{2}$}
                \psfrag{c}{$r=0.1$}
                \psfrag{d}{$r=1$}
                \psfrag{e}{$r=2$}
                \psfrag{f}{$r=10$}
                \psfrag{g}{$\theta_{1}=\theta_{2}$}
                \psfrag{h}{$r=0$}
                \psfrag{i}{\large$\dfrac{\partial q}{\partial \theta_1} = r \dfrac{\theta_1^{r-1}\theta_2^r}{(\theta_1^r + \theta_2^r)^2}$}
                \psfrag{l}{\large$ = \dfrac{r}{4\theta_1}$ for $\theta_2=\theta_1$}
                \psfrag{j}{\gcol{$\theta_1<\theta_2$}}
                \psfrag{k}{$\theta_1>\theta_2$}
                \scalebox{.6}{\includegraphics[width=\textwidth]{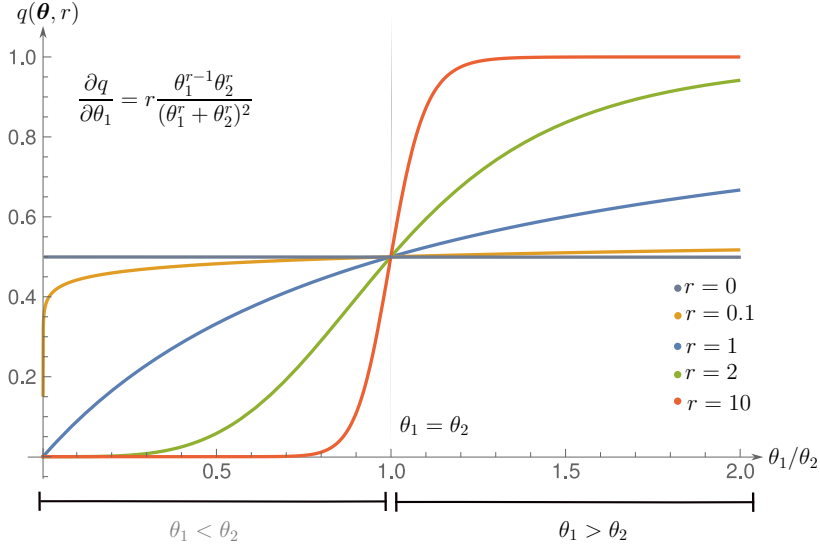}}
            \end{psfrags}
        \end{center}
        \caption{Reaction of Tullock ranking probabilities \eqref{def:ranking-tullock} to quality-ratio and precision.}\label{Ex-01-intuition-fig}
    \end{figure}

    In this first example, the Tullock ranking technology \citep{Tullock80} is employed, specifying the probabilities with which a firm is ranked first as:\footnote{Formulation \eqref{def:ranking-tullock} is a simplified but equivalent form of the Logistic contest success function defined in footnote \ref{fn:fn-q-def}.}
    \begin{dmath}\label{def:ranking-tullock}
        q_1(\bm{\theta},r)\hiderel{=} 1/\left(1+x^{-r}\right),\ q_2(\bm{\theta},r)\hiderel{=} 1/\left(1+x^{r}\right),
    \end{dmath}
    for $x \hiderel{=} \theta_1/\theta_2$. Note that the Tullock ranking function---illustrated in Figure \ref{Ex-01-intuition-fig}---satisfies assumptions $\mathcal{Q}$, as well as the regularity conditions \eqref{eq:IHR} and \eqref{eq:IHRt} from the existence appendix \ref{existence_app}, are satisfied for any $r>0$.
    
    %drawn where?
    
    A firm $i=1,2$ chooses i) the resources $\rho_i$ it wishes to expend on influencing the overall ranking precision and consumer demand and ii) announces the price of her product conditional on the observed precision and product ranks.
    
    We start by modeling the demand side.  Expected qualities are obtained using \eqref{cons-exp1_equ_dr}.  For the uniform distribution on the $\bar{\theta}$-scaled unit interval, the joint density \eqref{eq-jointorder_highlow-simplified} is
    \begin{dmath}\label{joint-density_equ}
        f_{(1,2:2)}(\tilde{\bm{\theta}}) \hiderel{=} 2 f(\tilde{\theta}_1) f(\tilde{\theta}_2) = 2 / \bar{\theta}^2.
    \end{dmath}
    The consumers assess the expected qualities of the first- and second-ranked products \eqref{cons-exp1_equ_dr}, given an observed ranking, for $\tilde{x}=\tilde{\theta}_1/\tilde{\theta}_2$ as:
    \begin{dmath}\label{eq:ex1-exp-cons1}\begin{array}{rcl}
            \Lambda^1(r) & = & 
            \dint_0^{\bar{\theta}} \dint_0^{\tilde{\theta}_1} \left( \dfrac{1}{1+\tilde{x}^{-r}}\tilde{\theta}_1 +\frac{1}{1+\tilde{x}^r}\tilde{\theta}_2 \right) \dfrac{2}{\bar{\theta}^2}d\tilde{\theta}_2  d\tilde{\theta}_1                                                                                                       \\
                         & = & \dfrac{2}{\bar{\theta}^2}\left(\dint_0^{\bar{\theta}} \dint_0^{\tilde{\theta}_1 } \dfrac{\tilde{\theta}_1}{1+\tilde{x}^{-r}}d\tilde{\theta}_2  d\tilde{\theta}_1+ \int_0^{\bar{\theta}} \dint_0^{\tilde{\theta}_1 } \frac{\tilde{\theta}_2}{1+\tilde{x}^r}d\tilde{\theta}_2  d\tilde{\theta}_1\right), \\
            \Lambda^2(r) & = & 
            \dint_0^{\bar{\theta}} \dint_0^{\tilde{\theta}_1} \left( \dfrac{1}{1+\tilde{x}^{r}}\tilde{\theta}_1+
            \frac{1}{1+\tilde{x}^{-r}} \tilde{\theta}_2 \right) \dfrac{2}{\bar{\theta}^2}d\tilde{\theta}_2  d\tilde{\theta}_1                                                                                                                                                                                                         \\
                         & = & 
            \dfrac{2}{\bar{\theta}^2}\left(\dint_0^{\bar{\theta}} \dint_0^{\tilde{\theta}_1 } \dfrac{\tilde{\theta}_1}{1+\tilde{x}^{r}}d\tilde{\theta}_2  d\tilde{\theta}_1+
            \dint_0^{\bar{\theta}} \int_0^{\tilde{\theta}_1 } \frac{\tilde{\theta}_2}{1+\tilde{x}^{-r}}d\tilde{\theta}_2  d\tilde{\theta}_1\right).
        \end{array}\end{dmath}
    Following the general derivation in subsection \ref{labeling_subsec} and using $\FO+\SO=\hat{\theta}=\bar{\theta}$, we obtain contest prizes \eqref{ex-prizes0_equ}.
    We turn to the supply side and state firm $i \in \mathcal{N}$'s maximization problem (under mutually known $x_i=\theta_i/\theta_j$, $j=3-i$) on the basis of \eqref{eqm-prices-def} as
    \begin{dmath}\label{ex1-obj_equ}
        \underset{\rho_i}{\max\; } \frac{1}{1+x_i^{-r}} P^1(r) + \frac{1}{1+x_i^{r}} P^2(r) - \frac{\rho_i^2}{2},
    \end{dmath}
    in which $r=|\rho_i+\rho_j|$.  In order to get numerical values, we fix $\theta_1=\fric{3}{4}$, $\theta_2 = \fric{1}{4}$ (i.e., $x=3$), and the distributional parameters at $\bar{\theta}=1$, $s=30$.  After taking derivatives with respect to $\rho_i$, we find the asymmetric equilibrium candidate\footnote{These are numerical results.  If qualities are Beta-distributed with parameters $\beta_1=2, \beta_2=2$, the asymmetric equilibrium is $\rho_1^\ast \approx 0.344,\ \rho_2^\ast\approx 0.156, \text{ implying that } r^\ast\approx0.5$.}
    \begin{dmath}
        \rho_1^\ast \hiderel{\approx} 1.083,\ \rho_2^\ast \hiderel{\approx} -0.042, \text{ implying that } r^\ast \hiderel{\approx} 1.041.
    \end{dmath}
    
    The sufficient condition for equilibrium existence \eqref{existence_p2_iq} for firm 2 is satisfied
    \begin{dmath}
        r^{\ast}\hiderel{\approx}1.041\hiderel{<}1.168\hiderel{\approx} r^{su}(\bm{\theta})
    \end{dmath}

    \begin{figure}[htb!]\large
        \begin{center}
            \begin{psfrags}
                \psfrag{a}{$\rho_1$}
                \psfrag{b}{\raisebox{2pt}{\hspace{-24pt}{$u_1(\rho_1,\rho_2^\ast)$}}}
                \psfrag{c}{$\rho_1^\ast$}
                \psfrag{d}{$\rho_2$}
                \psfrag{e}{\raisebox{2pt}{\hspace{-24pt}{$u_2(\rho_1^\ast,\rho_2)$}}}
                \psfrag{f}{$\rho_2^\ast$}
                \psfrag{g}{$\rho_2$}
                \resizebox{\dimexpr .49 \textwidth}{!}{\includegraphics{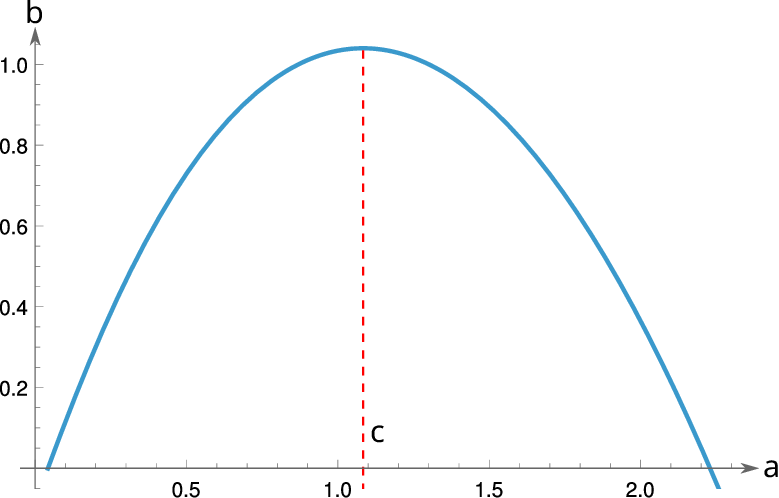}}\hspace{1mm}
                \resizebox{\dimexpr .49 \textwidth}{!}{\includegraphics{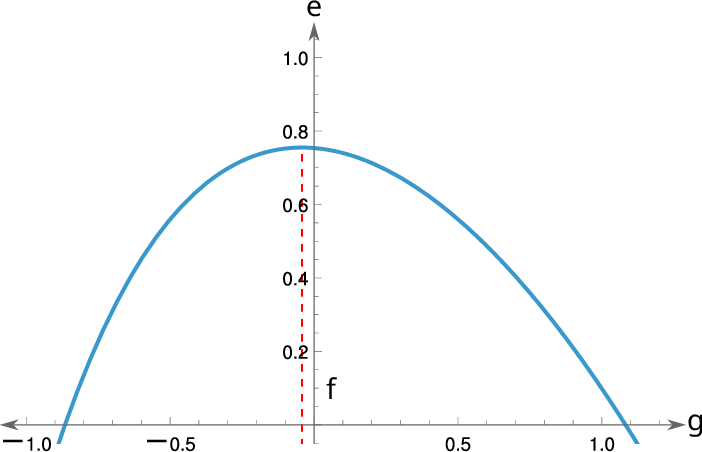}}
            \end{psfrags}
        \end{center}
        \caption{The two players' optimal choice of $\rho_i$ in asymmetric equilibrium for ratio-based ranking.}\label{Ex-01-Epsilons-fig}
    \end{figure}%computed in ps-ratex-01.nb
    
\begin{figure}[!htb]\fontsize{11}{15} \selectfont
        \begin{center}
            \psfrag{a}{$\theta_1$}
            \psfrag{b}{\hspace{-6pt}{$\rho_i^\ast(x)$}}
            \psfrag{c}{$\rho_2^\ast(x)$}
            \psfrag{d}{$\rho_1^\ast(x)$}
            \psfrag{e}{$\tilde{x}$}
            \psfrag{f}{$r^\ast(x)$}
            \psfrag{g}{$s$}
            \psfrag{h}{\hspace{-6pt}{$\rho_i^\ast(s)$}}
            \psfrag{i}{$\rho_2^\ast(s)$}
            \psfrag{j}{$\rho_1^\ast(s)$}
            \psfrag{k}{$\tilde{s}$}
            \psfrag{l}{$r^\ast(s)$}
            \resizebox{\dimexpr .49 \textwidth}{!}{\includegraphics{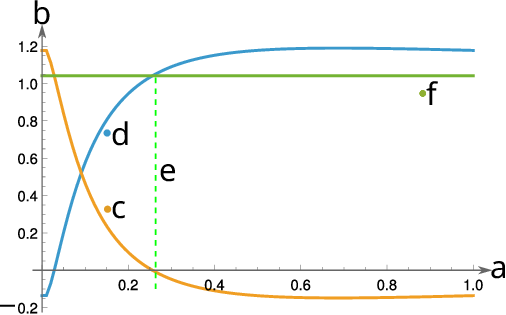}}
            \resizebox{\dimexpr .49 \textwidth}{!}{\includegraphics{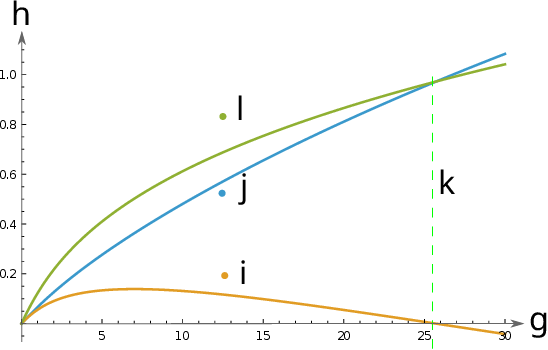}}
            \caption{Left: information dissemination $\rho_i^\ast$ for the ratio-form example, for heterogeneity $s=20$, as functions of $\theta_1 \in [0,1]$ for fixed $\theta_2=1/10$: $\rho_1^\ast(x)$ is plotted blue, $\rho_2^\ast(x)$ is shown gold, and aggregate information $r^\ast(x)=\rho_1^\ast(x)+\rho_2^\ast(x)$ is green.  The vertical dashed line indicates the ratio $\tilde{x}$ at which $\rho_2^\ast(x)=0$.  Right: information dissemination $\rho_i^\ast(s)$ as a function of the consumer heterogeneity $s$, for $\bar{\theta}=1$.  The dashed green line shows the critical heterogeneity $\tilde{s}\approx 25.5$ at which $\rho_2^\ast=0$.}\label{Resp02-1-fig}
        \end{center}%computed in faking-prz02.nb and ps-
    \end{figure}

    Notice that both firms choose to \emph{improve} the ranking precision by supplying positive amounts of information in equilibrium.  Hence, the market-supplied ranking precision is an informational improvement over the status quo of the unranked market with information $r=0$.  For other parameterizations, however, it turns out that this behavior depends on both the ratio $x$ and the consumer heterogeneity, $s$.  The left-hand panel of Figure~\ref{Resp02-1-fig} fixes the heterogeneity at $s=20$ and shows information emission as a function of $x$.
    For sufficiently wide type-difference, there exists a point $\tilde{x}$ at which the lower-quality firm turns to obfuscation.  Moreover, the figure illustrates that the symmetric case of $\theta_1=\theta_2$ (and therefore $\rho_1=\rho_2>0$) is sufficient to determine market information $r$.  The right-hand panel of Figure~\ref{Resp02-1-fig} shows that such an obfuscation-switching strategy also exists for fixed $x$ when the consumer heterogeneity $s$ is varied (Proposition~\ref{critical_prp}).

    \subsection{Difference-based ranking}\label{linear_example_sec}
    
\begin{figure}[htb!]
        \begin{center}\large
            \begin{psfrags}
                \psfrag{a}{$q(\bm{\theta},r)$}
                \psfrag{b}{$\theta_{1} - \theta_2$}
                \psfrag{c}{$r$}
                \psfrag{d}{$r/2$}
                \psfrag{e}{$2r$}
                \psfrag{g}{$\theta_{1}=\theta_{2}$}
                \psfrag{h}{$r=0$}
                \psfrag{j}{\gcol{$\theta_1<\theta_2$}}
                \psfrag{k}{$\theta_1>\theta_2$}
                \scalebox{.6}{\includegraphics[width=\textwidth]{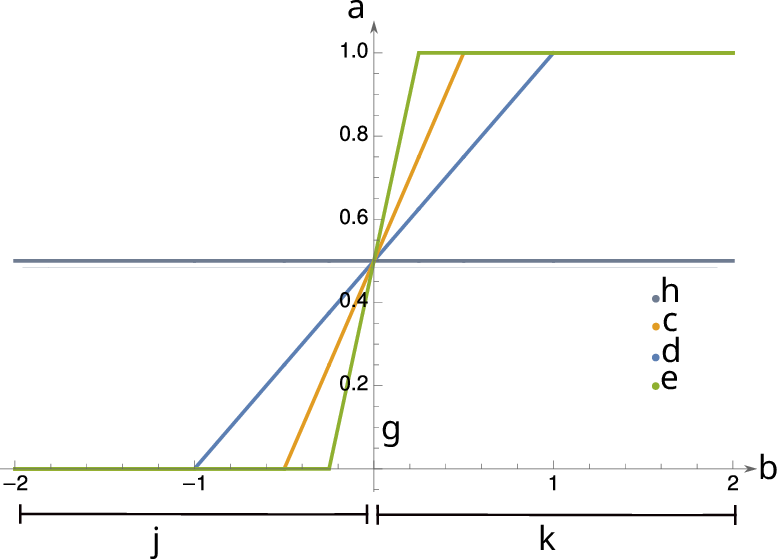}}
            \end{psfrags}
        \end{center}
        \caption{Reaction of ranking probabilities \eqref{def:difference-form} to quality-difference and precision.}\label{Ex-02-intuition-fig}
    \end{figure}

    Consider the same uniform two-firms example as in the previous subsection, but governed now by a difference-form ranking technology in a piece-wise linear form, as discussed, e.g., by \cite{Che00}.\footnote{Formulation \eqref{def:difference-form} is a simplified but equivalent form of the Logistic contest success function defined in footnote \ref{fn:fn-q-def}.  A very similar analysis can be performed (with similar results) for a ranking based on 1) the ``serial'' contest success function introduced by \citet{Alcalde_Dahm07} that combines aspects of both the ratio form \eqref{def:ranking-tullock} and the difference form \eqref{def:difference-form} and 2) a ranking based on the exponential type difference.}
    \begin{dmath}\label{def:difference-form}
        q_1(\bm{\theta},r)=\text{max}\left[\text{min}\left[1/2+r\left(\theta_1-\theta_2\right),1\right],0\right].
    \end{dmath}
    We tentatively simplify this ranking technology---illustrated in Figure \ref{Ex-02-intuition-fig}---using $\theta_1>\theta_2$ to define:
    \begin{dmath}\label{def:ranking-difference}
        q_1(\bm{\theta},r)\hiderel{=} \text{min}\left[1/2+r x,1\right],\ q_2(\bm{\theta},r) \hiderel{=} \text{max}\left[1/2-r x,0\right]
    \end{dmath}
    in which $x \hiderel{=} \theta_1-\theta_2$ and we tentatively guess that $xr^\ast<1/2$.\footnote{Since the relevant equilibrium candidates fall into this region, this guess turns out to be correct in the present example.} This allows \eqref{def:ranking-difference} to locally satisfy assumptions (Q2)---(Q6), as well as the regularity conditions \eqref{eq:IHR} and \eqref{eq:IHRt} from the existence appendix \ref{existence_app}, for any $x>0$ and $r>0$ whenever $xr^\ast<1/2$.
    
    %drawn in ps-diffex-01.nb?
    
    We obtain expected qualities \eqref{cons-exp1_equ_dr}, given the observed ranking, for $\tilde{x}=\tilde{\theta}_1-\tilde{\theta}_2$ as:
    \begin{dmath}\label{eq:ex2-exp-cons1}\begin{array}{rcl}
            \Lambda^1(r) & = & 
            \dfrac{2}{\bar{\theta}^2}\left(\dint_0^{\bar{\theta}} \dint_0^{\tilde{\theta}_1} \left(1/2+r\tilde{x}\right)\tilde{\theta}_1d\tilde{\theta}_2  d\tilde{\theta}_1+
            \dint_0^{\bar{\theta}} \dint_0^{\tilde{\theta}_1 } \left(1/2-r\tilde{x}\right) \tilde{\theta}_2d\tilde{\theta}_2  d\tilde{\theta}_1\right) \\
                         & = & \dfrac{\bar{\theta}}{6}\left(3+r\bar{\theta}\right),                                                                    \\%[12pt]
            \Lambda^2(r) & = & 
            \dfrac{2}{\bar{\theta}^2}\left(\dint_0^{\bar{\theta}} \dint_0^{\tilde{\theta}_1 } \left(1/2-r\tilde{x}\right) \tilde{\theta}_1d\tilde{\theta}_2  d\tilde{\theta}_1+
            \dint_0^{\bar{\theta}} \dint_0^{\tilde{\theta}_1 } \left(1/2+r\tilde{x}\right) \tilde{\theta}_2d\tilde{\theta}_2  d\tilde{\theta}_1\right) \\
                         & = & \dfrac{\bar{\theta}}{6}\left(3-r\bar{\theta}\right).
        \end{array}\end{dmath}
    Following the steps of the general derivation in subsection \ref{labeling_subsec} results in the same prizes \eqref{ex-prizes0_equ}, parameterized by $\Lambda^i(r)$ and for $\hat{\theta}=\bar{\theta}$, as in the previous subsection.
    The resulting supply-side firms' maximization problem (under mutually known $x=\theta_1-\theta_2$) is
    \begin{dmath}
        \underset{\rho_i}{\max\; } q_i(\bm{\theta},r) P^1(r) + \left(1-q_i(\bm{\theta},r)\right) P^2(r) - \frac{\rho_i^2}{2}.
    \end{dmath}
    Under the simplified ranking and assumed $xr^\ast<1/2$ in \eqref{def:ranking-difference}, the first-order conditions for this problem are
    \begin{equation}\label{ex2-foc-diff}\parbox{.9\linewidth}{%
            \setlength{\belowdisplayskip}{0pt} \setlength{\belowdisplayshortskip}{0pt}
            \setlength{\abovedisplayskip}{0pt} \setlength{\abovedisplayshortskip}{0pt}
            \begin{dgroup*}\begin{dmath*}
                    \bar{\theta}^2 s\left(\frac{\bar{\theta} r \left(\bar{\theta} r \left(5\bar{\theta} r+27\right)+69\right)+135}{2 \left(5 \bar{\theta} r+9\right)^3}+\frac{2 r x \left(\bar{\theta} r \left(5 \bar{\theta} r+21\right)+27\right)}{3 \left(5 \bar{\theta} r+9\right)^2} \right) = \rho_1,
                \end{dmath*}
                \begin{dmath*}
                    \bar{\theta}^2 s\left(\frac{\bar{\theta} r \left(\bar{\theta} r \left(5 \bar{\theta} r+27\right)+69\right)+135}{2 \left(5 \bar{\theta} r+9\right)^3}-\frac{2 r x \left(\bar{\theta} r \left(5 \bar{\theta} r+21\right)+27\right)}{3 \left(5 \bar{\theta} r+9\right)^2}\right) = \rho_2.
                \end{dmath*}\end{dgroup*}}
    \end{equation}
    Guessing affine solutions $\rho_i=\alpha \pm \beta x$ to \eqref{ex2-foc-diff} for $s=4$ and $\bar{\theta}=1$ results in the two functions
    \begin{dmath}%calculated in ps-diffex-01.nb veryfied in faking-info-daniel.nb
        \rho_1^\ast(x) \approx 0.237 + 0.373 x,\ \rho_2^\ast(x) \hiderel{\approx} 0.237 - 0.373 x, \text{ implying that } r^\ast \hiderel{\approx} 0.474,
    \end{dmath}
    which are indeed affine. Hence, the critical difference $\tilde{x}=\theta_1-\theta_2$ at which $\rho_2^\ast(x)=0$ equals%\footnote{Although homogeneity of degree zero fails for difference-form rankings, i.e., $q_1(x,r) \neq q_1(\lambda x,r)$ for all $\lambda>0$ \citep{Skaperdas96}, our formulation still implies that only the difference $x=\theta_1-\theta_2$ matters for the critical $\tilde{x}$, not absolute quality realizations.}
    \begin{dmath}
        \tilde{x} = \alpha / \beta \hiderel{\approx} 0.634.
    \end{dmath}
    For the same example values of $\theta_1=3/4$, $\theta_2=1/4$ as in the previous subsection, this yields the asymmetric equilibrium candidate\footnote{If both qualities are Beta-distributed with parameters $\beta_1=2, \beta_2=2$, the asymmetric equilibrium is $\rho_1^\ast \approx 0.799,\ \rho_2^\ast\approx-0.068, \text{ implying that } r^\ast\approx0.731$.}
    %The example is calculated in the file faking-ex-02-beta.nb.}
    \begin{dmath}
        \rho_1^\ast \hiderel{\approx} 0.423,\quad \rho_2^\ast \hiderel{\approx} 0.050.
    \end{dmath}
    This candidate satisfies the assumed $xr^\ast=0.237 < 1/2$.
    The sufficient condition for equilibrium existence \eqref{existence_p2_iq} for firm 2 is satisfied
    \begin{dmath}
        r^{\ast}\hiderel{\approx}0.474\hiderel{<}1/2\hiderel{=} r^{su}(\bm{\theta})
    \end{dmath}

\begin{figure}[ht!]\large
        \begin{center}\begin{psfrags}
                \psfrag{a}{$\rho_1$}
                \psfrag{b}{\raisebox{2pt}{\hspace{-24pt}{$u_1(\rho_1,\rho_2^\ast)$}}}
                \psfrag{c}{$\rho_1^\ast$}
                \psfrag{d}{$\rho_2$}
                \psfrag{e}{\raisebox{2pt}{\hspace{-18pt}{$u_2(\rho_1^\ast,\rho_2)$}}}
                \psfrag{f}{$\rho_2^\ast$}
                \psfrag{g}{$r^\ast$}
                \resizebox{\dimexpr .49 \textwidth}{!}{\includegraphics{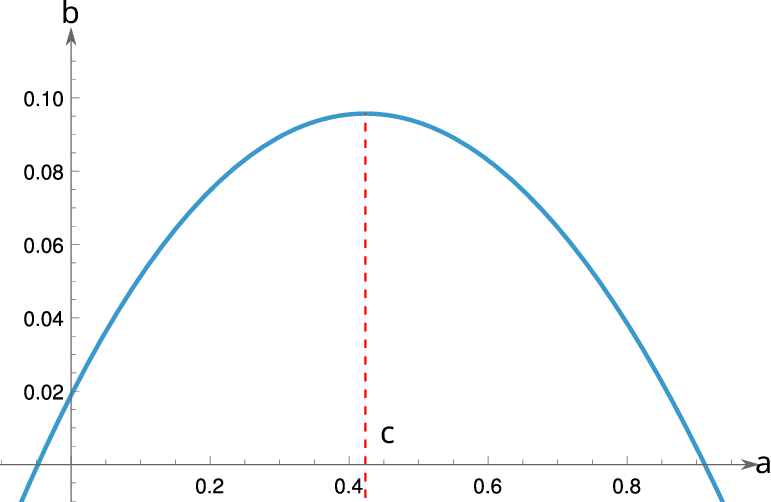}}
                \resizebox{\dimexpr .49 \textwidth}{!}{\includegraphics{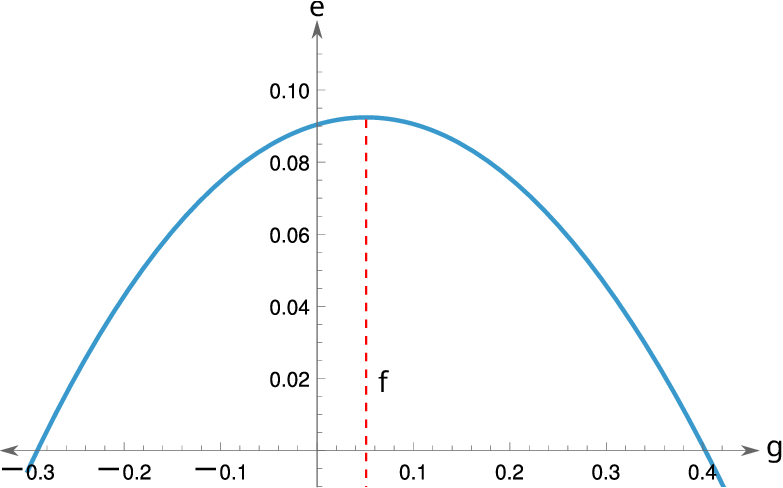}}
            \end{psfrags}\end{center}%calculated in faking-prz01.nb
        \caption{The two players' optimal choice of $\rho_i$ in asymmetric equilibrium for difference-based ranking.}\label{Ex-02-Epsilons-fig}
    \end{figure}

    Together with global cost convexity, Figure~\ref{Ex-02-Epsilons-fig} verifies this candidate as pure strategy equilibrium.
    While both players choose to improve the ranking precision in the Tullock-example of the previous subsection, the lower-quality firm obfuscates the ranking in this second example.  The (indirect) positive influence of aggregate information $r$ on the prizes outweighs the (direct) competitive effect in the difference-form example.
    
\begin{figure}[!ht]\fontsize{11}{15} \selectfont
        \begin{center}
            \psfrag{a}{$\theta_1$}
            \psfrag{b}{$\rho_i^\ast(x)$}
            \psfrag{c}{$\rho_2^\ast(x)$}
            \psfrag{d}{$\rho_1^\ast(x)$}
            \psfrag{e}{$\tilde{x}$}
            \psfrag{f}{$r^\ast(x)$}
            \psfrag{g}{$s$}
            \psfrag{h}{$\rho_i^\ast(s)$}
            \psfrag{i}{$\rho_2^\ast(s)$}
            \psfrag{j}{$\rho_1^\ast(s)$}
            \psfrag{k}{$\tilde{s}$}
            \psfrag{l}{$r^\ast(s)$}
            \resizebox{\dimexpr .49 \textwidth}{!}{\includegraphics{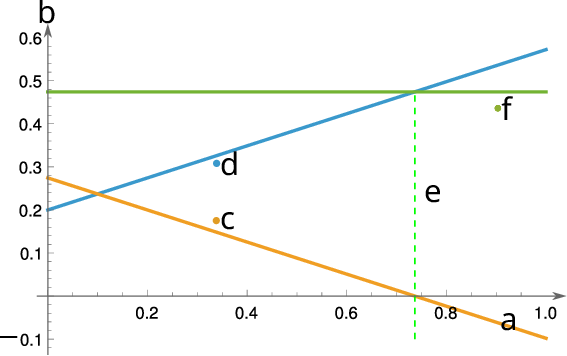}}
            \resizebox{\dimexpr .49 \textwidth}{!}{\includegraphics{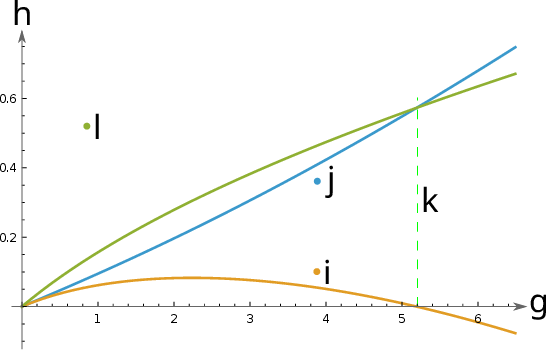}}
            \caption{Left: information dissemination $\rho_i^\ast$ for the difference-form example, for heterogeneity $s=10$, as functions of $\theta_1 \in [0,1]$ for fixed $\theta_2=1/10$: $\rho_1^\ast(x)$ is plotted blue, $\rho_2^\ast(x)$ is shown gold, and aggregate information $r^\ast(x)=\rho_1^\ast(x)+\rho_2^\ast(x)$ is green.  The vertical dashed line indicates the difference $\tilde{x}$ at which $\rho_2^\ast(x)=0$.  Right: information dissemination $\rho_i^\ast(s)$ as a function of the consumer heterogeneity $s$, for $\bar{\theta}=1$.  The dashed green line shows the critical heterogeneity $\tilde{s}\approx 5.192$ at which $\rho_2^\ast=0$.}\label{Resp02-2-fig}
        \end{center}%computed in faking-prz02.nb and ps-diffex-01-nb
    \end{figure}

    The left-hand panel of Figure~\ref{Resp02-2-fig} fixes the consumer heterogeneity at $s=4$ and shows equilibrium information emission as a function of $x$.
    The left-hand panel of Figure~\ref{Resp02-2-fig} shows that for sufficiently wide quality-difference there exists a point $\tilde{x}$ at which the lower-quality firm turns to obfuscation. A similar effect is illustrated in the right-hand panel of Figure~\ref{Resp02-2-fig} for a sufficiently large heterogeneity $s$.  Notice that $\rho_1^\ast(s)$ is convex in this example.

    \subsection{Piecewise-constant ranking}\label{Stepwise-constant_example_sec}
    
\begin{figure}[htb!]
        \begin{center}\large
            \begin{psfrags}
                \psfrag{a}{$q(\bm{\theta},r)$}
                \psfrag{b}{$\theta_{1} - \theta_2$}
                \psfrag{c}{$r$}
                \psfrag{d}{$r/2$}
                \psfrag{e}{$2r$}
                \psfrag{g}{$\theta_{1}=\theta_{2}$}
                \psfrag{h}{$r=0$}
                \psfrag{j}{\gcol{$\theta_1<\theta_2$}}
                \psfrag{k}{$\theta_1>\theta_2$}
                \scalebox{.6}{\includegraphics[width=\textwidth]{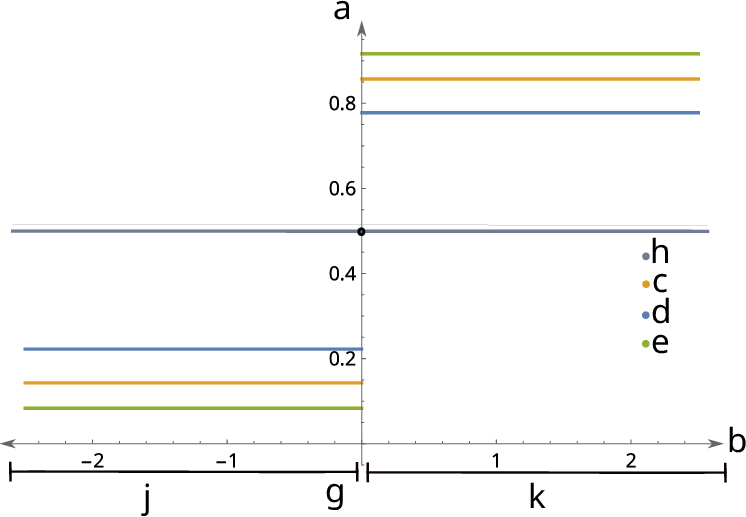}}
            \end{psfrags}
        \end{center}
        \caption{Reaction of ranking probabilities \eqref{def:ranking-constant} to quality-difference and precision.}\label{Ex-03-intuition-fig}
    \end{figure}

\begin{figure}[ht!]\fontsize{11}{15} \selectfont
        \begin{center}
            \psfrag{a}{$\rho_1$}
            \psfrag{b}{\raisebox{2pt}{\hspace{-20pt}{$u_1(\rho_1,\rho_2^\ast)$}}}
            \psfrag{c}{$\rho_1^\ast$}
            \psfrag{d}{$\rho_2$}
            \psfrag{e}{\raisebox{2pt}{\hspace{-20pt}{$u_2(\rho_1^\ast,\rho_2)$}}}
            \psfrag{f}{$\rho_2^\ast$}
            \psfrag{g}{$r^\ast$}
            \resizebox{\dimexpr .49 \textwidth}{!}{\includegraphics{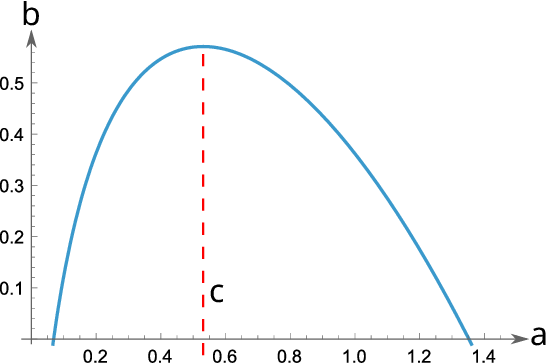}}
            \resizebox{\dimexpr .49 \textwidth}{!}{\includegraphics{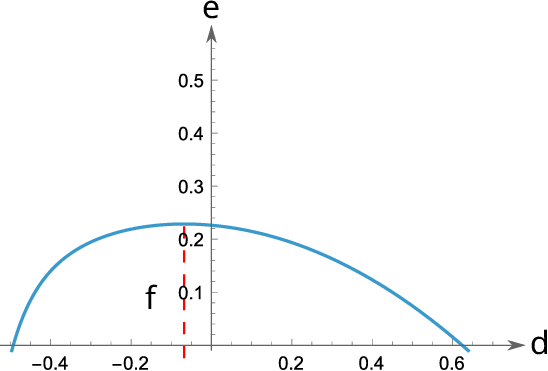}}
            \caption{The two players' optimal choice of $\rho_i$ in asymmetric equilibrium for piecewise-constant ranking.}\label{Ex-03-Epsilons-fig}
        \end{center}%calculated in faking-prz01.nb
    \end{figure}

    In our third example we give up the stipulation that probabilities must strictly increase in the quality spread (Q2) and the differentiability requirement (Q5).  We define a---to our knowledge novel---ranking technology $q_i(\bm{\theta},r)$ that is only weakly increasing in $\theta_i$ and weakly decreasing in $\theta_j$.  Inspired by the all-pay auction, we use a piecewise constant function to define\footnote{Formulation \eqref{def:ranking-constant} is a simplified but equivalent form of the Logistic contest success function defined in footnote \ref{fn:fn-q-def}.}
    \begin{dmath}\label{def:ranking-constant}
        q_i(\bm{\theta},r)\hiderel{=}\begin{cases}
            d(r),   & x>0 \\
            1/2,    & x=0 \\
            1-d(r), & x<0 \\
        \end{cases},
    \end{dmath}
    for $x=\theta_i-\theta_j$, in which $d(r)$ is constant in $x$, satisfying $\lim_{r\to\infty} d(r)=1$ and $\lim_{r\to0} d(r)=1/2$.  The responsiveness of this ranking technology is illustrated in Figure \ref{Ex-03-intuition-fig}.  To fix ideas we set
    \begin{dmath}\label{def:ranking-constant2}
        d(r)=\frac{\delta r+1}{\delta r+2}\condition{$\delta \in \mathds{R}_+$}.
    \end{dmath}
    The regularity condition \eqref{eq:IHR} from the existence appendix \ref{existence_app} is violated, since for $x>0$ we have
    \begin{dmath}
        h'(r)+s'(r)= -\frac{1}{(r+1)^2}.
    \end{dmath}
    %drawn where?
    Consumers assess expected qualities of the first- and second-ranked products \eqref{cons-exp1_equ_dr}, given the observed ranking, as:
    \begin{dmath}\label{eq:ex2-exp-cons2}\begin{array}{rcl}
            \Lambda^1(r) & = & 
            \dfrac{2}{\bar{\theta}^2}\left(\dint_0^{\bar{\theta}} \dint_0^{\tilde{\theta}_1} q_1(\tilde{\bm{\theta}},r)\tilde{\theta}_1d\tilde{\theta}_2  d\tilde{\theta}_1+
            \dint_0^{\bar{\theta}} \dint_0^{\tilde{\theta}_1 } \left(1-q_1(\tilde{\bm{\theta}},r)\right) \tilde{\theta}_2d\tilde{\theta}_2  d\tilde{\theta}_1\right) \\
                         & = & \dfrac{\bar{\theta}\left(2\delta r+3\right)}{3r+6},                                                                                   \\
            \Lambda^2(r) & = & 
            \dfrac{2}{\bar{\theta}^2}\left(\dint_0^{\bar{\theta}} \dint_0^{\tilde{\theta}_1 } \left(1-q_1(\tilde{\bm{\theta}},r)\right) \tilde{\theta}_1d\tilde{\theta}_2d\tilde{\theta}_1+
            \dint_0^{\bar{\theta}} \dint_0^{\tilde{\theta}_1 } q_1(\tilde{\bm{\theta}},r) \tilde{\theta}_2d\tilde{\theta}_2  d\tilde{\theta}_1\right)                \\
                         & = & \dfrac{\bar{\theta}  \left(r+\delta 3\right)}{3r+6}.
        \end{array}\end{dmath}
    The same steps as in the previous subsections, with parameters $\bar{\theta}=1,s=10$ and $\delta=10$, result in the two information release functions:
    \begin{dmath}%calculated in ps-diffex-01.nb veryfied in faking-info-daniel.nb
        \rho_1^\ast(x) \approx \begin{cases}
            0.531,                      & x>0  \\
            \left(0.531-0.068\right)/2, & x=0; \\
            -0.068,                     & x<0  \\
        \end{cases}	\ \rho_2^\ast(x) \hiderel{\approx} \begin{cases}
            -0.068,                     & x>0  \\
            \left(0.531-0.068\right)/2, & x=0. \\
            0.531,                      & x<0  \\
        \end{cases}
    \end{dmath}
    For the same parameter values of $\theta_1=3/4$, $\theta_2=1/4$ as in the previous subsection (and all other pairs for which $\theta_1>\theta_2$), this yields the asymmetric equilibrium candidate
    \begin{dmath}
        \rho_1^\ast \hiderel{\approx} 0.531,\ \rho_2^\ast \hiderel{\approx} -0.068, \text{ resulting in } r^\ast \hiderel{\approx} 0.463,
    \end{dmath}
    which is verified in Figure~\ref{Ex-03-Epsilons-fig}. Due to the fact that the regularity condition is not satisfied for this class, we do not verify the sufficient condition from Proposition \ref{existence_p2}.
    As in the previous subsection for the difference ranking, and in contrast to the ratio-form example, the lower-quality firm obfuscates the ranking in this example.
    
\begin{figure}[!ht]\fontsize{11}{15} \selectfont
        \begin{center}
            \psfrag{a}{$\theta_1$}
            \psfrag{b}{\raisebox{3pt}{$\rho_i^\ast(x)$}}
            \psfrag{c}{$\rho_2^\ast(x)$}
            \psfrag{d}{$\rho_1^\ast(x)$}
            \psfrag{e}{$\tilde{x}$}
            \psfrag{f}{$r^\ast(x)$}
            \psfrag{g}{$s$}
            \psfrag{h}{\raisebox{4pt}{$\rho_i^\ast(s)$}}
            \psfrag{i}{\raisebox{2pt}{$\rho_2^\ast(s)$}}
            \psfrag{j}{$\rho_1^\ast(s)$}
            \psfrag{k}{$\tilde{s}$}
            \psfrag{l}{$r^\ast(s)$}
            \resizebox{\dimexpr .49 \textwidth}{!}{\includegraphics{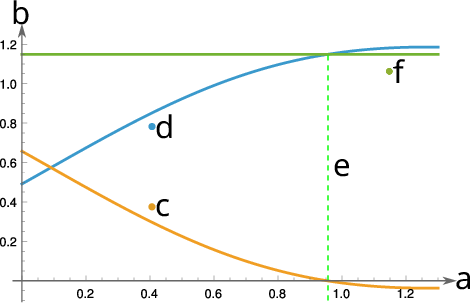}}
            \resizebox{\dimexpr .49 \textwidth}{!}{\includegraphics{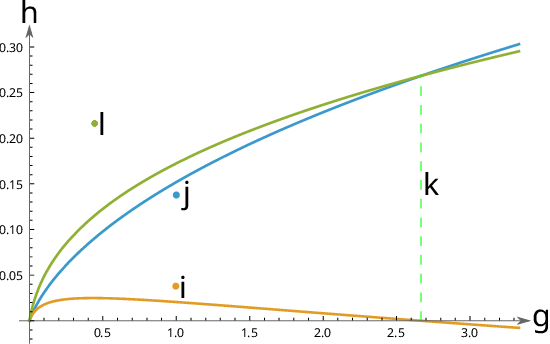}}
            \caption{Left: information dissemination $\rho_i^\ast$ for the piecewise-constant ranking, for parameters $d=10$ and $s=10$, as functions of $\theta_1 \in [0,1]$ for fixed $\theta_2=1/10$: $\rho_1^\ast(x)$ is plotted blue, $\rho_2^\ast(x)$ is shown gold, and aggregate information $r^\ast(x)=\rho_1^\ast(x)+\rho_2^\ast(x)$ is green.  The vertical dashed line indicates the step at which qualities swap ranks.  Right: information dissemination $\rho_i^\ast(s)$ as a function of the consumer heterogeneity $s$, $d=10$, $\bar{\theta}=1$.  The dashed green line shows the critical heterogeneity $\tilde{s}\approx 2.667$ at which $\rho_2^\ast=0$.}\label{Resp02-3-fig}
        \end{center}%computed in faking-prz02.nb and ps-constex-01.nb
    \end{figure}

    The left panel of Figure~\ref{Resp02-3-fig} fixes the heterogeneity at $s=10$ and shows the firms' (almost everywhere constant) information emission as a function of $x$.
    Although information dissemination is piecewise constant in types, the right-hand panel of Figure~\ref{Resp02-3-fig} illustrates that information release is still a function of heterogeneity $s$.  As with the quality dependency in the other examples (Figures~\ref{Resp02-1-fig}, \ref{Resp02-2-fig}, and \ref{Resp02-4-fig}), there is a critical heterogeneity $\tilde{s} \approx 2.667$ beyond which $\rho_2^\ast$ turns negative.

    \subsection{Noise-based ranking}\label{lnr_example_sec}
    
\begin{figure}[htb!]
        \begin{center}\large
            \begin{psfrags}
                \psfrag{a}{$q(\bm{\theta},r)$}
                \psfrag{b}{$\theta_{1} - \theta_2$}
                \psfrag{c}{$r$}
                \psfrag{d}{$r/2$}
                \psfrag{e}{$2r$}
                \psfrag{g}{$\theta_{1}=\theta_{2}$}
                \psfrag{h}{$r=0$}
                \psfrag{j}{\gcol{$\theta_1<\theta_2$}}
                \psfrag{k}{$\theta_1>\theta_2$}
                \scalebox{.6}{\includegraphics[width=\textwidth]{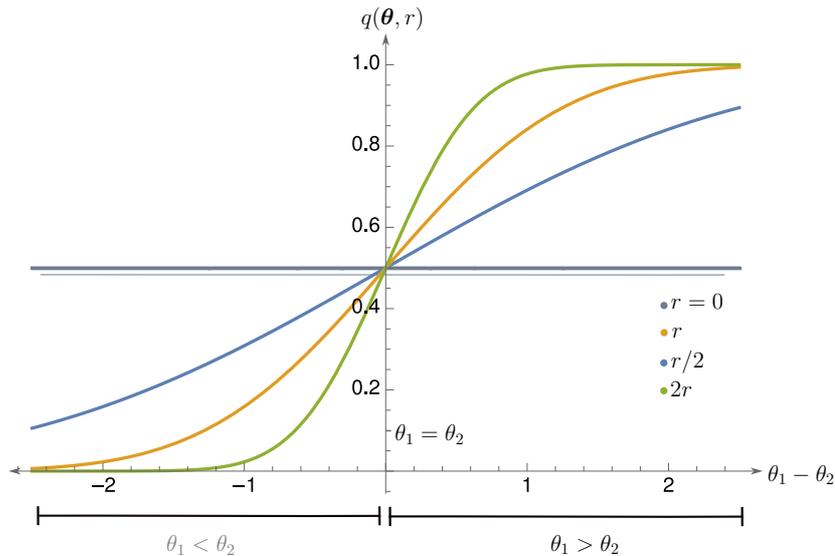}}
            \end{psfrags}
        \end{center}
        \caption{Reaction of noise-based ranking probabilities \eqref{def:ranking-lazear-rosen} to quality-difference and precision.}\label{Ex-04-intuition-fig}
    \end{figure}

    In a final example of a ranking function satisfying our assumptions $\mathcal{Q}$, we now consider a form of noise-based assignment \`a la \citet{Lazear_Rosen81}.  In this case, the idea is that random ``noise'' in addition to qualities determines the ranking and aggregate market-emitted information controls the variance of this noise.  More precisely, the two firms' probabilities of ranking first depend on the difference of qualities $x$ (as in subsection \ref{linear_example_sec}) and independent normally distributed noise with expectation zero and variance\footnote{Notice that Proposition~\ref{proposition_1} ensures that $r>0$ in equilibrium.}
    \begin{dmath}
        r = \frac{1}{|\rho_1+\rho_2|}.
    \end{dmath}
    Under such a ranking, firm $i$ is ranked first with probability
    \begin{dmath}\label{def:ranking-lazear-rosen}
        \Pr\left[\theta_i + \varepsilon_i \geq \theta_j + \varepsilon_j\right] \hiderel{=}
        \Pr\left[\varepsilon_j - \varepsilon_i \leq \theta_i - \theta_j\right]
    \end{dmath}
    which can be expressed using the normally distributed random variable $\eps = \varepsilon_j - \varepsilon_i$ with expectation zero and variance $r$ for i.i.d.\ noise terms denoted by $N_r$ \citep{Lazear_Rosen81}.  The responsiveness of this technology is illustrated in Figure \ref{Ex-04-intuition-fig}. %https://online.stat.psu.edu/stat500/book/export/html/572
    
    %drawn where?
    
    This yields firm 1's probability to be ranked first as\footnote{Formulation \eqref{def:ranking-lazear-rosen2} is a simplified but equivalent form of the Logistic contest success function defined in footnote \ref{fn:fn-q-def}.}
    \begin{dmath}\label{def:ranking-lazear-rosen2}
        q_1(\bm{\theta},r) \hiderel{=} \Pr\left[\theta_i + \varepsilon_i \geq \theta_j + \varepsilon_j\right] \hiderel{=} N_{r}(x),
    \end{dmath}
    for $x=\theta_1-\theta_{2}$. Under the current assumptions this simplifies to
    \begin{dmath}\label{def:ranking-lazear-rosen3}
        q_1(\bm{\theta},r) \hiderel{=} \frac{1}{2}\left(1-\text{erf}\left(\frac{-r x}{\sqrt{2}}\right)\right) \text{ for error function } \text{erf}(z) =\frac{2}{\sqrt{\pi}} \int_0^z e^{-t^2}dt.
    \end{dmath}
    The regularity condition \eqref{eq:IHR} from the existence appendix \ref{existence_app} is a transcendental inequality, which can be numerically confirmed for $r>0$ and $x>0$.
    For the usual $q_1=q$ and $q_2=1-q$ in \eqref{cons-exp1_equ_dr}, $\tilde{x}=\tilde{\theta}_1-\tilde{\theta}_2$, and joint order statistic $f_{(1,2:2)}$ from \eqref{eq-jointorder_highlow-simplified} simplifying to $2/\bar{\theta}^2$, consumer expectations under the noise-based ranking are:
    \begin{equation}\label{eq:exp-cons-lnr2}\parbox{.9\linewidth}{%
            \begin{dgroup*}
                \begin{dmath*}
                    \Lambda^1(r) = \dfrac{4\sqrt{\dfrac{2}{\pi}}+3r^3\bar{\theta}^3+e^{-\frac{1}{2}r^2\bar{\theta}^2}\sqrt{\dfrac{2}{\pi}}\left(r^2\bar{\theta}^2-4\right)+r\bar{\theta}\left(r^2\bar{\theta}^2-3\right)\text{erf}\left(\dfrac{r\bar{\theta}}{\sqrt{2}}\right)}{6r^3\bar{\theta}^2},
                \end{dmath*}
                \begin{dmath*}
                    \Lambda^2(r) =
                    \dfrac{-4\sqrt{\dfrac{2}{\pi}}+3r^3\bar{\theta}^3+e^{-\frac{1}{2}r^2\bar{\theta}^2}\sqrt{\dfrac{2}{\pi}}\left(4-r^2\bar{\theta}^2\right)+r\bar{\theta}\left(3-r^2\bar{\theta}^2\right)\text{erf}\left(\dfrac{r\bar{\theta}}{\sqrt{2}}\right)}{6r^3\bar{\theta}^2}.
                \end{dmath*}
            \end{dgroup*}}
    \end{equation}
    Similar steps as for the other ranking functions lead to the same demand side behavior and the resulting supply-side firms' maximization problem (under mutually known $x=\theta_1-\theta_2$) is
    \begin{dmath}
        \underset{\rho_i}{\max\; } q_i(\bm{\theta},r) P^1(r) + \left(1-q_i(\bm{\theta},r)\right) P^2(r) - \frac{\rho_i^2}{2}.
    \end{dmath}
    Solving the resulting first-order conditions numerically in the same  example as for the ratio-form for $\bar{\theta}=1$, $s=30$, $\theta_1=3/4$, and $\theta_2=1/4$ gives
    \begin{dmath}
        \rho_1^\ast \hiderel{\approx} 0.987,\ \rho_2^\ast \hiderel{\approx} 0.161, \text{ implying that } r^\ast \hiderel{\approx} 1.148.
    \end{dmath}
    
    The sufficient condition for equilibrium existence \eqref{existence_p2_iq} for firm 2 is satisfied
    \begin{dmath}
        r^{\ast}\hiderel{\approx}1.148.\hiderel{<}1.504\hiderel{\approx} r^{su}(\bm{\theta})
    \end{dmath}
    
\begin{figure}[ht!]\fontsize{16}{18} \selectfont
        \begin{center}
            \begin{psfrags}
                \psfrag{a}{$\rho_1$}
                \psfrag{b}{\raisebox{3pt}{\hspace{-24pt}{$u_1(\rho_1,\rho_2^\ast)$}}}
                \psfrag{c}{$\rho_1^\ast$}
                \psfrag{d}{$\rho_2$}
                \psfrag{e}{\raisebox{4pt}{\hspace{-24pt}{$u_2(\rho_1^\ast,\rho_2)$}}}
                \psfrag{f}{$\rho_2^\ast$}
                \psfrag{g}{$r^\ast$}
                \resizebox{\dimexpr .49 \textwidth}{!}{\includegraphics{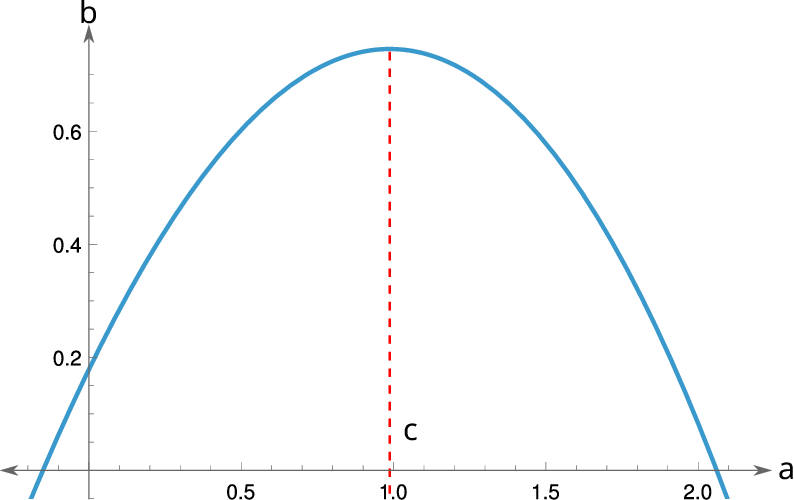}}
                \resizebox{\dimexpr .49 \textwidth}{!}{\includegraphics{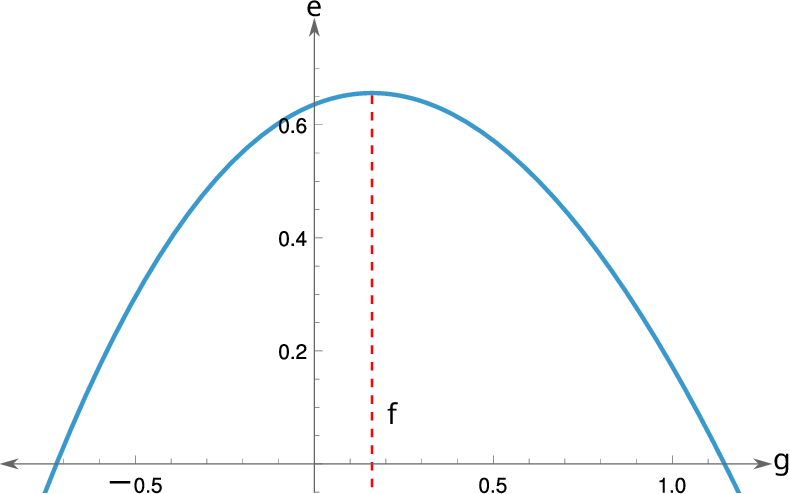}}
            \end{psfrags}%computed in ps-lnrex-02.nb
        \end{center}
        \caption{The two players' optimal choice of $\rho_i$ in asymmetric equilibrium for noise-based ranking.}\label{Ex-04-Epsilons-fig}
    \end{figure}

    Together with global cost convexity, Figure \ref{Ex-04-Epsilons-fig} verifies this candidate as pure strategy equilibrium.
    
\begin{figure}[!ht]\fontsize{11}{15} \selectfont
        \begin{center}
            \psfrag{a}{$\theta_1$}
            \psfrag{b}{\raisebox{4pt}{$\rho_i^\ast(x)$}}
            \psfrag{c}{$\rho_2^\ast(x)$}
            \psfrag{d}{$\rho_1^\ast(x)$}
            \psfrag{e}{$\tilde{x}$}
            \psfrag{f}{$r^\ast(x)$}
            \psfrag{g}{$s$}
            \psfrag{h}{\raisebox{4pt}{$\rho_i^\ast(s)$}}
            \psfrag{i}{$\rho_2^\ast(s)$}
            \psfrag{j}{$\rho_1^\ast(s)$}
            \psfrag{k}{$\tilde{s}$}
            \psfrag{l}{$r^\ast(s)$}
            \resizebox{\dimexpr .49 \textwidth}{!}{\includegraphics{ps-RespX04.eps}}
            \resizebox{\dimexpr .49 \textwidth}{!}{\includegraphics{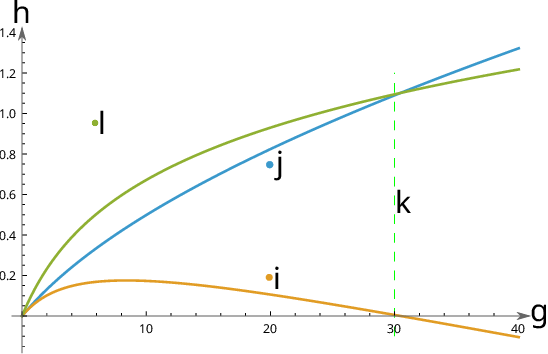}}
            \caption{Left: information dissemination $\rho_i^\ast$ for the noise-based ranking, heterogeneity $s=30$, as functions of $\theta_1 \in [0,1]$ for fixed $\theta_2=1/10$: $\rho_1^\ast(x)$ is plotted blue, $\rho_2^\ast(x)$ is shown gold, and aggregate information $r^\ast(x)=\rho_1^\ast(x)+\rho_2^\ast(x)$ is green.  The vertical dashed line indicates the difference $\tilde{x}$ at which $\rho_2^\ast(x)=0$.  Right: information dissemination $\rho_i^\ast(s)$ as a function of the consumer heterogeneity $s$, and  $\bar{\theta}=1$.  The dashed green line shows the critical heterogeneity $\tilde{s}\approx 31$ at which $\rho_2^\ast=0$.}\label{Resp02-4-fig}
        \end{center}%computed in faking-prz02.nb and ps-lnrex-02.nb
    \end{figure}

    As for the ratio-form, and in contrast to the difference ranking example, the lower-quality firm improves the ranking in this example.  The left panel of Figure \ref{Resp02-4-fig} fixes the heterogeneity at $s=35$ and shows information emission as a function of $x$.
    
    As in the other examples, for sufficiently wide type-difference, there exists a point $\tilde{x}$ at which the lower-quality firm turns to obfuscation.  A similar effect is illustrated in the right-hand panel of Figure \ref{Resp02-4-fig} for a sufficiently large heterogeneity $s$.
    
    \subsection{Other precision contests}\label{others_example_sec}
    
    Although the paper's main economic interest lies in the labeled credence goods application, we also want to point out some general properties of the precision contests we define.  Proposition \ref{proposition_1}, \eqref{lemma_1_equ}, describes an equilibrium fixed point relationship which is responsible for many of the complications in the full labeling application.  If we disentangle the underlying precision contest from the market reaction, however, analytic equilibrium strategies can be easily obtained that confirm some intuition of the general case.
    
    \begin{remark}[Exogenous Prizes]
        If we take demand to be exogenous of information, i.e., $P^1>P^2>0$, a direct consequence of \eqref{lemma_1_equ} is
        \begin{dmath}
            c'(|\rho_1|) + c'(|\rho_2|) \hiderel{=} 0 \hiderel{=} r^\ast
        \end{dmath}
        and the information emitted by the two firms in equilibrium exactly cancels out.  Individual information dissemination is determined by the first-order conditions
        \begin{dmath}
            c'(|\rho_1|) \hiderel{=} \frac{\partial q_1(\bm{\theta},r^\ast=0)}{\partial \rho_i} \left(P^1-P^2\right) \hiderel{=} -c'(|\rho_2|).
        \end{dmath}
        Since costs are invertible, this directly determines the equilibrium information strategies as $\rho^\ast_i = \pm {c'}^{-1}(q'(\bm{\theta},0))$, with the higher-quality firm emitting positive information while the lower-quality firm obfuscates the ranking by choosing $\rho_2<0$.
    \end{remark}
    
    \begin{remark}[Simple endogenous Prizes]
        We now allow prizes $P^1(r),P^2(r)$ to react to aggregate information, but in a less involved fashion than in subsection \ref{labeling_subsec}.  In particular, we assume prizes to be monomials of the form
        \begin{dmath}
            P^1(r) \hiderel{=} \alpha(r)^\beta,\quad P^2(r) \hiderel{=} \frac{\alpha(r)^\beta}{\gamma},\quad \alpha \hiderel{>} 0,\ \beta \hiderel{>} 1,\ \gamma \hiderel{>} 1,
        \end{dmath}
        and costs to be quadratic $c(\rho)=(\rho^2)/2$.
        These functional forms fix the left-hand side of \eqref{lemma_1_equ} as
        \begin{dmath}
            {P^1}'(r) + {P^2}'(r) = \alpha \beta \frac{\gamma+1}{\gamma} r^{\beta-1}
        \end{dmath}
        implying that aggregate available market information is
        \begin{dmath}
            r^\ast = \left(\frac{\alpha \beta + \alpha \beta \gamma}{\gamma}\right)^{\frac{1}{2-\beta}}.
        \end{dmath}
        Given this aggregate precision, individual maximization \eqref{gen-foc01} then leads to the pair of equilibrium information dissemination strategies
        \begin{equation}\label{SimpleG2-inf}\begin{array}{c}
                \rho_1^\ast =\displaystyle \alpha (r^\ast)^{\beta-1} \left( \beta + \beta (\gamma-1) q_1(\bm{\theta},r^\ast) +(\gamma-1) r^\ast q'(\bm{\theta},r^\ast)\right)/\gamma, \\
                \rho_2^\ast =\displaystyle \alpha (r^\ast)^{\beta-1} \left( \beta \gamma - \beta (\gamma-1) q_1(\bm{\theta},r^\ast) -(\gamma-1) r^\ast q'(\bm{\theta},r^\ast)\right)/\gamma.
            \end{array}\end{equation}%calculated in faking-gen01.nb
    \end{remark}
    
\begin{figure}[!ht]
        \begin{center}\footnotesize
            \psfrag{b}{$\theta_1$}
            \psfrag{c}{$q_2(\bm{\theta},r^\ast)$}
            \psfrag{d}{$q_1(\bm{\theta},r^\ast)$}
            \psfrag{e}{$\rho_2^\ast(\theta_1)$}
            \psfrag{f}{$\rho_1^\ast(\theta_1)$}
            \psfrag{g}{$r^\ast(\theta_1)$}
            \psfrag{h}{$\theta_2$}
            \scalebox{1}{\includegraphics{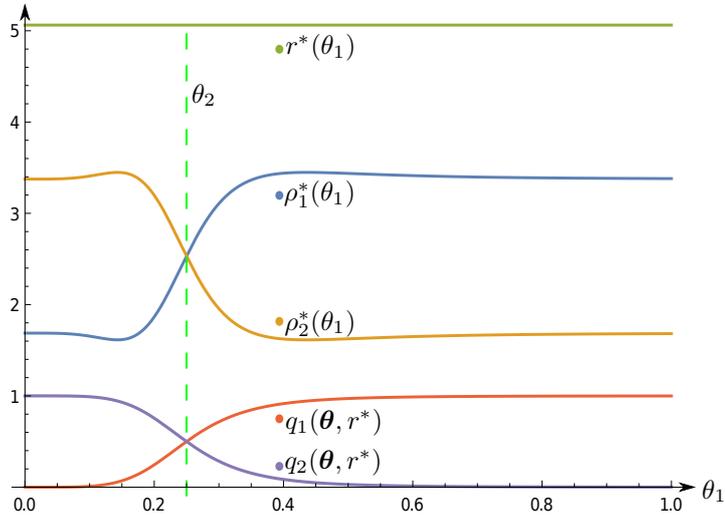}}
            \caption{Information dissemination \eqref{SimpleG2-inf} for $\theta_2=1/4$ as a function of $\theta_1 \in [0,1]$ is non-monotonic: $\rho_1^\ast$ is plotted blue, $\rho_2^\ast$ gold, aggregate information $r^\ast=\rho_1^\ast+\rho_2^\ast$ green.  For comparison, ranking probabilities $q_1(\bm{\theta},r^\ast)$ and $q_2(\bm{\theta},r^\ast)$ are drawn red and purple, respectively.}\label{SimpleG2-fig}
        \end{center}%computed in faking-gen01.nb
    \end{figure}

    Figure~\ref{SimpleG2-fig} illustrates this result for a simple Tullock-ranking example \eqref{def:ranking-tullock} with parameters $\theta_2=1/4$, $\alpha=1$, $\beta=3/2$, and $\gamma=2$.  Notice that---even in this simple case---information dissemination strategies $\rho_i$ are non-monotonic and, for any realization $\bm{\theta}$, aggregate market-provided information $r^\ast$ is constant.
    
    \section{Appendix: Equilibrium existence}\label{existence_app}
    
    This appendix provides formal arguments establishing the existence of a pure-strategy equilibrium in the firms' information release choices.  While Proposition~\ref{existence_p1} shows that the utility function of the high-quality firm is well-behaved and concave, the strategic problem for the low-quality firm is more complex. The tension between the beneficial ``differentiation effect'' and the detrimental ``categorization effect'' can introduce non-concavities into the low-quality firm's utility function, which invalidates standard existence proofs. The following analysis overcomes this challenge by demonstrating that the low-quality firm's utility function is quasi-concave under a set of sufficient conditions. This is achieved by leveraging the single-crossing property and advanced aggregation results from \cite{Quah_Strulovici12}. The existence results rely on additional assumptions on the class of rankings. To formalize these assumptions, we first define the ranking's hazard and success rates:
    \begin{dmath}\label{eq:HR}
        h(\bm{\theta},r)=\frac{\dfrac{\partial q_1(\bm{\theta},r)}{\partial r}}{1-q_1(\bm{\theta},r)}\text{ and } s(\bm{\theta},r)\hiderel{=}\frac{\dfrac{\partial q_1(\bm{\theta},r)}{\partial r}}{q_1(\bm{\theta},r)}.
    \end{dmath}
    We assume that the ranking's hazard rate is (sufficiently) increasing in both arguments, i.e.,
    \begin{equation}\label{eq:IHR}\tag{R1}
        \frac{\partial h(\bm{\theta},r)}{\partial r}+\frac{\partial s(\bm{\theta},r)}{\partial r}\geq 0,
    \end{equation}
    \begin{equation}\label{eq:IHRt}\tag{R2}
        \frac{\partial h(\bm{\theta},r)}{\partial \theta_i}\geq 0 \quad \text{ for } \theta_i>\theta_j.
    \end{equation}
    
    These two regularity conditions impose intuitive restrictions on the ranking technology to ensure it is well-behaved. The ranking's hazard rate, $h(\bm{\theta},r)$, can be interpreted as the instantaneous gain in the probability of being correctly ranked, conditional on being misranked at precision $r$. Assumption \eqref{eq:IHR} requires that this conditional probability does not decrease too quickly as precision increases, ensuring a predictable relationship between information and ranking accuracy \citep[see, e.g.,][]{Barlow_Proschan65}. Assumption \eqref{eq:IHRt} stipulates that for the higher-quality firm, this hazard rate is increasing in its own quality. This means that as the quality gap between firms widens, the ranking technology becomes more effective at discriminating between them, making an increase in precision more impactful.
    
    \begin{proposition}\label{existence_p1}
        Firm 1's utility function \eqref{gen-max3_equ} is concave in $\rho_1$ for any fixed $\rho_2^\ast$.
    \end{proposition}
    The proof shows that firm 1's benefit function---i.e., utility \eqref{gen-max3_equ} without costs---is concave and increasing in $\rho_1$.  Conceptually, the monotonicity of firm $1$'s benefits can be attributed to the dual interplay of effects: 1) \emph{differentiation:} prizes $P^i(r)$ increase monotonically in $r$, with ${P^1}(r)>{P^2}(r)>0$, and 2) \emph{categorization:} the ranking function $q(\bm{\theta},r)$ can discriminate more correctly between qualities as precision $r$ increases.  The combined effect allows firm 1 to assert the higher prize $P^1$, as precision increases.  Together with the assumed concavity of $-c(|\rho_1|)$ the result follows.
    Since firm 1's benefit function is concave and increasing for any $\rho_2^\ast$, standard methods imply that an optimal best response $\rho_1^\ast$ exists to any of firm 2's choices of $\rho_2$ \citep{Rosen65}.
    
    \begin{proposition}\label{existence_p2}
        A sufficient condition for quasi-concavity of firm 2's utility function in $r>0$ is that 
        \begin{dmath}\label{existence_p2_iq}
            r^{\ast}(s,\delta)<r^{su}(\bm{\theta})
        \end{dmath}
        in which: 
        \begin{enumerate}
            \item $r^{\ast}(s,\delta)$ is determined by equation \eqref{lemma_1_equ}, strictly increasing in $s$, strictly decreasing in $\delta$.
            \item $r^{su}(\bm{\theta})$ is strictly increasing for vanishing type spreads and determined by equation
                  \begin{dmath}\label{existence_p2_iq2}
                      h(\bm{\theta},r) = 1/r.
                  \end{dmath}
        \end{enumerate}
    \end{proposition}
    This proposition establishes a sufficient condition for the existence of a pure-strategy equilibrium by ensuring that the utility function of the low-quality firm is quasi-concave in its choice of information release, $\rho_2$.  Condition \eqref{existence_p2_iq} ensures that the equilibrium level of aggregate information $r^\ast$---which is increasing in consumer heterogeneity $s$ and decreasing in the cost parameter $\delta$---remains below a critical threshold $r^{su}(\bm{\theta})$. This threshold is determined by the properties of the ranking technology $q$ and the firms' quality spread $\bm{\theta}$. The intuition for this condition stems from the strategic tension faced by firm 2: while higher market precision $r$ is beneficial as it softens price competition by increasing the size of both contest prizes (``differentiation effect''), it is also detrimental as it increases the probability of being correctly identified as the lower-quality firm, thus reducing its chance of winning the larger prize (``categorization effect''). This trade-off can create non-concavities in firm 2's utility function, precluding standard existence arguments that rely on global concavity.
    
    The proof therefore proceeds with a more general technique, showing that the first derivative of firm 2's utility function satisfies the single-crossing property, which is sufficient for quasi-concavity. The main steps of the proof are as follows: first, the marginal utility is decomposed into a sum of four functions, each reflecting one of the underlying economic effects. Second, the core of the argument leverages the aggregation theorem of \cite{Quah_Strulovici12}, which provides conditions under which a sum of single-crossing functions is itself single-crossing. This condition, known as pairwise signed-ratio monotonicity (SRM), must be verified for every pair of functions in the sum that have opposite signs. The verification of this technical condition relies on a series of supporting lemmas. Specifically, Lemma A provides crucial bounds related to the cost function and the concavity of the ranking technology, while Lemma B establishes a key inequality concerning the prize differential. Lemma C connects the assumed regularity conditions on the ranking technology to the behavior of its second derivative. The regularity assumptions \eqref{eq:IHR} and \eqref{eq:IHRt} are technical conditions on the ranking's hazard rate that ensure it behaves predictably with respect to changes in precision and quality, which is essential for both the aggregation argument and for characterizing the properties of the stability threshold. Together, these lemmas provide the necessary mathematical bounds to simplify the SRM checks. This verification process ultimately distills the complex, multi-part SRM requirement into a single, interpretable condition: that the ranking's hazard rate, $h(\bm{\theta},r)$, is not too large, specifically satisfying \eqref{existence_p2_iq2}. The threshold $r^{su}(\bm{\theta})$ is then defined as the unique value of $r$ that solves this inequality with equality, thus defining the upper bound on permissible equilibrium precision for which existence is guaranteed.

\begin{remark}[Off-equilibrium path beliefs]\label{rem:off-eqm-beliefs}
    The underlying Bayesian Perfect Nash equilibrium requires uninformed consumers to hold beliefs $\tilde{q}(\bm{\theta},r,\bm{p})$ whenever they make their product choices.  These beliefs are discussed in section \ref{beliefs_sec} and defined in \eqref{equ:beliefs-gen}.  We have shown that, along the equilibrium path, the observed signal $r>0$ is uninformative of qualities, as Proposition \ref{proposition_1} ensures that firms with quadratic einformation costs \eqref{ass:q-costs} always emit the same aggregate information, whatever the realization of $\bm{\theta}$.
    
    Hence, rational consumers expect the equilibrium labeling precision $r^\ast$ and (identity-independent) equilibrium price announcements $\bm{p}^\ast$.  If some realization is incompatible with these expectations, we assume that consumers \emph{ignore} any deviations and retain their unchanged priors \eqref{equ:priors} as consumers cannot rationally make any credible precision- or price-based inferences: if a consumer were to interpret a deviation as an additional signal of quality, the same announcement could be made by the competing firm (with arbitrary quality), thus eliminating firms' ability to signal quality.
    As we use the priors $q(\bm{\theta},r) = q(\bm{\theta},r^\ast,\bm{p}^\ast)$ in our existence arguments, Propositions \ref{existence_p1} and \ref{existence_p2}, this belief structure discourages deviations.
\end{remark}

\section{Appendix: Proofs}\label{appendix}

\begin{proof}[\textnormal{\textbf{Proof of Proposition~\ref{comp_prop}}}]\hspace{1mm}\\
    The proof is derived from \citet{Gabszewicz_Shaked_Sutton_Thisse81}.  Since our setup is slightly different, we state the full proof here which establishes that $(p^1)^\ast=p^1(r)^\ast$ and $(p^2)^\ast=p^2(r)^\ast$ are maximizers of the revenue functions $P^1(r),P^2(r)$ defined in \eqref{eqm-profits-def}.
    
    \begin{enumerate}
        \item Consider the pricing problem \eqref{eqm-prices-def} for rank $k=1$ and tentatively denote by $\hat{\mu}^1_2=\hat{\mu}^1_2(p^1,(p^2)^\ast)$ and by $(p^1)^\ast$ a solution to the corresponding first-order condition
              \begin{dmath}\label{eq:p1-existence-foc}
                  \frad{\partial P^1(r)}{\partial p^1}= -\frac{p^1 G'\left(\hat{\mu}_2^1\right)}{\Lambda _1(r)-\Lambda _2(r)}-G\left(\hat{\mu}_2^1\right)+G(s)\hiderel{=}0.
              \end{dmath}
              Consider the second derivative of $P^1$ with respect to $p^1$
              \begin{dmath}
                  \frad{\partial P^1(r)}{\partial^2 p^1}= -\frac{p^1 G''\left(\hat{\mu}_2^1\right)}{\left(\Lambda^1(r)-\Lambda^2(r)\right)^2}-\frac{2 G'\left(\hat{\mu}_2^1\right)}{\Lambda^1(r)-\Lambda^2(r)}.
              \end{dmath}
              This is smaller than $0$ whenever
              \begin{dmath}
                  -\frac{2 G'\left(\hat{\mu}_2^1\right)}{\Lambda^1(r)-\Lambda^2(r)}<\frac{p^1 G''\left(\hat{\mu}_2^1\right)}{\left(\Lambda^1(r)-\Lambda^2(r)\right)^2}.
              \end{dmath}
              Multiplication with $(\Lambda^1(r)-\Lambda^2(r))^2>0$ gives
              \begin{dmath}\label{eq:p1-existence-1}
                  -2 G'\left(\hat{\mu}_2^1\right)\left(\Lambda^1(r)-\Lambda^2(r)\right)<p^1 G''\left(\hat{\mu}_2^1\right).
              \end{dmath}
              For distributions $G$ with strictly increasing hazard rate \eqref{ex-cond_equ} we have
              \begin{dmath}\label{def:cons-IHR}
                  \frad{d}{d \mu}\left(\frac{G'(\mu)}{1-G(\mu)}\right)\hiderel{>}0\iff-\frac{G'(\mu )^2}{1-G(\mu )}\hiderel{<}G''(\mu ).
              \end{dmath}
              Bounding the right-hand side of \eqref{eq:p1-existence-1} and simplification gives
              \begin{dmath}\label{eq:p1-existence-2}
                  2\left(\Lambda^1(r)-\Lambda^2(r)\right)>p^1\frac{G'(\mu)}{1-G(\mu)}.
              \end{dmath}
              Set $p^1=(p^1)^\ast$. Solving $\eqref{eq:p1-existence-foc}$ for $(p^1)^\ast$ gives
              \begin{dmath}
                  (p^1)^\ast=\frac{\left(\Lambda _1(r)-\Lambda _2(r)\right) \left(G(s)-G\left(\hat{\mu }_2^1\right)\right)}{G'\left(\hat{\mu }^1_2\right)}.
              \end{dmath}
              Inserting this back into \eqref{eq:p1-existence-2}, together with $G(s)=1$, gives
              \begin{dmath}\label{eq:p1-existence-3}
                  2\left(\Lambda^1(r)-\Lambda^2(r)\right)>\left(\Lambda^1(r)-\Lambda^2(r)\right),
              \end{dmath}
              which is true since $\left(\Lambda^1(r)-\Lambda^2(r)\right)>0$. Hence, at every critical point determined by \eqref{eq:p1-existence-foc}, $P^1(r)$ is strictly concave. Since $P^1(r)$ is continuous, this establishes the claimed result.
              
        \item Consider consider the pricing problem \eqref{eqm-prices-def} for rank $k=2$ and tentatively denote by $\hat{\mu}^1_2=\hat{\mu}^1_2((p^1)^\ast,{p}^2)$, $\hat{\mu}^2_3=\hat{\mu}^2_3(p^2)$ and by $(p^2)^\ast$ a solution to the corresponding first-order condition
              \begin{dmath}\label{def:p2-first-der}
                  \frad{\partial P^2(r)}{\partial p^2}=G\left(\hat{\mu}_2^1\right)-G\left(\hat{\mu}_3^2\right)-p^2 \left(\frac{G'\left(\hat{\mu}_2^1\right)}{\Lambda^1(r)-\Lambda^2(r)}+\frac{G'\left(\hat{\mu}_3^2\right)}{\Lambda^2(r)}\right)\hiderel{=}0.
              \end{dmath}
              The second derivative equals
              \begin{dmath}\label{def:p2-second-der}
                  \frad{\partial P^2(r)}{\partial p^2} =p^2\left(\frac{G''\left(\hat{\mu}_2^1\right)}{\left(\Lambda^1(r)-\Lambda^2(r)\right)^2}-\frac{G''\left(\hat{\mu}_3^2\right)}{\Lambda^2(r)^2}\right)-2\left(\frac{G'\left(\hat{\mu}_2^1\right)}{\Lambda^1(r)-\Lambda^2(r)}+\frac{G'\left(\hat{\mu}_3^2\right)}{\Lambda^2(r)}\right).
              \end{dmath}
              Define
              \begin{dmath}\label{def:p2-help-func}
                  M=G\left(\hat{\mu}_2^1\right)-G\left(\hat{\mu}_3^2\right),\quad \chi_1\hiderel{=}\frac{-1}{\Lambda^1(r)-\Lambda^2(r)}, \quad \chi_2\hiderel{=}\frac{1}{\Lambda^2(r)}
              \end{dmath}
              and observe that $\chi_1<0$ and $\chi_2>0$ by Lemma~\ref{lemma_consumerexp}. Inserting these definitions into \eqref{def:p2-first-der}, we simplify the first order condition to
              \begin{dmath}\label{eq:p2-existence-foc}
                  M+p^2 \chi _1 G'\left(\hat{\mu}_2^1\right)-p^2 \chi _2 G'\left(\hat{\mu}_3^2\right)=0.
              \end{dmath}
              Solving for $p^2$ establishes
              \begin{dmath}\label{eq:p2-pos}
                  (p^2)^\ast=\frac{M}{\chi _2 G'\left(\hat{\mu}_3^2\right)-\chi _1 G'\left(\hat{\mu}_2^1\right)}\hiderel{>}0
              \end{dmath}
              by the fact that $M>0,\chi_1<0,\chi_2>0$ and $G'>0$.
              Setting $p^2=(p^2)^\ast$, the second derivative \eqref{def:p2-second-der} can be represented using \eqref{def:p2-help-func} by
              \begin{dmath}\label{eq:p2-existence-1}
                  \chi_1\left(2 G'\left(\hat{\mu}_2^1\right)+(p^2)^\ast\chi_1 G''\left(\hat{\mu}_2^1\right)\right)-\chi _2\left(2 G'\left(\hat{\mu}_3^2\right)+(p^2)^\ast \chi_2 G''\left(\hat{\mu}_3^2\right)\right).
              \end{dmath}
              Consider the two expressions in parentheses. We want to establish that:
              \begin{enumerate}
                  \item$\left(2 G'\left(\hat{\mu}_2^1\right)+(p^2)^\ast\chi_1 G''\left(\hat{\mu}_2^1\right)\right)>0$.
                  By $\chi_1<0$, and \eqref{eq:p2-pos}, we have
                  \begin{dmath}\label{eq:p2-existence-2}
                      2 G'\left(\hat{\mu}_2^1\right)+(p^2)^\ast\chi_1 G''\left(\hat{\mu}_2^1\right)\geq2 G'\left(\hat{\mu}_2^1\right)+(p^2)^\ast\chi_1 \left\rvert G''\left(\hat{\mu}_2^1\right)\right\rvert\hiderel{>}0.
                  \end{dmath}
                  Solving the first-order condition \eqref{eq:p2-existence-foc} for $\chi_1$ gives
                  \begin{dmath}
                      \chi_1=\frac{(p^2)^\ast \chi_2 G'\left(\hat{\mu}_3^2\right)-M}{(p^2)^\ast G'\left(\hat{\mu}_2^1\right)}.
                  \end{dmath}
                  Plugging this back into \eqref{eq:p2-existence-2} gives
                  \begin{equation}\label{eq:p2-existence-st1}\parbox{.9\linewidth}{%
                          \setlength{\belowdisplayskip}{0pt} \setlength{\belowdisplayshortskip}{0pt}
                          \setlength{\abovedisplayskip}{0pt} \setlength{\abovedisplayshortskip}{0pt}
                          \begin{dgroup*}
                              \begin{dmath*}
                                  2 G'\left(\hat{\mu}_2^1\right)+(p^2)^\ast\left\rvert G''\left(\hat{\mu}_2^1\right)\right\rvert\frac{(p^2)^\ast \chi_2 G'\left(\hat{\mu}_3^2\right)-M}{(p^2)^\ast G'\left(\hat{\mu}_2^1\right)}=
                              \end{dmath*}
                              \begin{dmath*}
                                  2G'\left(\hat{\mu}_2^1\right)-M\left\rvert G''\left(\hat{\mu}_2^1\right)\right\rvert\frac{ 1}{G'\left(\hat{\mu}_2^1\right)}+(p^2)^\ast\chi_2\left\rvert G''\left(\hat{\mu}_2^1\right)\right\rvert\frac{G'\left(\hat{\mu}_3^2\right)}{G'\left(\hat{\mu}_2^1\right)}>
                              \end{dmath*}
                              \begin{dmath*}
                                  2G'\left(\hat{\mu}_2^1\right)-M\left\rvert G''\left(\hat{\mu}_2^1\right)\right\rvert\frac{1}{G'\left(\hat{\mu}_2^1\right)}>0
                              \end{dmath*}
                          \end{dgroup*}}
                  \end{equation}
                  by the fact that $\chi_2>0,\left\rvert G''\left(\hat{\mu}_2^1\right)\right\rvert>0$, and $G'>0$ together with \eqref{eq:p2-pos}. Take the last inequality, rearranging terms, multiplying by $G'\left(\hat{\mu}_2^1\right)>0$, and applying the definition of $M$ yields
                  \begin{dmath}\label{eq:p2-existence-st2}
                      2G'\left(\hat{\mu}_2^1\right)^2>\left(G\left(\hat{\mu}_2^1\right)-G\left(\hat{\mu}_3^2\right)\right)\left\rvert G''\left(\hat{\mu}_2^1\right)\right\rvert.
                  \end{dmath}
                  We simplify to
                  \begin{dmath}
                      2G'\left(\hat{\mu}_2^1\right)^2>G\left(\hat{\mu}_2^1\right)\left\rvert G''\left(\hat{\mu}_2^1\right)\right\rvert,
                  \end{dmath}
                  since, $G\left(\hat{\mu}_3^2\right)\left\rvert G''\left(\hat{\mu}_2^1\right)\right\rvert>0$. The above inequality then follows again using \eqref{def:cons-IHR}.
                  
                  \item $2 G'\left(\hat{\mu}_3^2\right)+(p^2)^\ast \chi_2 G''\left(\hat{\mu}_3^2\right)>0$.
                        Observe that
                        \begin{dmath}\label{eq:p2-existence-3}
                            G'\left(\hat{\mu}_3^2\right)+(p^2)^\ast \chi_2 G''\left(\hat{\mu}_3^2\right)>0
                        \end{dmath}
                        whenever $G''\left(\hat{\mu}_3^2\right)\geq0$, by the fact that $G'\left(\hat{\mu}_3^2\right)>0$, $(p^2)^\ast>0$, and $\chi_2>0$. For the case of $G''\left(\hat{\mu}_3^2\right)<0$ take again the first-order condition \eqref{eq:p2-existence-foc} and solve for $\chi_2$, giving in return
                        \begin{dmath}
                            \chi_2=\frac{(p^2)^\ast \chi _1 G'\left(\hat{\mu}_2^1\right)+M}{(p^2)^\ast G'\left(\hat{\mu}_3^2\right)}.
                        \end{dmath}
                        Plugging this back into \eqref{eq:p2-existence-3} gives
                        \begin{dgroup*}
                            \begin{dmath*}
                                G'\left(\hat{\mu}_3^2\right)+(p^2)^\ast G''\left(\hat{\mu}_3^2\right)\frac{(p^2)^\ast \chi_1 G'\left(\hat{\mu}_2^1\right)+M}{(p^2)^\ast G'\left(\hat{\mu}_3^2\right)} > 0
                            \end{dmath*}
                            \begin{dmath}
                                G'\left(\hat{\mu}_3^2\right)+M\frac{G''\left(\hat{\mu}_3^2\right)}{G'\left(\hat{\mu}_3^2\right)}+\frac{(p^2)^\ast G''\left(\hat{\mu}_3^2\right)\chi_1 G'\left(\hat{\mu}_2^1\right)}{G'\left(\hat{\mu}_3^2\right)} > 0
                            \end{dmath}
                        \end{dgroup*}
                        The left-hand side is again greater than
                        \begin{dmath}
                            G'\left(\hat{\mu}_3^2\right)+G''\left(\hat{\mu}_3^2\right)\frac{M}{G'\left(\hat{\mu}_3^2\right)}>0
                        \end{dmath}
                        whenever $G''\left(\hat{\mu}_3^2\right)<0$, by the fact that $\chi_1<0$ and $G'>0,(p^2)^\ast>0$. The above inequality then again follows by using the definition $M$ and the increasing hazard rate property \eqref{def:cons-IHR}.
              \end{enumerate}
              The above properties, together with $\chi_1<0,\chi_2>0$, establish \eqref{eq:p2-existence-1}. Thus, at any critical point determined by \eqref{eq:p2-existence-foc}, $P^2(r)$ is strictly concave. Since $P^2(r)$ is continuous, this completes the proof.\qedhere
    \end{enumerate}
\end{proof}

\begin{proof}[\textnormal{\textbf{Proof of Lemma~\ref{lemma_consumerexp}}}]\hspace{1mm}\\
    \begin{enumerate}
        \item We start by proving \eqref{lemma-1-1}, $\Lambda^1(r)+\Lambda^2(r)=\Exp[\Theta_{(1:2)}+\Theta_{(2:2)}]$.
              This follows immediately from $q_1(\tilde{\bm{\theta}},r)+q_2(\tilde{\bm{\theta}},r)=1$ (Q1), through
              \begin{equation}\everymath{\displaystyle}\begin{array}{rcl}
                      \Lambda^1(r)+\Lambda^2(r) & = & 
                      \int_0^{\bar{\theta}} \int_0^{\tilde{\theta}_1 }
                      \left(q_1(\tilde{\bm{\theta}},r)\tilde{\theta}_1 + (1-q_1(\tilde{\bm{\theta}},r)) \tilde{\theta}_2\right)f_{(1,2:2)}(\tilde{\bm{\theta}})
                      d\tilde{\theta}_2  d\tilde{\theta}_1                                                                            \\ &&+
                      \int_0^{\bar{\theta}} \int_0^{\tilde{\theta}_1 }
                      \left(q_2(\tilde{\bm{\theta}},r)\tilde{\theta}_1+ (1-q_2(\tilde{\bm{\theta}},r)) \tilde{\theta}_2\right)f_{(1,2:2)}(\tilde{\bm{\theta}})
                      d\tilde{\theta}_2  d\tilde{\theta}_1                                                                            \\
                                                & = & \int_0^{\bar{\theta}} \int_0^{\tilde{\theta}_1 }

                      \left(\left(q_1(\tilde{\bm{\theta}},r)+ q_2(\tilde{\bm{\theta}},r)\right) \tilde{\theta}_1 \right.              \\ &&+ \left.\left(q_1(\tilde{\bm{\theta}},r)+q_2(\tilde{\bm{\theta}},r)\right) \tilde{\theta}_2 \right)
                      f_{(1,2:2)}(\tilde{\bm{\theta}}) d\tilde{\theta}_2  d\tilde{\theta}_1                                           \\
                                                & = & \int_0^{\bar{\theta}} \int_0^{\tilde{\theta}_1 }

                      \left(\tilde{\theta}_1 + \tilde{\theta}_2\right)
                      f_{(1,2:2)}(\tilde{\bm{\theta}}) d\tilde{\theta}_2  d\tilde{\theta}_1                                           \\
                                                & = & \Exp[\Theta_{(1:2)}+\Theta_{(2:2)}]=\Exp[\Theta_{(1:2)}]+\Exp[\Theta_{(2:2)}],
                  \end{array}\end{equation}
              and the linearity of the expectation.
              
        \item Consider \eqref{lemma-1-2}, $\FO \geq \Lambda^1(r) \geq \hat{\theta}/2 > \Lambda^2(r) \geq \SO > 0$:
              \begin{itemize}
                  \item We start with $\FO \geq \Lambda^1(r)$. Using the joint density \eqref{eq-jointorder_highlow-simplified} to calculate $\FO$, the first inequality reduces to
                        \begin{dmath}
                            \int_0^{\bar{\theta}} \int_0^{\tilde{\theta}_1 }
                            \tilde{\theta}_1f_{(1,2:2)}(\tilde{\bm{\theta}})
                            d\tilde{\theta}_2  d\tilde{\theta}_1\\ \geq
                            {\int_0^{\bar{\theta}} \int_0^{\tilde{\theta}_1 }
                                \left(q_1(\tilde{\bm{\theta}},r)\tilde{\theta}_1+ \left(1-q_1(\tilde{\bm{\theta}},r)\right) \tilde{\theta}_2\right)f_{(1,2:2)}(\tilde{\bm{\theta}})
                                d\tilde{\theta}_2  d\tilde{\theta}_1.}
                        \end{dmath}
                        By monotonicity of the integral it is sufficient to show
                        \begin{dmath}
                            \tilde{\theta}_1f_{(1,2:2)}(\tilde{\bm{\theta}})\geq
                            \left(q_1(\tilde{\bm{\theta}},r)\tilde{\theta}_1+ \left(1-q_1(\tilde{\bm{\theta}},r)\right) \tilde{\theta}_2\right)f_{(1,2:2)}(\tilde{\bm{\theta}}).
                        \end{dmath}
                        Since $f_{(1,2:2)}>0$, $q_1(\tilde{\bm{\theta}},r)+q_2(\tilde{\bm{\theta}},r)=1$, and $\tilde{\theta}_1>\tilde{\theta}_2$, this confirms the inequality.
                        
                  \item For $\Lambda^1(r) \geq \hat{\theta}/2$ we use again the joint density \eqref{eq-jointorder_highlow-simplified} to calculate $\hat{\theta}/2$ and, following the same steps as above, we need to confirm
                        \begin{dmath}
                            \left(q_1(\tilde{\bm{\theta}},r)\tilde{\theta}_1+ \left(1-q_1(\tilde{\bm{\theta}},r)\right) \tilde{\theta}_2\right)\geq\frac{\tilde{\theta}_1+\tilde{\theta}_2}{2}.
                        \end{dmath}
                        Since $q_1(\tilde{\bm{\theta}},r)\geq1/2$ for $\tilde{\theta}_1\geq\tilde{\theta}_2$ (Q2 \& Q3), this confirms the inequality.
                        
                  \item For $\hat{\theta}/2 \geq \Lambda^2(r)$, the same reasoning leads us to confirm
                        \begin{dmath}
                            \frac{\tilde{\theta}_1+\tilde{\theta}_2}{2}\geq\left(q_2(\tilde{\bm{\theta}},r)\tilde{\theta}_1+ \left(1-q_2(\tilde{\bm{\theta}},r)\right) \tilde{\theta}_2\right)
                        \end{dmath}
                        in which substituting $q_2(\tilde{\bm{\theta}},r)=1-q_1(\tilde{\bm{\theta}},r)$ (Q1) together with $\tilde{\theta}_1\geq\tilde{\theta}_2$ confirms the inequality.
                        
                  \item $\Lambda^2(r)>0$ follows again from monotonicity. It suffices to show that
                        \begin{dmath}
                            q_2(\tilde{\bm{\theta}},r) \tilde{\theta}_1 + \left(1-q_2(\tilde{\bm{\theta}},r)\right) \tilde{\theta}_2>0,
                        \end{dmath}
                        which obviously holds since $\tilde{\theta}_1\geq\tilde{\theta}_2>0$.
              \end{itemize}
        \item Considering \eqref{lemma-1-3}, the limits
              \begin{dmath}
                  \lim_{r\to\infty}\Lambda^1(r)\hiderel{=}\FO,\ \lim_{r\to\infty}\Lambda^2(r)\hiderel{=}\SO
              \end{dmath}
              follow directly from (Q1) and (Q3) and application of the monotone convergence theorem.
              
        \item Consider \eqref{lemma-1-4},
              \begin{dmath}
                  (\Lambda^1)^{(n)}(r) =\int_0^{\bar{\theta}} \int_0^{\tilde{\theta}_1 }
                  \left(\tilde{\theta}_1-\tilde{\theta}_2\right)\frad{\partial^n q_1(\tilde{\bm{\theta}},r)}{\partial r^n}
                  f_{(1,2:2)}(\tilde{\bm{\theta}}) d\tilde{\theta}_2  d\tilde{\bm{\theta}} \condition{$\forall n \hiderel{\in} \mathds{N}^+$}.
              \end{dmath}
              ${\Lambda^1}'(r)$ is computed using the Leibniz integral rule on fixed integration limits, i.e.,
              \begin{dmath}
                  {\Lambda^1}'(r) =	\dfrac{\partial}{\partial r}\left(\int_0^{\bar{\theta}} \int_0^{\tilde{\theta}_1 }	\left( q_1(\tilde{\bm{\theta}},r)\tilde{\theta}_1 +(1-q_1(\tilde{\bm{\theta}},r)) \tilde{\theta}_2 \right)	f_{(1,2:2)}(\tilde{\bm{\theta}}) d\tilde{\theta}_2  d\tilde{\theta}_1\right)=
                  \int_0^{\bar{\theta}} \int_0^{\tilde{\theta}_1 }\dfrac{\partial}{\partial r}\left(\left( q_1(\tilde{\bm{\theta}},r)\tilde{\theta}_1 +(1-q_1(\tilde{\bm{\theta}},r)) \tilde{\theta}_2 \right)	f_{(1,2:2)}(\tilde{\bm{\theta}})\right) d\tilde{\theta}_2  d\tilde{\theta}_1=
                  \int_0^{\bar{\theta}} \int_0^{\tilde{\theta}_1 }
                  \left( \dfrac{\partial q_1(\tilde{\bm{\theta}},r)}{\partial r} \tilde{\theta}_1 -\dfrac{\partial q_1(\tilde{\bm{\theta}},r)}{\partial r} \tilde{\theta}_2 \right)
                  f_{(1,2:2)}(\tilde{\bm{\theta}}) d\tilde{\theta}_2  d\tilde{\theta}_1=
                  \int_0^{\bar{\theta}} \int_0^{\tilde{\theta}_1 }
                  \left(\tilde{\theta}_1-\tilde{\theta}_2\right)\dfrac{\partial q_1(\tilde{\bm{\theta}},r)}{\partial r}
                  f_{(1,2:2)}(\tilde{\bm{\theta}}) d\tilde{\theta}_2  d\tilde{\theta}_1.
              \end{dmath}
              Any higher derivatives $(\Lambda^1)^{(n)}(r)$ then follow inductively
              \begin{dmath}
                  (\Lambda^1)^{(n)}(r) =\int_0^{\bar{\theta}} \int_0^{\tilde{\theta}_1 }
                  \left(\tilde{\theta}_1-\tilde{\theta}_2\right)\dfrac{\partial^n q_1(\tilde{\bm{\theta}},r)}{\partial r^n}
                  f_{(1,2:2)}(\tilde{\bm{\theta}}) d\tilde{\theta}_2  d\tilde{\theta}_1.
              \end{dmath}
              
        \item Concerning \eqref{lemma-1-5}, $\Lambda^1(0)=\Lambda^2(0)\hiderel{=}\hat{\theta}/2$ follows directly from (Q3).\qedhere
    \end{enumerate}
\end{proof}

\begin{proof}[\textnormal{\textbf{Proof of Lemma~\ref{lemma_spacings}}}]\hspace{1mm}\\
    The inequality \eqref{lemma_spacings_equ1} is an extension of the statement that $\Exp[\Theta_{i:n}]$ is bounded, whatever the form of its distribution $F$. %This result is due to \cite{Moriguti53}, \cite{Gumbel54}, and  \cite{Hartley&David54}, elaborated in condensed form in \cite{David_Nagaraja03}.
    By \citet[Inequalities 4.2.6]{David_Nagaraja03}, for any distribution exhibiting a finite mean-variance pair $(m,\sigma^2)$, it is the case that
    \begin{dmath}
        \Exp[\Theta_{1:n}]\leq m+\dfrac{\left(n-1\right)\sigma}{\sqrt{\left(2n-1\right)}}\quad\text{and likewise}\quad\Exp[\Theta_{n:n}]\hiderel{\geq}m-\dfrac{\left(n-1\right)\sigma}{\sqrt{\left(2n-1\right)}.}
    \end{dmath}
    Since the support of $F$ is bounded, its mean and variance are finite. For the case of $n=2$, combining inequalities gives the desired property \eqref{lemma_spacings_equ1}.
\end{proof}

\begin{proof}[\textnormal{\textbf{Proof of Lemma~\ref{lemma_prizescon}}}]\hspace{1mm}\\
    We restrict attention to the case of uniformly distributed consumer tastes $G(\mu)=\mu/s$ and write ${\Lambda^1}'(r) =\partial \Lambda^1(r)/\partial r$ as well as ${\Lambda^1}''(r) = \partial^2 \Lambda^1(r)/\partial r^2$.
    \begin{enumerate}
        \item Consider \eqref{lemma-3-1}, $P^1(r) > P^2(r) > 0$ and $P^1(0)=P^2(0)=0$.
              From \eqref{ex-prizes0_equ}, we have to show that
              \begin{dmath}
                  P^1(r) = 4s\dfrac{(2 \Lambda^1(r)-\hat{\theta}) \Lambda^1(r)^2}{(\hat{\theta}-5 \Lambda^1(r))^2}>
                  s\dfrac{(2 \Lambda^1(r)-\hat{\theta}) (\hat{\theta}-\Lambda^1(r)) \Lambda^1(r)}{(\hat{\theta}-5 \Lambda^1(r))^2}\hiderel{=} P^2(r)\hiderel{>}0.
              \end{dmath}
              Since $s>0$, we know that $(2 \Lambda^1(r)-\hat{\theta})>0$ and $(\hat{\theta}-5 \Lambda^1(r))^2>0$ by \eqref{lemma-1-1} \& \eqref{lemma-1-2}. Simplifying on both sides yields
              \begin{dmath}
                  4\Lambda^1(r)^2\hiderel{>}(\hat{\theta}-\Lambda^1(r)) \Lambda^1(r)\hiderel{>}0
              \end{dmath}
              which is indeed true by \eqref{lemma-1-1} \& \eqref{lemma-1-2}.
              
              To see that $P^1(0)=P^2(0)=0$, take again \eqref{ex-prizes0_equ} and substitute $\Lambda^1(0)=\hat{\theta}/2$, from Lemma~\ref{lemma_consumerexp}, resulting in the denominator $9\hat{\theta}/4>0$. Substituting into the first term in the numerator of both prizes, i.e, $(2 \Lambda^1(r)-\hat{\theta})=0$, yields $P^1(0)=P^2(0)=0$.
              
        \item Consider \eqref{lemma-3-2}, ${P^1}'(r)>0, {P^1}''(r)>0$ as well as ${P^1}'(r) > {P^2}'(r)$.
              \begin{itemize}
                  \item ${P^1}'(r)>0$: The derivative of $P^1(r)$ with respect to $r$ is
                        \begin{dmath}\label{lemma-3-2-eq1}
                            {P^1}'(r) \hiderel{=} 8 s \dfrac{\Lambda^1(r) \left(\hat{\theta}^2-3 \hat{\theta} \Lambda^1(r)+5 \Lambda^1(r)^2\right)}{(5 \Lambda^1(r)-\hat{\theta})^3} {\Lambda^1}'(r).
                        \end{dmath}
                        Dividing by $8s>0$ and multiplying by $(5 \Lambda^1(r)-\hat{\theta})^3>0$ by \eqref{lemma-1-2} gives
                        \begin{dmath}
                            \Lambda^1(r) \left(\hat{\theta}^2-3 \hat{\theta} \Lambda^1(r)+5 \Lambda^1(r)^2\right){\Lambda^1}'(r)>0.
                        \end{dmath}
                        We have $\Lambda^1(r)>0$ by \eqref{lemma-1-2} and ${\Lambda^1}'(r)>0$ for $r>0$ by \eqref{lemma-1-4} together with (Q2). Hence, it is sufficient to show that
                        \begin{dmath}\label{lemma-3-2-eq2}
                            \hat{\theta}^2-3 \hat{\theta} \Lambda^1(r)+5 \Lambda^1(r)^2>0,
                        \end{dmath}
                        which holds by \eqref{lemma-1-2}.
                        
                  \item ${P^1}''(r)<0$: Take $P^1(r)$, differentiating twice with respect to $r$ gives
                        \begin{dmath}
                            {P^1}''(r)=-8s\left(\frac{\hat{\theta }^2 (\hat{\theta }+4 \Lambda^1(r))}{(\hat{\theta }-5 \Lambda^1(r))^4}{\Lambda^1}'(r)^2+\frac{\Lambda^1(r) (\hat{\theta }-5 \Lambda^1(r)) \left(\hat{\theta }^2-3 \hat{\theta } \Lambda^1(r)+5 \Lambda^1(r)^2\right)}{(\hat{\theta }-5 \Lambda^1(r))^4}{\Lambda^1}''(r)\right)\hiderel{<}0.
                        \end{dmath}
                        Dividing by $-8s<0$ and multiplying by $(\hat{\theta}-5 \Lambda^1(r))^4>0$ by \eqref{lemma-1-2} gives
                        \begin{dmath}
                            \hat{\theta }^2 (\hat{\theta }+4 \Lambda^1(r)) {\Lambda^1}'(r)^2 +\\ \Lambda^1(r) (\hat{\theta }-5 \Lambda^1(r)) \left(\hat{\theta }^2-3 \hat{\theta } \Lambda^1(r)+5 \Lambda^1(r)^2\right) {\Lambda^1}''(r) > 0.
                        \end{dmath}
                        Clearly, $\hat{\theta }^2 (\hat{\theta }+4 \Lambda^1(r)){\Lambda^1}'(r)^2>0$, since $\hat{\theta}>0,\Lambda^1(r)>0$ and ${\Lambda^1}'(r)>0$ for $r>0$ by Lemma~\ref{lemma_consumerexp} and (Q1) and (Q2). Hence, it is sufficient to show that
                        \begin{dmath}
                            \Lambda^1(r) \left(\hat{\theta }-5 \Lambda^1(r)\right) \left(\hat{\theta }^2-3 \hat{\theta } \Lambda^1(r)+5 \Lambda^1(r)^2\right){\Lambda^1}''(r)\geq0.
                        \end{dmath}
                        Observe that $\Lambda^1(r)>0$ and $(\hat{\theta }^2-3 \hat{\theta } \Lambda^1(r)+5 \Lambda^1(r)^2)>0$ again by \eqref{lemma-1-2}. Thus, we need to show that
                        \begin{dmath}
                            \left(\hat{\theta }-5 \Lambda^1(r)\right){\Lambda^1}''(r)\geq0.
                        \end{dmath}
                        Since $(\hat{\theta }-5 \Lambda^1(r))<0$ follows again from \eqref{lemma-1-2} and  ${\Lambda^1}''(r)\leq0$ follows from \eqref{lemma-1-4} together with (Q6), this proves the inequality.
                        
                  \item ${P^1}'(r)>{P^2}'(r)$: Take $P^1(r)$ and $P^2(r)$, differentiating with respect to $r$ gives
                        \begin{dmath}\label{lemma-3-4-eq1}
                            {P^1}'(r) = 8 s \dfrac{\Lambda^1(r) \left(\hat{\theta}^2-3 \hat{\theta} \Lambda^1(r)+5 \Lambda^1(r)^2\right)}{(5 \Lambda^1(r)-\hat{\theta})^3} {\Lambda^1}'(r) > s\dfrac{\hat{\theta}^3 - \Lambda^1(r) \left(\hat{\theta}^2-6 \hat{\theta} \Lambda^1(r)+10 \Lambda^1(r)^2\right)}{(5 \Lambda^1(r)-\hat{\theta})^3} {\Lambda^1}'(r)\hiderel{=}{P^2}'(r).
                        \end{dmath}
                        Multiplying both sides by $(5\Lambda^1(r)-\hat{\theta})^3/s \Lambda'_{1}(r)>0$ (established by Lemma~\ref{lemma_consumerexp}) gives
                        \begin{dmath}
                            8\Lambda^1(r) \left(\hat{\theta}^2-3 \hat{\theta} \Lambda^1(r)+5 \Lambda^1(r)^2\right)>\hat{\theta}^3 - \Lambda^1(r) \left(\hat{\theta}^2-6 \hat{\theta} \Lambda^1(r)+10 \Lambda^1(r)^2\right).
                        \end{dmath}
                        Adding $\Lambda^1(r) (\hat{\theta}^2-6 \hat{\theta} \Lambda^1(r)+10 \Lambda^1(r)^2)$ on both hand sides simplifies the above expression to
                        \begin{dmath}
                            \Lambda^1(r) \left(7\hat{\theta}^2-18\hat{\theta} \Lambda^1(r)+30 \Lambda^1(r)^2\right)>\hat{\theta}^3
                        \end{dmath}
                        which is always satisfied since $\Lambda^1(r)>\hat{\theta}/2>0$ for any $r>0$, by Lemma~\ref{lemma_consumerexp}.
              \end{itemize}
              
        \item Consider \eqref{lemma-3-3}: ${P^2}'(r)>0, {P^2}''(r)>0$. First observe that by \eqref{ass1} and \eqref{lemma_spacings_equ1} we have
              \begin{dmath}\label{lemma-3-eq1}
                  \frac{\FO}{\SO}\leq 2.
              \end{dmath}
              Adding $\FO$ on both sides gives
              \begin{dmath}
                  \frac{3}{2}\FO\leq \hat{\theta}.
              \end{dmath}
              By \eqref{lemma-1-2} we substitute $\Lambda^1(r)<\FO$ on the left-hand side and obtain
              \begin{dmath}\label{lemma-3-eq2}
                  3\Lambda^1(r)\leq 2\hat{\theta}.
              \end{dmath}
              This fact establishes
              \begin{dmath}\label{Lemma2-implification}
                  3\Lambda^1(r)\hiderel{\leq} 2\hat{\theta} \hiderel{\Rightarrow} \hat{\theta }^3-\hat{\theta }^2 \Lambda^1(r)+6 \hat{\theta } \Lambda^1(r)^2-10 \Lambda^1(r)^3\hiderel{>}0.
              \end{dmath}
              \begin{itemize}
                  \item ${P^2}'(r)>0$: Differentiating $P^2(r)$ with respect to $r$ gives
                        \begin{dmath}\label{lemma-3-3-eq1}
                            {P^2}'(r) \hiderel{=}s\dfrac{\hat{\theta}^3 - \Lambda^1(r) \left(\hat{\theta}^2-6 \hat{\theta} \Lambda^1(r)+10 \Lambda^1(r)^2\right)}{(5 \Lambda^1(r)-\hat{\theta})^3} {\Lambda^1}'(r)>0.
                        \end{dmath}
                        Dividing by $s>0$ and ${\Lambda^1}'(r)>0$ for $r>0$ by \eqref{lemma-1-4} together with (Q2), and multiplying by $(5 \Lambda^1(r)-\hat{\theta})^3>0$ by \eqref{lemma-1-2} gives
                        \begin{dmath}
                            \hat{\theta}^3>\Lambda^1(r) \left(\hat{\theta}^2-6 \hat{\theta} \Lambda^1(r)+10 \Lambda^1(r)^2\right).
                        \end{dmath}
                        Rearranging and multiplying out gives
                        \begin{dmath}\label{lemma-3-3-eq2}
                            \hat{\theta }^3-\hat{\theta }^2 \Lambda^1(r)+6 \hat{\theta } \Lambda^1(r)^2-10 \Lambda^1(r)^3 \hiderel{>} 0
                        \end{dmath}
                        which is implied by \eqref{Lemma2-implification}.
                        
                  \item ${P^2}''(r)<0$: Take $P^2(r)$, differentiating twice with respect to $r$ gives
                        \begin{dmath}
                            {P^2}''(r)=-2s\frac{\hat{\theta }^2 (7 \hat{\theta }+\Lambda^1(r))}{(\hat{\theta }-5 \Lambda^1(r))^4}{\Lambda^1}'(r)^2-
                            s\frac{(\hat{\theta }-5 \Lambda^1(r)) \left(\hat{\theta }^3-\hat{\theta }^2 \Lambda^1(r)+6 \hat{\theta } \Lambda^1(r)^2-10 \Lambda^1(r)^3\right)}{(\hat{\theta }-5 \Lambda^1(r))^4}{\Lambda^1}''(r)\hiderel{<}0.
                        \end{dmath}
                        Dividing by $-s<0$ and multiplying by $(\hat{\theta}-5 \Lambda^1(r))^4>0$ by \eqref{lemma-1-2} gives
                        \begin{dmath}
                            2\hat{\theta }^2 (7 \hat{\theta }+\Lambda^1(r)){\Lambda^1}'(r)^2 +
                            (\hat{\theta }-5 \Lambda^1(r))\\ \left(\hat{\theta }^3-\hat{\theta }^2 \Lambda^1(r)+6 \hat{\theta } \Lambda^1(r)^2-10 \Lambda^1(r)^3\right){\Lambda^1}''(r)>0.
                        \end{dmath}
                        For the first term we have $2\hat{\theta }^2 (7 \hat{\theta }+\Lambda^1(r)){\Lambda^1}'(r)^2>0$, since $\hat{\theta}>0,\Lambda^1(r)>0$ and ${\Lambda^1}'(r)>0$ for $r>0$ by Lemma~\ref{lemma_consumerexp}, (Q1) and (Q2). It is thus sufficient to show that
                        \begin{dmath}
                            \left(\hat{\theta }-5 \Lambda^1(r)\right) \left(\hat{\theta }^3-\hat{\theta }^2 \Lambda^1(r)+6 \hat{\theta } \Lambda^1(r)^2-10 \Lambda^1(r)^3\right){\Lambda^1}''(r)\hiderel{\geq}0.
                        \end{dmath}
                        Since $(\hat{\theta }-5 \Lambda^1(r))<0$ follows again from \eqref{lemma-1-2}, and ${\Lambda^1}''(r)\leq0$ follows from \eqref{lemma-1-4} together with (Q6), this gives
                        \begin{dmath}
                            \hat{\theta }^3-\hat{\theta }^2 \Lambda^1(r)+6 \hat{\theta } \Lambda^1(r)^2-10 \Lambda^1(r)^3\hiderel{\geq}0.
                        \end{dmath}
                        Implication \eqref{Lemma2-implification} completes the proof.
                  \item ${P^1}''(r)<{P^2}''(r)$. Take $P^i(r)$, differentiating twice with respect to $r$ and simplifying gives
                        \begin{dmath}
                            \frac{6 s \bar{\theta }^2 {\Lambda^1}'(r)^2}{\bar{\theta }-5 {\Lambda^1}(r)}+s {\Lambda^1}''(r) \left(\bar{\theta }^2-4 \bar{\theta } {\Lambda^1}(r)+10 {\Lambda^1}(r)^2\right)\leq 0
                        \end{dmath}
                        The first term is negative by the fact that $s>0,\bar{\theta }>0$, ${\Lambda^1}'(r)>0$ for $r>0$ by \eqref{lemma-1-4} and  $(\hat{\theta }-5 \Lambda^1(r))<0$ by \eqref{lemma-1-2}. The second term is negative by the fact that $s>0$ and ${\Lambda^1}''(r)\leq0$ by \eqref{lemma-1-4} together with (Q6), and $\left(\bar{\theta }^2-4 \bar{\theta } {\Lambda^1}(r)+10 {\Lambda^1}(r)^2\right)\geq0$ by \eqref{lemma-3-eq2}.
                        \qedhere
              \end{itemize}
    \end{enumerate}
\end{proof}

\begin{proof}[\textnormal{\textbf{Proof of Proposition~\ref{proposition_1}}}]\hspace{1mm}\\
    Firm $i$ chooses information $\rho_i$.  We drop the subscripts on rankings, writing $q_1=q$ and $q_2=1-q$, and abuse notation by writing $q'(\bm{\theta},\rho_1+\rho_2)$ for the partial derivative of $q(\bm{\theta},\rho_1+\rho_2)$ with respect to $\rho_1$ or $\rho_2$.  The first-order condition of firm 1's objective \eqref{gen-max3_equ} with respect to $\rho_1$ is then
    \begin{dmath}\label{gen-foc1}
        q(\bm{\theta},r) {P^1}'(r) + \left(1-q(\bm{\theta},r)\right){P^2}'(r) + \frac{\partial q(\bm{\theta},r)}{\partial \rho_1} P^1(r) - \frac{\partial q(\bm{\theta},r)}{\partial \rho_1} P^2(r) - c'(|\rho_1|) = 0
    \end{dmath}
    and the same condition for firm 2 is
    \begin{dmath}\label{gen-foc2}
        \left(1-q(\bm{\theta},r)\right) {P^1}'(r) + q(\bm{\theta},r){P^2}'(r) - \frac{\partial q(\bm{\theta},r)}{\partial \rho_2} P^1(r) + \frac{\partial q(\bm{\theta},r)}{\partial \rho_2} P^2(r) - c'(|\rho_2|) = 0.
    \end{dmath}
    Summing the two for $q'(\bm{\theta},r) = \partial q(\bm{\theta},r)/\partial \rho_1 = \partial q(\bm{\theta},r)/\partial \rho_2$ gives \eqref{lemma_1_equ}.
\end{proof}

\begin{proof}[\textnormal{\textbf{Proof of Proposition~\ref{critical_prp}}}]\hspace{1mm}\\
    Notice that prizes \eqref{ex-prizes0_equ} depend linearly on $s$ while none of the other utility-components depend directly on $s$.  Similarly, ${P^1}'(r)$ and ${P^2}'(r)$ (explicitly determined in Lemma \ref{lemma_prizescon}, \eqref{lemma-3-2-eq1} and \eqref{lemma-3-3-eq1}, respectively) are linear in $s$.  Hence, firm 2's necessary condition \eqref{gen-foc2} can be rewritten as a function of consumer heterogeneity $s$ at the point of interest $r^\ast=\rho_1^\ast>0$ with $\rho_2^\ast=0$ (yielding $c'(|\rho_2^\ast|)=0$, by assumption) as:
    \begin{dmath}\label{eq:prop5-foc2-1}
        %s(\rho_1^\ast)=
        (1-q(\bm{\theta},\rho_1^\ast))\frac{{P^1}'(\rho_1^\ast)}{s}+q(\bm{\theta},\rho_1^\ast)\frac{{P^2}'(\rho_1^\ast)}{s}-q'(\bm{\theta},\rho_1^\ast)\frac{P^1(\rho_1^\ast)}{s}+q'(\bm{\theta},\rho_1^\ast)\frac{P^2(\rho_1^\ast)}{s}=0.
    \end{dmath}
    Define
    \begin{dgroup*}
        \begin{dmath}
            \phi(\rho_1^\ast)=(1-q(\bm{\theta},\rho_1^\ast))\frac{{P^1}(\rho_1^\ast)}{s}\text{ and }
            \psi(\rho_1^\ast) \hiderel{=} q(\bm{\theta},\rho_1^\ast)\frac{{P^2}(\rho_1^\ast)}{s}.
        \end{dmath}
    \end{dgroup*}
    Inserting these definitions, \eqref{eq:prop5-foc2-1} holds whenever
    \begin{dmath}\label{eq:prop5-foc2-2}
        \frac{\phi'(\rho_1^\ast)}{\psi'(\rho_1^\ast)}=-1.
    \end{dmath}
    At this point we apply Cauchy's mean value theorem \citep[see, e.g.,][Theorem 5.12]{Apostol74}, which states that, for functions $\phi$ and $\psi$, both continuous on some closed interval $[a,b]$ and differentiable on the open interval $(a,b)$ with $a<b$, there exists some $\rho_1^\ast\in(a,b)$, such that
    \begin{dmath}
        \frac{\phi'(\rho_1^\ast)}{\psi'(\rho_1^\ast)}=\frac{\phi(b)-\phi(a)}{\psi(b)-\psi(a)},
    \end{dmath}
    provided that $\psi(b)\neq\psi(a)$ and $\psi'(\rho_1^\ast)\neq0$. Note that $\psi$ is strictly increasing and positive by Lemma~\ref{lemma_prizescon} and assumptions $\mathcal{Q}$. Moreover, both functions are continuous and differentiable on any closed interval $[a,b]$, with $0\leq a<b$. For \eqref{eq:prop5-foc2-1} to hold for some $\rho_1^\ast$, it is thus sufficient to provide some interval $[a,b]$, with $a<b$, such that
    \begin{dmath}\label{p5-step-3}
        \xi(a,b)=\frac{\phi(a)-\phi(b)}{\psi(b)-\psi(a)}\hiderel{=}1.
    \end{dmath}
    As next step, note that $\phi(\rho_1^\ast)$ attains a maximum since ${P^1}(\rho_1^\ast)/s$ is concave and increasing, as a consequence of Lemma~\ref{lemma_prizescon} and the fact that $(1-q(\bm{\theta},\rho_1^\ast))$ is strictly decreasing and converges to $0$ as $\rho_1^\ast$ increases (by assumptions $\mathcal{Q}$). Denoting the maximizers of $\phi$ by $\hat{\rho}_1=\argmax_{\rho_1^\ast} \phi(\rho_1^\ast)$, we establish the following claims for $\hat{\rho}_1(\theta_{2})$ and $\phi(\hat{\rho}_1(\theta_{2}),\theta_{2})$ for arbitrarily fixed $\theta_{1}$ with $\theta_{1}>\theta_{2}$:
    \begin{itemize}
        \item[(C1)] The value function $v(\theta_{2})=\phi(\hat{\rho}_1(\theta_{1}),\theta_{1})$ is strictly increasing in $\theta_{2}$.
            Applying the envelope theorem \citep[][Theorem M.L.1]{Mas-Colell_Whinston_Green95} gives
            \begin{dmath}
                \frac{\mathrm{d}v(\theta_{2})}{\mathrm{d}\theta_{2}}=\left.{\frac{\partial \phi(\rho_1,\theta_{2})}{\partial \theta_{2}}}\right|_{\rho_1=\hat{\rho}_1(\theta_{2})}\hiderel{=}-\left.{\frac{\partial q(\bm{\theta},\rho_1)}{\partial \theta_{2}} \frac{P^1(\rho_1)}{s}}\right|_{\rho_1=\hat{\rho}_1(\theta_{2})}
            \end{dmath}
            which is strictly increasing in $\theta_{2}$ for $\theta_{2}<\theta_{1}$, by the fact that $\partial q(\bm{\theta},\rho_1)/\partial \theta_{2}<0$ by assumptions $\mathcal{Q}$, and $P^1(\rho_1)>0$ by Lemma~\ref{lemma_prizescon}.
            
            \item[(C2)]We claim that
            \begin{dmath}
                \lim_{\theta_2\to\theta_1}\hat{\rho}_1(\theta_{2})=\infty
            \end{dmath}
            which follows from the fact that
            \begin{dmath}
                \lim_{\theta_{2}\to\theta_{1}}\phi(\hat{\rho}_1(\theta_{2}),\theta_{2})= \frac{1}{2} \frac{P^1(\hat{\rho}_1)}{s},
            \end{dmath}
            which is strictly increasing in $\hat{\rho}_1$ by Lemma~\ref{lemma_prizescon}.
            
        \item[(C3)] We claim that
            \begin{dmath}
                \lim_{\theta_2\to\theta_1}\phi(\hat{\rho}_1(\theta_{2}),\theta_{2}) = \lim_{\rho_1 \to \infty} \frac{1}{2} \frac{P^1(\rho_1)}{s}
            \end{dmath}
            which follows as a consequence of (C1) and (C2), by the fact that the maximizer $\hat{\rho}_1(\theta_{2})$ converges to infinity.%Here we are somehow unsure about notation.
    \end{itemize}
    
    Now set $a=\hat{\rho}_1$. By continuity of $\phi(b)$ and $\psi(b)$, respectively, we apply the intermediate value theorem \citep[see, e.g.,][Theorem 4.33]{Apostol74}. It is sufficient for $\xi(\hat{\rho}_1,b)=1$ to establish that for an adequately chosen parameter pair $\theta_{1}>\theta_{2}$ within $[0,\bar{\theta}]$
    \begin{dgroup*}
        \begin{dmath}\label{prop5-int-1}
            \exists b: \xi(\hat{\rho}_1,b)\hiderel{<}1,
        \end{dmath}
        \begin{dmath}\label{prop5-int-2}
            \exists b: \xi(\hat{\rho}_1,b)\hiderel{>}1.
        \end{dmath}
    \end{dgroup*}
    For the case of \eqref{prop5-int-1}, we take the limit of $b\to\hat{\rho}_1$ in \eqref{p5-step-3}, resulting in
    \begin{dmath}\label{eq:prop5-limit-1}
        \lim_{b \to \hat{\rho}_1}\frac{\phi(\hat{\rho}_1)-\phi(b)}{\psi(b)-\psi(\hat{\rho}_1)}=0,
    \end{dmath}
    directly following from the application of L'H\^{o}pital's rule with $$\lim_{b \to \hat{\rho}_1}\phi'(b)=0,$$ by definition of the maximum in the numerator, and $$\lim_{b \to \hat{\rho}_1}\psi'(b)>0$$ in the denominator, by the fact that $\psi$ is strictly increasing by $\mathcal{Q}$ together with Lemma~\ref{lemma_prizescon}.
    
    For the case of \eqref{prop5-int-2}, we take the limit of $b\to \infty$. Note that
    \begin{dmath}\label{eq:prop5-limit-2}
        \lim_{b \to \infty}\frac{\phi(\hat{\rho}_1)-\phi(b)}{\psi(b)-\psi(\hat{\rho}_1)}>\lim_{b \to \infty}\frac{\phi(\hat{\rho}_1)-\phi(b)}{\psi(b)},
    \end{dmath}
    by the fact that $\psi$ is strictly increasing. Hence for \eqref{prop5-int-2}, we only need to establish
    \begin{dmath}\label{p5-step-1}
        \lim_{b \to \infty}\frac{\phi(\hat{\rho}_1)-\phi(b)}{\psi(b)}\geq1.
    \end{dmath}
    Note that $$\lim_{b \to \infty}\phi(b)=0,$$ by the fact that $P^1(r)/s$ is bounded and $(1-q(\bm{\theta},r))$ converges to zero. Furthermore, note that $$\lim_{b \to \infty}\psi(b)=\lim_{b \to \infty}\frac{P^2(b)}{s}.$$ Hence, \eqref{p5-step-1} simplifies to
    \begin{dmath}
        \phi(\hat{\rho}_1)\geq\lim_{b \to \infty}\frac{P^2(b)}{s}.
    \end{dmath}
    But since $\phi(\hat{\rho}_1(\theta_{2}),\theta_{2})$ converges to $$\lim_{\rho_1 \to \infty} \frac{1}{2} \frac{P^1(\rho_1)}{s}$$ for diminishing type spreads $\bm{\theta}$ by (C3), it is sufficient to show that
    \begin{dmath}\label{p5-step-2}
        \lim_{\rho_1 \to \infty}\frac{1}{2}\frac{P^1(\rho_1)}{s}\geq\lim_{b \to \infty}\frac{P^2(b)}{s}.
    \end{dmath}
    Taking this limit for the demand-side expectations \eqref{cons-exp1_equ_dr} gives
    \begin{dgroup*}\begin{dmath}
            \lim_{\rho_1 \to \infty}\Lambda^1(\rho_1)\hiderel{=}\FO \text{ and }
            \lim_{\rho_1 \to \infty}\Lambda^2(\rho_1)\hiderel{=}\SO,
        \end{dmath}\end{dgroup*}
    by assumptions $\mathcal{Q}$ and the definition of order statistics. Inserting these into prizes \eqref{ex-prizes1_equ} simplifies \eqref{p5-step-2} to
    \begin{dmath}
        \frac{2\left(\FO\right)^2\left(\FO-\SO\right)}{\left(\SO-4\FO\right)^2}\geq\frac{\FO\SO\left(\FO-\SO\right)}{\left(\SO-4\FO\right)^2},
    \end{dmath}
    which is true by the fact that $\FO>\SO>0$, completing the proof.
\end{proof}

\begin{proof}[\textnormal{\textbf{Proof of Proposition~\ref{no-cartel_prp}}}]\hspace{1mm}\\
    The maximization of joint firm (cartel) benefits over costs solves
    \begin{dmath}
        \underset{\rho_1,\rho_2}{\max\;}
        q_1(\bm{\theta},r)\left(P^1(r) + P^2(r)\right)+(1-q_1(\bm{\theta},r))\left(P^1(r) + P^2(r)\right) - c(|\rho_1|) - c(|\rho_2|)
    \end{dmath}
    which simplifies, using (Q1), to
    \begin{dmath}\label{soc-max_equ}
        \underset{\rho_1,\rho_2}{\max\;}P^1(r) + P^2(r)- c(|\rho_1|) - c(|\rho_2|).
    \end{dmath}
    The first-order conditions of this problem are
    \begin{dmath}\label{eq_pwfoc_rho}
        {P^1}'(r) + {P^2}'(r) \hiderel{=} c'(|\rho_1|),\
        {P^1}'(r) + {P^2}'(r) \hiderel{=} c'(|\rho_2|).
    \end{dmath}
    The implied symmetry in costs simplifies the analysis since in a joint producer utility maximum \eqref{soc-max_equ} it must be the case that $\rho_1=\rho_2$. We therefore rewrite \eqref{soc-max_equ} as the simplified symmetric problem
    \begin{dmath}\label{eq:pwelfare-max}
        \underset{r}{\max\;} W_P(r)=P^1(r) + P^2(r)- 2 c\left(|r/2|\right)
    \end{dmath}
    with the first-order condition
    \begin{dmath}\label{eq:pwelfare-foc}
        {P^1}'(r) + {P^2}'(r) = c'\left(|r/2|\right).
    \end{dmath}
    Since information emissions differ in equilibrium, \eqref{remark_3_equ}, strict convexity of $c(|\cdot|)$ implies that the right-hand side of \eqref{eq:pwelfare-foc} is strictly smaller than the right-hand side of \eqref{lemma_1_equ} for non-zero $r$.
\end{proof}

\begin{proof}[\textnormal{\textbf{Proof of Proposition~\ref{consumer-welfare_prp}}}]\hspace{1mm}\\
    Consider the welfare of served consumer segments $W_H(r),W_L(r)$ in \eqref{def:consumer-welfare-segments}. We first show that equilibrium pricing strategies $p^k(r)$, $k=1,2$, are both strictly increasing in $r$. To see this, take equilibrium prices \eqref{ex-prices_equ} and apply Lemma~\ref{lemma_consumerexp} to obtain
    \begin{dmath}
        p^1(r)^\ast \hiderel{=} 2s\dfrac{\Lambda^1(r)(2\Lambda^1(r)-\hat{\theta}) }{5\Lambda^1(r)-\hat{\theta}},\
        p^1(r)^\ast \hiderel{=} s\dfrac{(2\Lambda^1(r)-\hat{\theta})(\hat{\theta}-\Lambda^1(r))}{5\Lambda^1(r)-\hat{\theta}}.
    \end{dmath}
    Taking the derivative of $p^1(r)^\ast$ with respect to $r$ gives
    \begin{dmath}
        \frac{\partial p^1(r)^\ast}{\partial r}=2s\dfrac{\left(\hat{\theta}^2-4\hat{\theta}\Lambda^1(r)+10\Lambda^1(r)^2\right){\Lambda^1}'(r)}{(\hat{\theta}-5\Lambda^1(r))^2}
    \end{dmath}
    which is positive by the fact that ${\Lambda^1}'(r)>0$ and $2\Lambda^1(r)>\hat{\theta}$, as a consequence of Lemma~\ref{lemma_consumerexp}.
    Taking the derivative of $p^1(r)^\ast$ with respect to $r$ gives
    \begin{dmath}
        \frac{\partial p^1(r)^\ast}{\partial r}=2s\dfrac{\left(\hat{\theta}^2+2\hat{\theta}\Lambda^1(r)-5\Lambda^1(r)^2\right){\Lambda^1}'(r)}{(\hat{\theta}-5\Lambda^1(r))^2}.
    \end{dmath}
    The same reasoning as above establishes the inequality
    \begin{dmath}
        \hat{\theta}\left(\hat{\theta}+2\Lambda^1(r)\right)>5\Lambda^1(r)^2
    \end{dmath}
    which is true by the fact that \eqref{ass1} together with Lemma~\ref{lemma_spacings} ensure $3\Lambda^1(r)\leq \hat{\theta}$.
    
    We proceed to show that the served consumer segments' welfare, defined over the equilibrium cutoff vector \eqref{uniform_cutoffs_equ}, also decrease in $r$.  For the high type segment, $W_H(r)$ from \eqref{def:consumer-welfare-segments}, the argument follows by the fact that the upper bound, $\hat{\mu}_1^0(r)=s$, is constant and the lower bound $\hat{\mu}_1^2(r)$ increases in $r$. Differentiating $\hat{\mu}_1^2(r)$ with respect to $r$ gives
    \begin{dmath}
        \frac{\partial \hat{\mu}_1^2(r)}{\partial r}=2s\dfrac{\hat{\theta}{\Lambda^1}'(r)}{(\hat{\theta}-5\Lambda^1(r))^2}
    \end{dmath}
    which is positive by the fact that ${\Lambda^1}'(r)>0$ and $2\Lambda^1(r)>\hat{\theta}$, as a consequence of Lemma~\ref{lemma_consumerexp}.
    The argument for $W_L(r)$ follows by the fact that ${\hat{\mu}_1^2}'(r)<{\hat{\mu}_2^3}'(r)$. Taking derivatives of ${\hat{\mu}_1^2}(r)$, ${\hat{\mu}_2^3}(r)$ with respect to $r$ gives:
    \begin{dmath}
        2s\dfrac{\hat{\theta}{\Lambda^1}'(r)}{(\hat{\theta}-5\Lambda^1(r))^2}<3s\dfrac{\hat{\theta}{\Lambda^1}'(r)}{(\hat{\theta}-5\Lambda^1(r))^2}
    \end{dmath}
    which is always true.
\end{proof}

\begin{proof}[\textnormal{\textbf{Proof of Proposition~\ref{total-welfare_prp}}}]\hspace{1mm}\\
    Computing the integrals of served consumer welfare \eqref{def:consumer-welfare} results in
    \begin{dmath}
        W_H(r) + W_L(r) = \dfrac{s \Lambda^1(r)^2\left(4\Lambda^1(r)+5\Lambda^2(r)\right)}{2\left(\Lambda^2(r)-4\Lambda^1(r)\right)^2}.
    \end{dmath}
    Using $\mathcal{Q}$, the firms' utilities $u_1+u_2$ simplify to
    \begin{dmath}
        P^1(r)+P^2(r)-c(|\rho_1|)-c(|\rho_2|).
    \end{dmath}
    Adding the above two expressions to obtain total welfare \eqref{def:total-welfare}, inserting contest prizes \eqref{ex-prizes1_equ}, and substituting $\Lambda^2=\hat{\theta}-\Lambda^1$ yields
    \begin{dmath}\label{total-welfare_prp-proof-step1}
        W(r) = \dfrac{s \Lambda^1(r)\left(11\Lambda^1(r)^2+3\Lambda^1(r)\hat{\theta}-2\hat{\theta}^2\right)}{2(\hat{\theta}-5\Lambda^1(r))^2}-c(|\rho_1|)-c(|\rho_2|).
    \end{dmath}
    We now demonstrate that both partial derivatives of \eqref{total-welfare_prp-proof-step1}, evaluated at $\rho_1=\rho_2=0=r$, are positive. Evaluating the first derivative
    \begin{dmath}
        \frac{\partial W(r)}{\partial \rho_1} = -s\dfrac{2\hat{\theta}^3+4\hat{\theta}^2\Lambda^1(r)-33\hat{\theta}\Lambda^1(r)^2+55\Lambda^1(r)}{2(\hat{\theta}-5\Lambda^1(r))^3}-c'(\rho_1)
    \end{dmath}
    at $\rho_1=\rho_2=0=r$, using the fact that $\Lambda^1(0)=\hat{\theta}/2$ by Lemma~\ref{lemma_consumerexp}, simplifies the derivative to
    \begin{dmath}
        \dfrac{7}{18}s {\Lambda^1}'(0)-c'(0)
    \end{dmath}
    which is positive by the fact that $s>0$, $c'(0)=0$, and ${\Lambda^1}'(0)>0$ by \eqref{lemma-1-4}.
    Since the derivative with respect to $\rho_2$ enters \eqref{total-welfare_prp-proof-step1} in exactly the same way as that for $\rho_1$, the same argument applies to firm 2, completing the proof.
\end{proof}

\begin{proof}[\textnormal{\textbf{Proof of Proposition~\ref{regulatory-welfare_prp}}}]\hspace{1mm}\\
    Regulating the lower-ranked firm to marginal cost, $\hat{p}_2=0$, simplifies the cutoff-vector \eqref{uniform_cutoffs_equ} to
    \begin{dmath}%\label{sep-types_equ2}
        \hat{\mu} \hiderel{=} \biggl(\hat{\mu}_1^0 \hiderel{=} s,\
        \hat{\mu}_{2}^1 \hiderel{=} \frac{p^1}{\Lambda^1(r) - \Lambda^2(r)},\
        \hat{\mu}_{3}^2 \hiderel{=}0\biggr),
    \end{dmath}
    resulting in the fully-served market welfare with consumer segments $W_H$ and $W_L$ from \eqref{def:consumer-welfare-segments} of
    \begin{dmath}
        W_H(r) \hiderel{=} \dint_{\hat{\mu}_2^1}^{s} \left(\mu \Lambda^1(r) - p^1\right) dG(\mu),\
        W_L(r) \hiderel{=} \dint_{0}^{\hat{\mu}_2^1} \mu \Lambda^2(r) dG(\mu)
    \end{dmath}
    summing to total consumer
    \begin{dmath}
        W_H(r)+W_L(r)=\frac{1}{2}\left(s\Lambda^1(r)+p^1\left(\frac{p^1}{s\left(\Lambda^1(r)-\Lambda^2(r)\right)}-2\right)\right).
    \end{dmath}
    Equating this with the unlabeled market benchmark $W_U$ from \eqref{def:unlabeled-welfare} and solving for $p^1$ yields $\bar{p}_1$ as \eqref{regulatory-welfare_equ} in which the claimed inequalities $0<\bar{p}_1(r)<p^1(r)^\ast$ follow from $\Exp[\Theta]=\left(\Lambda^1(r)+\Lambda^2(r)\right)/2$ and $\Lambda^1(r)>\Exp[\Theta]>\Lambda^2(r) \forall r>0$ (Lemma~\ref{lemma_consumerexp}). Since the regulator knows aggregate information, $r$, emitted under any realization of $\bm{\theta}$, the announced price cap $\bar{p}_1$ is a simple number.  Finally, as the regulator announces prices before the firms choose their individual information emission, $\rho_i$, non-negative firm participation utility is ensured.
\end{proof}

\begin{proof}[\textnormal{\textbf{Proof of Proposition~\ref{existence_p1}}}]\hspace{1mm}\\
    As in the previous proofs, we drop the ranking-subscripts, writing $q_1=q$ and $q_2=1-q$, and abuse notation by writing $q'(\bm{\theta},\rho_1+\rho_2)$ and ${P^i}'(r,s) = \partial P^i(r,s)/\partial r$ for the partial derivatives with respect to $\rho_1$, $\rho_2$, and $r$.  We first show that firm 1's benefit function, defined as
    \begin{dmath}\label{f1-benefits}
        {\phi_1}(\rho_1)=q(\bm{\theta},\rho_1+\rho_2^\ast) P^1(\rho_1+\rho_2^\ast) + \left(1-q(\bm{\theta},\rho_1+\rho_2^\ast)\right) P^2(\rho_1+\rho_2^\ast),
    \end{dmath}
    is strictly increasing in $r$. Taking the derivative with respect to $r$ gives
    \begin{dmath}\label{f1p-benefits}
        \frac{\partial{\phi_1}(\rho_1)}{\partial \rho_1}=q(\bm{\theta},\rho_1+\rho_2^\ast) {P^1}'(\rho_1+\rho_2^\ast) + \left(1-q(\bm{\theta},\rho_1+\rho_2^\ast)\right) {P^2}'(\rho_1+\rho_2^\ast)+q'(\bm{\theta},\rho_1+\rho_2^\ast)\left(P^1(r)-P^2(r)\right)\hiderel{>}0,
    \end{dmath}
    in which the inequality follows from ${P^1}'(r)>{P^2}'(r)$ and $P^1(r)>P^2(r)$, for any $r>0$, as a consequence of Lemma~\ref{lemma_prizescon}, together with assumptions $\mathcal{Q}$.
    
    Next, we show that player 1's objective \eqref{gen-max3_equ} is concave in $\rho_1$, for any given $\rho_2^\ast$.  Recall that \eqref{remark_3_equ} establishes $\rho_1\geq|\rho_2|$, yielding $r\geq 0$ in equilibrium. By definition, $-c(|\rho_1|)$ is strictly concave. Since the sum of two concave functions is again concave, it is sufficient to establish concavity of firm 1's benefit function \eqref{f1-benefits} on the interval $r\in[0,\infty)$.
    Arbitrarily fixing $\rho_2^\ast \in \mathds{R}$, this is the case iff:
    \begin{dmath}\label{eq:concavity}
        \lambda {\phi_1}(u)+\left(1-\lambda\right){\phi_1}(v) \leq {\phi_1}(\lambda u +\left(1-\lambda\right)v)\condition{for $u,v\in [0,\infty)$, $\lambda\in [0,1]$}.
    \end{dmath}
    We use the fact that the consumers' expectations sum up to a constant---Lemma~\ref{lemma_consumerexp}, \eqref{lemma-1-1}---and replace  $\Lambda^2(r)=\hat{\theta}-\Lambda^1(r)$. For any $\theta_1 \geq \theta_2$, (Q6) together with property \eqref{lemma-1-4} establishes concavity of $\Lambda^1(r)$, i.e.,
    \begin{dmath}\label{iq:Lambda}
        \lambda \Lambda^1(u)+\left(1-\lambda\right)\Lambda^1(v)\leq\Lambda^1(\lambda u+\left(1-\lambda\right)v).
    \end{dmath}
    We redefine contest prizes as functions of the consumers' expectation $\Lambda^1(r)$ on the basis of \eqref{ex-prizes0_equ}:
    \begin{dmath}\label{def:P-unif}
        P^1(\Lambda^1(r)) \hiderel{=} 4s\dfrac{\Lambda^1(r)^2(2\Lambda^1(r)-\hat{\theta})}{(\hat{\theta}-5\Lambda^1(r))^2},\ P^2(\Lambda^1(r)) \hiderel{=} s\dfrac{(\hat{\theta}-\Lambda^1(r))\Lambda^1(r)(2\Lambda^1(r)-\hat{\theta})}{(\hat{\theta}-5\Lambda^1(r))^2}.
    \end{dmath}
    As $r>0$ and $\Lambda^1$ is bounded by $\hat{\theta}/2<\Lambda^1(r)<2\hat{\theta}/3$, in which the lower bound follows from \eqref{lemma-1-2} and the upper bound is a consequence of \eqref{ass1}, obtained by repeating steps \eqref{lemma-3-eq1}--\eqref{lemma-3-eq2}. Differentiation with respect to $\Lambda^1(r) \in [\hat{\theta}/2,2\hat{\theta}/3)$ yields the inequalities:
    \begin{dmath}\label{iq:P1-pos}
        \frac{\partial P^1(\Lambda^1(r))}{\partial \Lambda^1(r)}=
        8 s \dfrac{\Lambda^1(r) \left(\hat{\theta}^2-3 \hat{\theta} \Lambda^1(r)+5 \Lambda^1(r)^2\right)}{(5 \Lambda^1(r)-\hat{\theta})^3}\hiderel{\geq}0,
    \end{dmath}
    \begin{dmath}\label{iq:P2-pos}
        \frac{\partial P^2(\Lambda^1(r))}{\partial \Lambda^1(r)}=
        s\dfrac{\hat{\theta}^3 - \Lambda^1(r) \left(\hat{\theta}^2-6 \hat{\theta} \Lambda^1(r)+10 \Lambda^1(r)^2\right)}{(5 \Lambda^1(r)-\hat{\theta})^3}\hiderel{\geq}0,
    \end{dmath}
    \begin{dmath}\label{iq:P-mon}
        \frac{\partial P^1(\Lambda^1(r))}{\partial \Lambda^1(r)}>\frac{\partial P^2(\Lambda^1(r))}{\partial \Lambda^1(r)}.
    \end{dmath}
    Since we have ${\Lambda^1}'(r)>0$ by \eqref{lemma-1-4} and (Q2), inequality \eqref{iq:P1-pos} follows directly by inspection of ${P^1}'(r)>0$ \eqref{lemma-3-2-eq1}, inequality \eqref{iq:P2-pos} follows by ${P^2}'(r)>0$ \eqref{lemma-3-3-eq1}, and inequality \eqref{iq:P-mon} follows by ${P^1}'(r)>{P^2}'(r)$ \eqref{lemma-3-4-eq1}. Moreover we have
    \begin{dmath}\label{iq:P1-conc}
        \frac{\partial^2 P^1(\Lambda^1(r))}{\partial \Lambda^1(r)^2}=
        -\dfrac{8 s \hat{\theta}^2(\hat{\theta}+4\Lambda^1(r))}{(\hat{\theta}-5\Lambda^1(r))^4}\hiderel{<}0
    \end{dmath}
    which follows from $s>0$, and $\hat{\theta}>\Lambda^1(r)>\hat{\theta}/2>0$ by Lemma~\ref{lemma_consumerexp}. Similarly, the same arguments establish
    \begin{dmath}\label{iq:P2-conc}
        \frac{\partial^2 P^2(\Lambda^1(r))}{\partial \Lambda^1(r)^2}=
        -\dfrac{2 s \hat{\theta}^2(7\hat{\theta}+\Lambda^1(r)^2)}{(\hat{\theta}-5\Lambda^1(r))^4}\hiderel{<}0
    \end{dmath}
    Hence, the inequalities \eqref{iq:P1-pos}--\eqref{iq:P2-conc} establish concavity of the form
    \begin{dgroup*}
        \begin{dmath}\label{iq:P}
            \lambda P^1(\Lambda^1(u))+(1-\lambda)P^1(\Lambda^1(v)) < P^1\left(\lambda \Lambda^1(u)+(1-\lambda)\Lambda^1(v)\right),
        \end{dmath}
        \begin{dmath}\label{iq:P2}
            \lambda P^2(\Lambda^1(u))+(1-\lambda)P^2(\Lambda^1(v)) < P^2\left(\lambda \Lambda^1(u)+(1-\lambda)\Lambda^1(v)\right).
        \end{dmath}
    \end{dgroup*}
    We similarly redefine firm 1's benefit function \eqref{f1-benefits} as ${\phi_1}(r,\Lambda^1(r))$. We use and establish two properties of the benefit functions. Firstly, ${\phi_1}(r,\Lambda^1(r))$ is strictly increasing in $\Lambda^1(r)$:
    \begin{dmath}\label{def:increasing-lambda}
        \frac{\partial {\phi_1}(r,\Lambda^1(r))}{\partial \Lambda^1(r)} \hiderel{=} q(\bm{\theta},r) {P^1}'(\Lambda^1(r)) + (1-q(\bm{\theta},r)){P^2}'(\Lambda^1(r)) \hiderel{>} 0
    \end{dmath}
    which is confirmed by the fact that ${P^1}'(\Lambda^1(r))>{P^2}'(\Lambda^1(r))>0$ and $\mathcal{Q}$. Secondly, ${\phi_1}(r,\Lambda^1(r))$ is strictly increasing in $r$:
    \begin{dmath}\label{def:increasing-r}
        \frac{\partial {\phi_1}(r,\Lambda^1(r))}{\partial r}= q'(\bm{\theta},r)\left(P^1(\Lambda^1(r))-P^2(\Lambda^1(r))\right)+(q(\bm{\theta},r){P^1}'(\Lambda^1(r))+(1-q(\bm{\theta},r){P^2}'(\Lambda^1(r))){\Lambda^1}'(r)\hiderel{>}0
    \end{dmath}
    which follows from \eqref{def:increasing-lambda} together with $P^1(\Lambda^1(r))>P^2(\Lambda^1(r))>0$, $\mathcal{Q}$, and the fact that ${\Lambda^1}'(r)>0$ as a consequence of Lemma~\ref{lemma_consumerexp}.
    Inserting the definition of ${\phi_1}(r,\Lambda^1(r))$ into \eqref{eq:concavity} yields on the lhs
    \begin{dmath}\label{iq:J1-1}
        \lambda q(\bm{\theta},u)P^1(\Lambda^1(u))+\left(1-\lambda\right)q(\bm{\theta},v)P^1(\Lambda^1(v))+\lambda  \left(1-q(\bm{\theta},u)\right)P^2(\Lambda^1(u))+\left(1-\lambda\right)\left(1-q(\bm{\theta},v)\right)P^2(\Lambda^1(v))
    \end{dmath}
    and on the right-hand side
    \begin{dmath}
        q(\bm{\theta},\lambda u+\left(1-\lambda\right)v)P^1\left(\Lambda^1(\lambda u+\left(1-\lambda\right)v)\right)\\ + \left(1-q(\bm{\theta},\lambda u+\left(1-\lambda\right)v)\right)P^2\left(\Lambda^1(\lambda u+\left(1-\lambda\right)v)\right).
    \end{dmath}
    We start on the right-hand side; since $q(\bm{\theta},r)$ is weakly concave for $\theta_1 \geq \theta_2$, we have
    \begin{dmath}\label{eq:concavity-q}
        \lambda q(\bm{\theta},u)+\left(1-\lambda\right)q(\bm{\theta},v)\leq q\left(\bm{\theta},(\lambda u+\left(1-\lambda\right)v)\right).
    \end{dmath}
    Combined with \eqref{def:increasing-r}, the right-hand side is thus greater than
    \begin{dmath}
        \left(\lambda q(\bm{\theta},u)+\left(1-\lambda\right)q(\bm{\theta},v)\right)P^1\left(\Lambda^1(\lambda u+\left(1-\lambda\right)v)\right)+\left(1-\lambda q(\bm{\theta},u)\left(1-\lambda\right)q(\bm{\theta},v)\right)P^2\left(\Lambda^1(\lambda u+\left(1-\lambda\right)v)\right).
    \end{dmath}
    Rearranging terms yields
    \begin{dmath}\label{iq:J1-2}
        \lambda q(\bm{\theta},u)P^1\left(\Lambda^1(\lambda u+\left(1-\lambda\right)v)\right)+ \left(1-\lambda\right)q(\bm{\theta},v)P^1\left(\Lambda^1(\lambda u+\left(1-\lambda\right)v)\right)+\lambda \left(1-q(\bm{\theta},u)\right)P^2\left(\Lambda^1(\lambda u+(1-\lambda)v)\right)+\left(1-\lambda\right)\left(1-q(\bm{\theta},v)\right)P^2\left(\Lambda^1(\lambda u+\left(1-\lambda\right)v)\right).
    \end{dmath}
    Using \eqref{iq:P} and \eqref{iq:P2} again on the right-hand side, substitution and simplification yields
    \begin{dmath}\label{iq:J1-3}
        \left(\lambda q(\bm{\theta},u)+q(\bm{\theta},v)-\lambda q(\bm{\theta},v)\right)\left(\lambda P^1(\Lambda^1(u))+\left(1-\lambda\right)P^1(\Lambda^1(v))\right)+
        \left(1-\lambda q(\bm{\theta},u)-q(\bm{\theta},v)+\lambda q(\bm{\theta},v)\right)\left(\lambda P^2(\Lambda^1(u))+\left(1-\lambda\right)P^2(\Lambda^1(v))\right).
    \end{dmath}
    \begin{dmath}\label{iq:J1-4}
        \lambda\left(1-\lambda\right)\left(q(\bm{\theta},u)-q(\bm{\theta},v)\right)
        \left(P^1(\Lambda^1(u))-P^1(\Lambda^1(v))+P^2(\Lambda^1(v))-P^2(\Lambda^1(u))\right)\leq0.
    \end{dmath}
    Assume $u<v$. Since $q(r,\theta)$ is strictly increasing in $r$ by $\mathcal{Q}$, we have $q(\bm{\theta},u)< q(\bm{\theta},v)$. Using $\lambda\in[0,1]$ and inserting definitions \eqref{def:P-unif} yields
    \begin{dmath}
        4s\dfrac{\Lambda^1(u)^2(2\Lambda^1(u)-\hat{\theta})}{(\hat{\theta}-5\Lambda^1(u))^2}+
        s\frac{(\hat{\theta}-\Lambda^1(v))\Lambda^1(v)(2\Lambda^1(v)-\hat{\theta})}{(\hat{\theta}-5\Lambda^1(v))^2}
        \geq
        4s\frac{\Lambda^1(v)^2(2\Lambda^1(v)-\hat{\theta})}{(\hat{\theta}-5\Lambda^1(v))^2}+
        s\frac{(\hat{\theta}-\Lambda^1(u))\Lambda^1(u)(2\Lambda^1(u)-\hat{\theta})}{(\hat{\theta}-5\Lambda^1(u))^2}.
    \end{dmath}
    Converting terms to a common denominator and factoring in the numerator yields
    \begin{dmath}\label{iq:J5}
        \frac{s\left(2\Lambda^1(u)(5\Lambda^1(v)-\hat{\theta})+\hat{\theta}(\hat{\theta}-2\Lambda^1(v))(\Lambda^1(u)-\Lambda^1(v))\right)}
        {(\hat{\theta}-5\Lambda^1(u))(\hat{\theta}-5\Lambda^1(v))}\geq 0.
    \end{dmath}
    Since $u> v\Rightarrow \Lambda^1(u)> \Lambda^1(v)$ and $s>0$, the above expression simplifies to
    \begin{dmath}\label{iq:J6}
        \dfrac{2\Lambda^1(u)(5\Lambda^1(v)-\hat{\theta})+\hat{\theta}(\hat{\theta}-2\Lambda^1(v))}
        {(\hat{\theta}-5\Lambda^1(u))(\hat{\theta}-5\Lambda^1(v))}\geq0.
    \end{dmath}
    Splitting the fraction and simplification gives
    \begin{dmath}\label{iq:J7}
        \dfrac{2}{5}+\frac{3\hat{\theta}^2}{5(\hat{\theta}-5\Lambda^1(u))(\hat{\theta}-5\Lambda^1(v))}\geq0.
    \end{dmath}
    Using $\hat{\theta}>\Lambda^1(u)\geq \hat{\theta}/2$ gives
    \begin{dmath}\label{iq:J8}
        2\Lambda^1(u)\geq\dfrac{\hat{\theta}(\hat{\theta}-2\Lambda^1(v))}{(\hat{\theta}-5\Lambda^1(v))}
    \end{dmath}
    which is confirmed by the fact that
    \begin{dmath}\label{iq:J9}
        2\Lambda^1(u)\geq\dfrac{2\hat{\theta}}{3}\hiderel{\geq}\dfrac{\hat{\theta}(\hat{\theta}-2\Lambda^1(v))}{(\hat{\theta}-5\Lambda^1(v))},
    \end{dmath}
    follows from $\hat{\theta}>\Lambda^1(v)\geq \hat{\theta}/2$. The case of $v<u$ is entirely symmetric and the special case of $u=v$ is immediate since $u=v\Rightarrow (q(\bm{\theta},u)-q(\bm{\theta},v))=0$ in \eqref{iq:J1-4}, completing the argument.%\qedhere
\end{proof}

\begin{proof}[\textnormal{\textbf{Proof of Proposition~\ref{existence_p2}}}]\hspace{1mm}\\
    We want to proof that a sufficient condition for firm 2's equilibrium existence is that 
    \begin{dmath*}
        r^{\ast}(s,\delta)<r^{su}(\bm{\theta})
    \end{dmath*}
    where $r^{\ast}(s,\delta)$ is determined by equation \eqref{lemma_1_equ} and $r^{su}(\bm{\theta})$ is determined by equation
    \begin{dmath}\label{eq:su_condition_eq}
        h(\bm{\theta},r)=\frac{1}{r}.
    \end{dmath}
    
    As in the previous proofs, we drop the subscripts on rankings, writing $q_1=q$ and $q_2=1-q$, and abuse notation by writing $q'(\bm{\theta},\rho_1+\rho_2)$ for the partial derivatives with respect to $r$. Firm 2's objective is given by:
    \begin{dmath}
        u_2(\bm{\theta},\rho_1^\ast+\rho_2) = (1-q(\bm{\theta},\rho_1^\ast+\rho_2))P^1(\rho_1^\ast+\rho_2) + q(\bm{\theta},\rho_1^\ast+\rho_2)P^2(\rho_1^\ast+\rho_2)-c(|\rho_2|)
    \end{dmath}
    in which $\rho_1^\ast$ denotes the optimal response of firm 1 from Proposition~\ref{existence_p1}.
    Following \cite{Quah_Strulovici12}, global strict quasi-concavity of firm 2's objective is ensured whenever
    \begin{dmath}\label{p2_mu}
        -u_2'(\bm{\theta},\rho_1^\ast+\rho_2)=-\frac{\partial u_2(\bm{\theta},\rho_1^\ast+\rho_2)}{\partial \rho_2} = -\left(1-q(\bm{\theta},\rho_1^\ast+\rho_2)\right) {P^1}'(\rho_1^\ast+\rho_2) - q(\bm{\theta},\rho_1^\ast+\rho_2){P^2}'(\rho_1^\ast+\rho_2)\\ + q'(\bm{\theta},\rho_1^\ast+\rho_2)P^1(\rho_1^\ast+\rho_2) - q'(\bm{\theta},r)P^2(\rho_1^\ast+\rho_2) + c'(|\rho_2|)
    \end{dmath}
    satisfies the strict single crossing (SC) condition in the first derivative:
    \begin{dmath*}
        -u_2'(\bm{\theta},\rho_1^\ast+\rho_2')>0\hiderel{\Rightarrow}-u'(\bm{\theta},\rho_1^\ast+\rho_2'')\hiderel{>}0 \text{ whenever } \rho_2'' \hiderel{>} \rho_2'.
    \end{dmath*}
    Recall again that \eqref{remark_3_equ} establishes $\rho_1^\ast\geq|\rho_2^\ast|$, and thus $r^\ast>0$ and $|\rho_2^\ast|\leq r^\ast/2$. For arbitrary but fixed $\rho_1^\ast$, it is thus sufficient to establish single crossing on the domain $0\leq r\leq r^\ast$ under the restriction that $|\rho_2|<r/2$. I.e., we check single crossing of the function
    \begin{dmath}
        -u_2'(\bm{\theta},r)= -\left(1-q(\bm{\theta},r)\right) {P^1}'(r) - q(\bm{\theta},r){P^2}'(r)\\ + q'(\bm{\theta},r)P^1(r) - q'(\bm{\theta},r)P^2(r) + c'(|\rho_2|)
    \end{dmath}
    Furthermore we consider the class of quadratic costs $c(\rho_i)=\delta \rho_i^2$, with $\delta>0$ obtained from Corollary~\ref{cor:quadratic}. 
    The proof builds upon five lemmas:
    \begin{enumerate}
        \item \textbf{Lemma A:} In the auxiliary Lemma A we establish that the inequality
              \begin{dmath} 
                  \frac{c''(|\rho_2|)}{c'(|\rho_2|)}\geq \frac{1}{r}\geq \frac{q'(\bm{\theta},r)}{q(\bm{\theta},r)}
              \end{dmath}
              holds for any $|\rho_2|<r/2$ the class of quadratic costs and the class of ranking functions (Q).
        \item \textbf{Lemma B:} In the auxiliary Lemma B we establish that the inequality
              \begin{dmath}
                  \frac{1}{r} \leq \frac{{P^1}'(r)-{P^2}'(r)}{{P^1}(r)-{P^2}(r)}-\frac{{P^1}''(r)-{P^2}''(r)}{{P^1}'(r)-{P^2}'(r)}.
              \end{dmath}
              holds for any $r>0$.
        \item \textbf{Lemma C:} In the auxiliary Lemma C we establish that the regularity condition \eqref{eq:IHR} implies that:
              \begin{enumerate}
                  \item The ranking's rate $h(\bm{\theta},r)$ is increasing in $r$, i.e.
                        \begin{dmath}
                            \frac{\partial h(\bm{\theta},r)}{\partial r}>0
                        \end{dmath}
                  \item The inequality
                        \begin{dmath}\label{eq:Lemma-C-2}
                            s(\bm{\theta},r)-h(\bm{\theta},r)\leq \frac{q''(\bm{\theta},r)}{q'(\bm{\theta},r)}
                        \end{dmath}
                        holds for any $r>0$.
              \end{enumerate}
        \item \textbf{Lemma D:} In this Lemma, we show that a sufficient condition for for quasi-concavity of firm 2's objective, is given by the inequality
              \begin{dmath}\label{eq:su_condition_1}
                  h(r)\leq \frac{1}{r}
              \end{dmath}
              We establish this claim using Lemma~1 in \cite{Quah_Strulovici12}, together with Lemma~A,B and Lemma~C.
              \item\textbf{Lemma E:} In this Lemma we show that there exists a unique solution $r^{su}$ to the equation \eqref{eq:su_condition_eq}. Furthermore, we establish that this solution, $r^{su}$, is increasing for diminishing type spreads $\bm{\theta}$.
    \end{enumerate}
    
    \begin{proof}[\textnormal{\textbf{Proof of Lemma A}}]\hspace{1mm}\\
        We want to prove the inequality 
        \begin{dmath} 
            \frac{c''(|\rho_2|)}{c'(|\rho_2|)}\geq \frac{1}{r}\geq \frac{q'(\bm{\theta},r)}{q(\bm{\theta},r)}
        \end{dmath}
        for any $|\rho_2|<r/2$, using the quadratic cost function: $c(r) = \delta r^2$, with parameter $\delta > 0$. Taking derivatives on the lhs we obtain,
        \begin{dmath*}
            \frac{c''(|\rho_2|)}{c'(|\rho_2|)} = \frac{1}{ |\rho_2|} > \frac{1}{r},
        \end{dmath*}
        where the last inequality follows by $|\rho_2|<r/2$. 
        Now, take
        \begin{dmath*}
            \frac{1}{r}\geq \frac{q'(\bm{\theta},r)}{q(\bm{\theta},r)}     \end{dmath*}
        By $q(\bm{\theta},r) > 1/2 > 0$, we can multiply by $r q(\bm{\theta},r)$ without changing the inequality direction:
        \begin{equation} 
            q(\bm{\theta},r) \geq r q'(\bm{\theta},r)
        \end{equation}
        Define the auxiliary function:
        \begin{equation*}
            f(r) = q(\bm{\theta},r) - r q'(\bm{\theta},r)
        \end{equation*}
        We examine the derivative of $f(r)$:
        \begin{equation*}
            f'(r) = \frac{d}{dr}(q(\bm{\theta},r) - r q'(\bm{\theta},r)) = q'(\bm{\theta},r) - (q'(\bm{\theta},r) + r q''(\bm{\theta},r)) = -r q''(\bm{\theta},r)
        \end{equation*}
        Implying that $f$ is non-decreasing on $(0, \infty)$, by the fact that $q''(\bm{\theta},r)$ and $r>0$. Consequently, for any $r > 0$, $f(r)$ must be greater than or equal to its limit as the argument approaches 0 from the right:
        \begin{equation*}
            f(r) \geq \lim_{t\to 0^+} f(t)
        \end{equation*}
        We evaluate this limit:
        \begin{equation*}
            \lim_{t\to 0^+} f(t) = \lim_{t\to 0^+} (q(\bm{\theta},t) - t q'(\bm{\theta},t)) = \lim_{t\to 0^+} q(\bm{\theta},t) - \lim_{t\to 0^+} (t q'(\bm{\theta},t)),
        \end{equation*}
        using the initial condition, $\lim_{t\to 0^+} q(\bm{\theta},t) = q(\bm{\theta},0) = 1/2$. Furthermore note that $\lim_{t\to 0^+} (t q'(\bm{\theta},t)) = 0$, by the fact that $q'(\bm{\theta},t)$ is bounded as $t \to 0^+$ as a consequence of $q$ being bounded and concave with $q'(\bm{\theta},t)>0$. Thus,
        $$\lim_{t\to 0^+} f(t) = 1/2 - 0 = 1/2$$
        Since $f(r)$ is non-decreasing, we have: $f(r) \geq 1/2$ for all  $r > 0$, establishing the claim.
    \end{proof}
    
    \begin{proof}[\textnormal{\textbf{Proof of Lemma B}}]\hspace{1mm}\\
        We prove the inequality:
        \begin{dmath} \label{eq:LemmaB_ineq}
            \frac{1}{r} \leq \frac{{P^1}'(r)-{P^2}'(r)}{{P^1}(r)-{P^2}(r)}-\frac{{P^1}''(r)-{P^2}''(r)}{{P^1}'(r)-{P^2}'(r)}
        \end{dmath}
        for $r > 0$. Define $F(r) = P^1(r) - P^2(r)$. From Lemma~\ref{lemma_prizescon}, we have for any $r>0$ that
        $F(r) > 0, F'(r) > 0, F''(r) < 0$, and $F(0) = 0$.
        So, $F(r)$ is positive, strictly increasing, strictly concave for $r>0$, with $F(0)=0$.
        Substitute $F(r)$, $F'(r)$, $F''(r)$ into \eqref{eq:LemmaB_ineq}:
        \begin{dmath}
            \frac{1}{r} \leq \frac{F'(r)}{F(r)} - \frac{F''(r)}{F'(r)}
        \end{dmath}
        Since $r, F(r), F'(r)$ are positive for $r>0$, multiply by $r F(r) F'(r)$ and rearrange:
        \begin{align*}
            F(r) F'(r) + r F(r) F''(r) & \leq r (F'(r))^2                               \\
            0                          & \leq r (F'(r))^2 - F(r) F'(r) - r F(r) F''(r)
        \end{align*}
        Let $\xi(r) = r (F'(r))^2 - F(r) F'(r) - r F(r) F''(r)$. The inequality is equivalent to $\xi(r) \ge 0$.
        Since $F(0)=0$ and $F''(r) < 0$, $F$ is strictly concave. 
        By the Mean Value Theorem, for $r>0$, $\exists c \in (0, r)$ such that
        \begin{dmath}
            F'(c) = \frac{F(r) - F(0)}{r - 0} \hiderel{=} \frac{F(r)}{r},
        \end{dmath}
        using the fact that $F(0)=0$. As $F''(r) < 0$, $F'(r)$ is strictly decreasing, so $F'(r) < F'(c)$ for $c < r$.
        Thus, $F'(r) < F(c)=\frac{F(r)}{r}$, which implies $F(r) > r F'(r)$. We use the non-strict form:
        \begin{equation} \label{eq:concave_prop}
            F(r) \ge r F'(r) \quad \text{for } r>0
        \end{equation}
        Recall $\xi(r) = r(F'(r))^2 - F(r)F'(r) - r F(r)F''(r)$.
        From \eqref{eq:concave_prop}, $F(r) \ge r F'(r)$. Since $F'(r)>0$, multiply by $-F'(r)$:
        \[
            -F(r)F'(r) \le -r (F'(r))^2
        \]
        Substitute into $\xi(r)$:
        \begin{align*}
            \xi(r) & = r(F'(r))^2 - F(r)F'(r) - r F(r)F''(r)    \\
                   & \ge r(F'(r))^2 - r(F'(r))^2 - r F(r)F''(r) \\
                   & = -r F(r)F''(r)
        \end{align*}
        Given $r>0$, $F(r)>0$, and $F''(r)<0$, the product $r F(r) F''(r)$ is negative. Hence, $-r F(r) F''(r) > 0$. Therefore, $\xi(r) \ge -r F(r) F''(r) > 0$.
    \end{proof}
    
    \begin{proof}[\textnormal{\textbf{Proof of Lemma C}}]\hspace{1mm}\\
        Take the regularity condition \eqref{eq:IHR}, 
        \begin{dmath}
            \frac{\partial h(\bm{\theta},r)}{\partial r}+\frac{\partial s(\bm{\theta},r)}{\partial r}\geq0
        \end{dmath}
        First compute 
        \begin{align*}
            h(\bm{\theta},r) & = \frac{q'(\bm{\theta},r)}{1-q(\bm{\theta},r)} \implies \frac{\partial h(\bm{\theta},r)}{\partial r} =\frac{q''(\bm{\theta},r)(1-q(\bm{\theta},r)) + (q'(\bm{\theta},r))^2}{(1-q(\bm{\theta},r))^2} \\
            s(\bm{\theta},r) & = \frac{q'(\bm{\theta},r)}{q(\bm{\theta},r)} \implies \frac{\partial s(\bm{\theta},r)}{\partial r} = \frac{q''(\bm{\theta},r)q(\bm{\theta},r) - (q'(\bm{\theta},r))^2}{(q(\bm{\theta},r))^2}
        \end{align*}
        note that $\frac{\partial s(\bm{\theta},r)}{\partial r}< 0$ follows by the fact that $q''\leq0,q'>0,q>1/2$. Hence the regularity condition implies that $\frac{\partial h(\bm{\theta},r)}{\partial r}>0$, which also establishes that (after simplifying terms):
        \begin{dmath}\label{eq:IHR2}
            -\frac{q'(\bm{\theta},r)}{(1-q(\bm{\theta},r))}< \frac{q''(\bm{\theta},r)}{q'(\bm{\theta},r)}
        \end{dmath}
        Take regularity condition \eqref{eq:IHR}, 
        \begin{dmath*}
            \frac{q''(\bm{\theta},r)(1-q(\bm{\theta},r)) + (q'(\bm{\theta},r))^2}{(1-q(\bm{\theta},r))^2} + \frac{q''(\bm{\theta},r)q(\bm{\theta},r) - (q'(\bm{\theta},r))^2}{(q(\bm{\theta},r))^2} \geq 0
        \end{dmath*}
        Since $1/2 < q < 1$, both $q^2 > 0$ and $(1-q)^2 > 0$. We can multiply by the common positive denominator $(q(\bm{\theta},r))^2(1-q(\bm{\theta},r))^2$:
        \begin{dmath*}
            \left(q''(\bm{\theta},r)(1-q(\bm{\theta},r)) + (q'(\bm{\theta},r))^2\right)(q(\bm{\theta},r))^2 + \left(q''(\bm{\theta},r)q(\bm{\theta},r) - (q'(\bm{\theta},r))^2\right)(1-q(\bm{\theta},r))^2 \geq 0
        \end{dmath*}
        Simplification yields
        \begin{dmath}\label{eq:Lemma-C-1}
            q''(\bm{\theta},r)q(\bm{\theta},r)(1-q(\bm{\theta},r)) \geq (q'(\bm{\theta},r))^2(1-2q(\bm{\theta},r)).
        \end{dmath}
        Take the target inequality \eqref{eq:Lemma-C-1}. Simplification yields:
        \begin{align*}
              & \frac{q'(\bm{\theta},r)}{q(\bm{\theta},r)}-\frac{q'(\bm{\theta},r)}{1-q(\bm{\theta},r)} \leq \frac{q''(\bm{\theta},r)}{q'(\bm{\theta},r)}    \\
            = & q'(\bm{\theta},r) \left( \frac{1}{q(\bm{\theta},r)} - \frac{1}{1-q(\bm{\theta},r)} \right) \leq \frac{q''(\bm{\theta},r)}{q'(\bm{\theta},r)} \\
            = & q'(\bm{\theta},r) \frac{1-2q(\bm{\theta},r)}{q(\bm{\theta},r)(1-q(\bm{\theta},r))}\leq \frac{q''(\bm{\theta},r)}{q'(\bm{\theta},r)}          \\
            = & q'(\bm{\theta},r))^2(1-2q(\bm{\theta},r))\leq q''(\bm{\theta},r)q(\bm{\theta},r)(1-q(\bm{\theta},r)),
        \end{align*}
        which is equivalent to \eqref{eq:Lemma-C-1}.
    \end{proof}
    
    \begin{proof}[\textnormal{\textbf{Proof of Lemma D}}]\hspace{1mm}\\
        In the following, we will use following definition of \textit{signed-ratio monotonicity} (SRM) from \cite{Quah_Strulovici12}:
        \begin{definition}[signed-ratio monotonicity]\label{def:srm}
            Two functions $f_i(r)$ and $f_j(r)$ obey signed-ratio monotonicity on $D$ if for any any $r \in D$  with $f_i(r)<0<f_j(r)$:
            \begin{itemize}
                \item [a.)]
                      \begin{dmath*}
                          -\frac{f_i(r)}{f_j(r)}\condition{ is decreasing.}
                      \end{dmath*}
                \item [b.)] for any any $r \in D$  with $f_j(r)<0<f_i(r)$:
                      \begin{dmath*}
                          -\frac{f_j(r)}{f_i(r)}\condition{ is decreasing.}
                      \end{dmath*}
            \end{itemize}
            If $f_i(r)$ and $f_j(r)$ never take opposite signs on $D$, then these conditions are vacuously true.
        \end{definition}
        We use their Lemma~1 which states that: 
        \begin{lemma*}[Lemma~1, \cite{Quah_Strulovici12}]
            Let $\mathcal{F}=\left\{f_i\right\}_{1 \leq i \leq M}$ be a family of SC functions such that any two members obey signed-ratio monotonicity. Then $\sum_{i=1}^M \alpha_i f_i$, where $\alpha_i \geq 0$ for all $i$, is a SC function.
        \end{lemma*}
        
        Define:
        \begin{equation}
            \begin{aligned}
                f_1(\rho_2) & := -\left(1 - q(\bm{\theta},r)\right){P^1}'(r), \quad     & f_2(\rho_2) & := -\,q(\bm{\theta},r){P^2}'(r), \\
                f_3(\rho_2) & := q'(\bm{\theta},r)\left({P^1}(r)-{P^2}(r)\right), \quad & f_4(\rho_2) & := c'(|\rho_2|).
            \end{aligned}
        \end{equation}
        We want to proof that $-u'(\bm{\theta},\rho_2)=\sum^{4}_{i=1}f_i$ satisfies SC by applying Lemma~1 of \cite{Quah_Strulovici12}, i.e., we verify \emph{pairwise signed-ratio monotonicity} (SRM) among $\{f_1, f_2, f_3, f_4\}$. In a first step we verify that the component functions $f_i$ are single crossing. Note that, $f_1,f_2<0$ and $f_3,f_4>0$ by the definitions of (Q), $P_i>0$, and ${P^i}'>0$, for any $r>0$. Hence, the component functions $f_i$ are trivially SC. In a next step, we need to establish SRM for $6$ pairs $(f_i,f_j)$: 
        \begin{enumerate}
            \item $(f_1,f_2)$ and $(f_3,f_4)$: By $f_1,f_2<0$ and $f_3,f_4>0$, the SRM condition is trivially satisfied for the pairs $(f_1,f_2)$ and $(f_3,f_4)$.
                  
            \item $(f_1,f_3)$: Here, $f_1<0$ and $f_3>0$. Condition (a) for SRM applies. We need to establish that
                  \begin{dmath*}
                      f_{1,3}(\rho_2) \hiderel{:=} \frac{-f_1(\rho_2)}{f_3(\rho_2)} = \frac{\left(1-q(\bm{\theta},r)\right){P^1}'(r)}{q'(\bm{\theta},r)\left({P^1}(r)-{P^2}(r)\right)},
                  \end{dmath*}
                  is strictly decreasing in $\rho_2$. Consider
                  \begin{dmath*}
                      \frac{d}{d \rho_2} \log f_{1,3}(\rho_2)=\left(-\frac{q'(\bm{\theta},r)}{1-q(\bm{\theta},r)}-\frac{q''(\bm{\theta},r)}{q'(\bm{\theta},r)} \right)+\frac{{P^1}''(r)}{{P^1}'(r)}-\frac{{P^1}'(r)-{P^2}'(r)}{{P^1}(r)-{P^2}(r)}\hiderel{\leq} 0.
                  \end{dmath*}
                  Note that the term in brackets is negative by \eqref{eq:IHR2}, the remaining terms are negative by ${P^i},{P^i}'>0$ and ${P^i}''<0$. Thus, $f_{1,3}$ is decreasing and SRM holds.
                  
            \item $(f_1,f_4)$: Here, $f_1<0$ and $f_3>0$. Condition (a) for SRM applies. We need to establish that
                  \begin{dmath*}
                      f_{1,4}(\rho_2) \hiderel{:=} \frac{-f_1(\rho_2)}{f_4(\rho_2)} = \frac{\left(1-q(\bm{\theta},r)\right){P^1}'(r)}{c'(|\rho_2|)},
                  \end{dmath*}
                  is strictly decreasing in $\rho_2$. Consider
                  \begin{dmath*}
                      \frac{d}{d \rho_2} \log f_{1,4}(\rho_2)=-\frac{q'(\bm{\theta},r)}{1-q(\bm{\theta},r)}+\frac{{P^1}''(r)}{{P^1}'(r)}-\frac{1}{2} \frac{c''(|\rho_2|)}{c'(|\rho_2|)}\hiderel{\leq} 0.
                  \end{dmath*}
                  All terms are negative by the fact that ${P^i}'>0,{P^i}''<0$, $q<1,q'>0$, and $c'>0,c''\geq0$. Thus, $f_{1,4}$ is decreasing and SRM holds.
                  
            \item $(f_2,f_3)$: Here, $f_2<0$ and $f_3>0$. Condition (a) for SRM applies, i.e., we establish that
                  \begin{dmath*}
                      f_{2,3}(\rho_2) \hiderel{:=} \frac{-f_2(\rho_2)}{f_3(\rho_2)} = \frac{q(\bm{\theta},r){P^2}'(r)}{q'(\bm{\theta},r)\left({P^1}(r)-{P^2}(r)\right)},
                  \end{dmath*}
                  is strictly decreasing in $\rho_2$. Consider
                  \begin{dmath*}
                      \frac{d}{d \rho_2} \log f_{2,3}(\rho_2)=\frac{q'(\bm{\theta},r)}{q(\bm{\theta},r)}-\frac{q''(\bm{\theta},r)}{q'(\bm{\theta},r)}+\frac{{P^2}''(r)}{{P^2}'(r)}-\frac{{P^1}'(r)-{P^2}'(r)}{{P^1}(r)-{P^2}(r)}\hiderel{\leq} 0.
                  \end{dmath*}
                  Rearranging yields
                  \begin{dmath*}
                      \frac{q'(\bm{\theta},r)}{q(\bm{\theta},r)}- \frac{q''(\bm{\theta},r)}{q'(\bm{\theta},r)}\leq \frac{{P^1}'(r)-{P^2}'(r)}{{P^1}(r)-{P^2}(r)}-\frac{{P^2}''(r)}{{P^2}'(r)}
                  \end{dmath*}
                  Note that on the rhs  
                  \begin{dmath}\label{eq:su-bound}
                      \frac{{P^1}''(r)-{P^2}''(r)}{{P^1}'(r)-{P^2}'(r)} \geq \frac{{P^2}''(r)}{{P^2}'(r)},
                  \end{dmath}
                  follows by applying definitions ${P^i}$, \eqref{ex-prizes0_equ}, and differentiating, giving in return:
                  \begin{dmath*}
                      \frac{\bar{\theta}^2 {\Lambda^1}'(r) \left(\bar{\theta }^3-6 \bar{\theta }^2 {\Lambda^1}(r)+12 \bar{\theta } {\Lambda^1}(r)^2+10 {\Lambda^1}(r)^3\right)}{\left(\bar{\theta }-5 {\Lambda^1}(r)\right) \left(\bar{\theta }^2-4 \bar{\theta } {\Lambda^1}(r)+10 {\Lambda^1}(r)^2\right) \left(\bar{\theta }^3-\bar{\theta }^2 {\Lambda^1}(r)+6 \bar{\theta } {\Lambda^1}(r)^2-10 {\Lambda^1}(r)^3\right)}\leq 0
                  \end{dmath*}
                  which, by $2{\Lambda^1}>\bar{\theta}$ and $\bar{\theta}>{\Lambda^1}$, \eqref{lemma-1-2}, simplifies to 
                  \begin{dmath*}
                      \frac{{\Lambda^1}'(r)}{\bar{\theta }^3-\bar{\theta }^2 {\Lambda^1}(r)+6 \bar{\theta } {\Lambda^1}(r)^2-10 {\Lambda^1}(r)^3}\geq 0
                  \end{dmath*}
                  which is true by $3\Lambda^1\leq 2\hat{\theta}$, \eqref{lemma-3-eq2}, and the fact that ${\Lambda^1}'>0$, \eqref{lemma-1-4}. Bounding \eqref{eq:su-bound} on the rhs yields, that a sufficient condition for SRM is:
                  \begin{dmath*}
                      \frac{q'(\bm{\theta},r)}{q(\bm{\theta},r)}- \frac{q''(\bm{\theta},r)}{q'(\bm{\theta},r)}\leq \frac{{P^1}'(r)-{P^2}'(r)}{{P^1}(r)-{P^2}(r)}-  \frac{{P^1}''(r)-{P^2}''(r)}{{P^1}'(r)-{P^2}'(r)} 
                  \end{dmath*}
                  Using Lemma~B on the rhs gives
                  \begin{dmath}\label{equ:suff1}
                      \frac{q'(\bm{\theta},r)}{q(\bm{\theta},r)}- \frac{q''(\bm{\theta},r)}{q'(\bm{\theta},r)}\leq \frac{1}{r}
                  \end{dmath}
                  Using the target inequality of Lemma~C, \eqref{eq:Lemma-C-2}, gives
                  \begin{dmath}\label{equ:suff2}
                      \frac{q'(\bm{\theta},r)}{1-q(\bm{\theta},r)}\leq \frac{1}{r}
                  \end{dmath}
                  giving the condition \eqref{eq:su_condition_1}.
            \item $(f_2,f_4)$: $f_2<0$ and $f_4>0$.  Condition (a) for SRM applies. We need to establish that
                  \begin{dmath*}
                      f_{2,4}(\rho_2) \hiderel{:=} \frac{-f_2(\rho_2)}{f_4(\rho_2)} = \frac{q(\bm{\theta},r){P^2}'(r)}{c'(|\rho_2|)},
                  \end{dmath*}
                  is strictly decreasing in $r\in D$. Consider
                  \begin{dmath*}
                      \frac{d}{d \rho_2} \log f_{2,4}(\rho_2)=\frac{q'(\bm{\theta},r)}{q(\bm{\theta},r)}+\frac{{P^2}''(r)}{{P^2}'(r)}-\frac{c''(|\rho_2|)}{c'(|\rho_2|)}\hiderel{\leq} 0.
                  \end{dmath*}
                  Rearranging terms yields
                  \begin{dmath*}
                      \frac{{P^2}''(r)}{{P^2}'(r)}\hiderel{\leq} \frac{c''(|\rho_2|)}{c'(|\rho_2|)}-\frac{q'(\bm{\theta},r)}{q(\bm{\theta},r)},
                  \end{dmath*}
                  where the lhs is negative by the fact that that ${P^i}'>0,{P^i}''<0$, and the rhs is positive by Lemma A.
        \end{enumerate}
    \end{proof}
    
    \begin{proof}[\textnormal{\textbf{Proof of Lemma E}}]\hspace{1mm}\\
        Take the equation \eqref{eq:su_condition_eq}:
        \begin{dmath}
            r \cdot h(r)=1.
        \end{dmath}
        The proof proceeds in three steps: 
        \begin{enumerate}
            \item  Note that $r \cdot h(r)$ is strictly increasing in $r$ by the fact that $h(r)>0, h'(r)>0$ and $r>0$. We want to establish existence and uniqueness of a solution $r^{su}$ to the equation \eqref{eq:su_condition_eq}. The proof follows by the intermediate value theorem. Take the limit from the right as $r$ goes to zero
                  \begin{dmath}\label{eq:su_initial}
                      \lim_{r \to 0^+} r \cdot h(r)=\lim_{r \to 0^+} r\frac{q'(\bm{\theta},r)}{1-q(\bm{\theta},r)}\hiderel{=}0
                  \end{dmath}
                  By the fact that $(1-q(\bm{\theta},0))=\frac{1}{2}$, $q'(\bm{\theta},0)$ is finite, and $r$ goes to zero. Now, take the limit as $r$ goes to $\infty$,
                  \begin{dmath*}
                      \lim_{r \to \infty} r \cdot h(r)=\infty.
                  \end{dmath*}
                  This follows by the fact that $q$ is bounded, and $h(r)$ is increasing (and potentially unbounded), and $r$ approaches $\infty$. By the fact that $r \cdot h(r)$ is continuous and increasing a unique solution $r^{su}$ to the equation \eqref{eq:su_condition_eq} must exist.
                  
            \item Denote by $r^{su}$ a solution to the equation \eqref{eq:su_condition_eq}, and fix $\theta_1$. Define $k(r,\bm{\theta})=r \cdot h(\bm{\theta},r) - 1$. We apply implicit differentiation to the equation
                  \begin{dmath*}
                      \frac{\partial}{\partial \theta_2} \left(r^{su} h(\bm{\theta},r^{su}) - 1\right) = 0
                  \end{dmath*}
                  Using the chain and product rule gives
                  \begin{dmath*}
                      \frac{\partial  r^{su}}{\partial \theta_2}h(\bm{\theta}, r^{su}) + r^{su} \frac{\partial h(\bm{\theta}, r^{su})}{\partial \theta_2} = 0
                  \end{dmath*}
                  The term $\frac{\partial  h(\bm{\theta}, r^{su})}{\partial \theta_2}$ involves the partial derivative of $h$ with respect to its second argument $\theta_2$ (direct effect) and its third argument $r$ (indirect effect, since $r^{su}$ depends on $\theta_2$):
                  \begin{dmath*}
                      \frac{\partial r^{su}}{\partial \theta_2} h(\bm{\theta}, r^{su}) + r^{su} \left( \frac{\partial h(\bm{\theta}, r^{su})}{\partial \theta_2} + \frac{\partial h(\bm{\theta}, r^{su})}{\partial r} \frac{\partial r^{su}}{\partial \theta_2} \right) = 0
                  \end{dmath*}
                  We rearrange terms to solve for $\frac{\partial r^{su}}{\partial \theta_2}$:
                  \begin{dmath*}
                      \frac{\partial r^{su}}{\partial \theta_2} \left( h(\bm{\theta}, r^{su}) + r^{su} \frac{\partial h(\bm{\theta}, r^{su}) }{\partial r}\right) + r^{su} \frac{\partial h(\bm{\theta}, r^{su})}{\partial \theta_2} = 0
                  \end{dmath*}
                  The term in the square brackets is precisely $\frac{\partial k (r, \bm{\theta})}{\partial r}$ evaluated at $r=r^{su}$.
                  \begin{dmath*}
                      \frac{\partial k (r, \bm{\theta})}{\partial r} = \frac{\partial}{\partial r} (r h(\bm{\theta}, r) - 1) \hiderel{=} h(\bm{\theta}, r) + r \frac{\partial h (\bm{\theta}, r)}{\partial r}		\end{dmath*}
                  So, the equation becomes:
                  \begin{dmath*}
                      \frac{\partial r^{su}}{\partial \theta_2} \left( \frac{\partial k(\bm{\theta}, r^{su})}{\partial r} \right) = -r^{su} \frac{\partial h(\bm{\theta}, r^{su})}{\partial \theta_2}
                  \end{dmath*}
                  For $r^{su}(\bm{\theta})$ to be a well-defined differentiable function of $\theta_2$, the Implicit Function Theorem requires that $\frac{\partial k(\bm{\theta}, r^{su})}{\partial r}\neq 0$. This satisfied, since $h(r, \bm{\theta})$ is strictly increasing in $r$. Hence,
                  \begin{dmath*}
                      \frac{\partial r^{su}}{\partial \theta_2} = - \frac{r^{su} \frac{\partial h(\bm{\theta}, r^{su})}{\partial \theta_2}}{\frac{\partial k(\bm{\theta}, r^{su})}{\partial r}}
                  \end{dmath*}
                  Since $r^{su} > 0$ (from step 1) and $\frac{\partial k (\bm{\theta}, r^{su})}{\partial r} > 0$, the sign of $\frac{\partial r^{su}}{\partial \theta_2}$ is opposite to the sign of $\frac{\partial h(\bm{\theta}, r^{su})}{\partial \theta_2}$. $\frac{\partial h(\bm{\theta}, r^{su})}{\partial \theta_2}<0$ is a direct consequence of regularity condition \eqref{eq:IHRt} and the symmetry of $q$. This establishes the claim that $\frac{\partial r^{su}}{\partial \theta_2}$ is increasing in $\theta_2$.
        \end{enumerate}
    \end{proof}
    Now, by Lemma~D, a sufficient condition for equilibrium existence is given by \eqref{eq:su_condition_1}. Note that by Lemma~E (more specifically by \eqref{eq:su_initial}), this condition is satisfied for small $r$. Moreover, Lemma~E establishes that this condition is satisfied for any $r<r^{su}$. We want to establish that the sufficient condition \eqref{eq:su_condition_1} holds for any $r<r^\ast$, where $r^\ast$ is determined by equation \eqref{lemma_1_equ}. Thus the sufficient condition holds for any $r<r^\ast$ whenever,
    \begin{dmath}\label{eq:}
        r^{\ast}(s,\delta)<r^{su}(\bm{\theta})
    \end{dmath}
    By inspection of $\eqref{lemma_1_equ}$, linearity of ${P^i}$ in $s$, and linearity of $c'(|\rho_2|)$ in $\delta$, establish the fact that $r^{\ast}(s,\delta)$ is increasing in $s$ and decreasing in $\delta$. The sufficient bound $r^{su}(\bm{\theta})$ is strictly increasing for diminishing type spreads $\bm{\theta}$ by Lemma~E. Hence the sufficient condition is satisfied for parameters $(\bm{\theta},s,\delta)$, when big type spreads, are off-set by low $s$ or big $\delta$. 
\end{proof}

%%%---------------------------------------------------------------------------------------
\singlespacing \setlength{\bibsep}{3.0pt}
\bibliographystyle{apalike}
\bibliography{MyBib.bib}

\end{document}